\documentclass[fleqn,twoside]{article}
\usepackage{amsmath,amssymb,espcrc2,color,cite}

\usepackage{graphicx,soul}
\usepackage[figuresright]{rotating}

\newcommand{\url}[1]{{\tt #1}}
\newcommand{\lsim}
{\;\raisebox{-.3em}{$\stackrel{\displaystyle <}{\sim}$}\;}

\newcommand{\gmt}{\ensuremath{(g-2)_\mu}}
\newcommand{\br}{{\rm BR}}
\newcommand{\bsg}{\ensuremath{\br(b \to s \gamma)}}
\newcommand{\btn}{\ensuremath{\br(B_u \to \tau \nu_\tau)}}

\newcommand{\bmm}{\ensuremath{\br(B_s \to \mu^+\mu^-)}}
\newcommand{\ssi}{\ensuremath{\sigma^{\rm SI}_p}}

\newcommand{\MW}{M_W}

\newcommand{\Mh}{M_h}

\newcommand{\MA}{M_A}

\newcommand{\mt}{m_t}
\newcommand{\mgl}{m_{\tilde g}}
\newcommand{\msq}{m_{\tilde q}}

\newcommand{\neu}[1]{\tilde \chi^0_{#1}}
\newcommand{\mneu}[1]{m_{\tilde \chi^0_{#1}}}

\newcommand{\mstaue}{m_{\staue}}
\newcommand{\staue}{\tilde \tau_1}

\newcommand{\tb}{\tan\beta}

\newcommand{\tev}{\,\, \mathrm{TeV}}
\newcommand{\gev}{\,\, \mathrm{GeV}}

\newcommand{\delphes}{{\it {\tt Delphes}}}
\newcommand{\atlasfive}{ATLAS 5/fb jets + $\ETslash$}
\newcommand{\lhco}{LHC$_{\rm 1/fb}$}
\newcommand{\lhcf}{LHC$_{\rm 5/fb}$}

\newcommand{\atlasf}{ATLAS$_{\rm 5/fb}$}

\newcommand{\htr}[1]{{\color{red}  #1}}

\newcommand{\ETslash}{/ \hspace{-.7em} E_T}

\graphicspath{{figs/}}

\title{\bf The CMSSM and NUHM1 in Light of 7 TeV LHC, $\mathbf{B_s \to \mu^+ \mu^-}$ and XENON100 Data \\ \vspace{0.5em}}

\author{
{\bf O.~Buchmueller}\address[Imperial]
   {High\,Energy\,Physics\,Group,\,Blackett\,Laboratory,\,Imperial\,College,\,Prince\,Consort\,Road,\,London\,SW7\,2AZ,\,UK},
{\bf R.~Cavanaugh}\address[FNAL]
   {Fermi National Accelerator Laboratory, P.O. Box 500, 
    Batavia, Illinois 60510, USA}\hbox{$^{\rm ,}$}\address[UIC]
   {Physics Department, University of Illinois at Chicago, Chicago, 
    Illinois 60607-7059, USA},
{\bf M.~Citron}\addressmark[Imperial],
{\bf A.~De Roeck}\address[CERN]
   {Physics Department, CERN, CH--1211 Gen\`eve 23, Switzerland}\hbox{$^{\rm ,}$}\address[Antwerpen]
   {Antwerp University, B--2610 Wilrijk, Belgium},
 {\bf M.J.~Dolan}\address[IPPP]
{Institute for Particle Physics
     Phenomenology,\,University\,of\,Durham,\,South 
     Road,\,Durham\,DH1\,3LE,\,UK},
{\bf J.R.~Ellis}\address[KCL]{Theoretical Particle Physics
  and Cosmology Group, Department of Physics, King's College London, London~WC2R~2LS, UK}\hbox{$^{\rm ,}$}\addressmark[CERN], 
{\bf H.~Fl\"acher}\address[Rochester]
   {H.H.~Wills Physics Laboratory, University of Bristol, Tyndall Avenue, Bristol BS8 1TL, UK},
{\bf S.~Heinemeyer}\address[Santander]
   {Instituto de F\'{\i}sica de Cantabria (CSIC-UC), 
    E--39005 Santander, Spain},
{\bf G.~Isidori}\address[Frascati]
{INFN, Laboratori Nazionali di Frascati, Via E. Fermi 40, 
I--00044 Frascati, Italy}\hbox{$^{\rm ,}$}\addressmark[CERN],
{\bf J.~Marrouche}\addressmark[Imperial],
{\bf D.~Mart\'inez~Santos}\addressmark[CERN],
{\bf S.~Nakach}\addressmark[Imperial],
{\bf K.A.~Olive}\address[Minnesota] 
{William I.\ Fine Theoretical Physics Institute, School of Physics and
 Astronomy, University of Minnesota, Minneapolis, Minnesota 55455, USA}, 
{\bf S.~Rogerson}\addressmark[Imperial],
{\bf F.J.~Ronga}\address[ETHZ]
   {Institute for Particle Physics, ETH Z\"urich, CH--8093 Z\"urich, 
   Switzerland},
{\bf K.J.~de~Vries}\addressmark[Imperial],
{\bf G.~Weiglein}\address[DESY]
   {DESY, Notkestrasse 85, D--22607 Hamburg, Germany}
}

\begin{document}

\begin{abstract}
We make a frequentist analysis of the parameter space of the CMSSM and NUHM1,
using a Markov Chain Monte Carlo (MCMC) with 95 (221) million points to
sample the CMSSM (NUHM1) parameter spaces. Our analysis
includes the ATLAS search for supersymmetric jets + $\ETslash$ 
signals using $\sim 5$/fb of LHC data at 7 TeV, which we apply using {\tt PYTHIA} and a {\tt Delphes}
implementation that we validate in the relevant parameter regions of the
CMSSM and NUHM1. Our analysis also includes the constraint imposed by
searches for \bmm\ by LHCb, CMS, ATLAS and CDF, and the
limit on spin-independent dark matter scattering from 225 live days of
XENON100 data. We assume $\Mh \sim
125 \gev$, and use a full set of electroweak precision and
other flavour-physics observables, as well as the cold dark matter density constraint. The
ATLAS$_{\rm 5/fb}$ constraint has relatively limited effects on the 68 and 95\% CL regions in the
$(m_0, m_{1/2})$ planes of the CMSSM and NUHM1. The new
\bmm\ constraint has greater impacts on these CL regions, and
also impacts significantly the  68 and 95\% CL regions in the $(\MA, \tb)$ planes of both
models, reducing the best-fit values of $\tb$. 
The recent XENON100 data eliminate the focus-point region in the CMSSM and affect the 68 and 95\% CL
regions in the NUHM1. 
In combination, these new constraints reduce the best-fit values of $m_0, m_{1/2}$ in the
CMSSM, and increase the global $\chi^2$ from 31.0 to 32.8, reducing the
$p$-value from 12\% to 8.5\%. In the case of the NUHM1, they have little effect on
the best-fit values of $m_0, m_{1/2}$, but increase the global $\chi^2$
from 28.9 to 31.3, thereby reducing the $p$-value from 15\% to 9.1\%.

\bigskip
\begin{center}
{\tt KCL-PH-TH/2012-26, LCTS/2012-13, CERN-PH-TH/2012-164, \\
DCPT/12/82, DESY 12-115, IPPP/12/41, FTPI-MINN-12/20, UMN-TH-3106/12}
\end{center}

\vspace{2.0cm}
\end{abstract}

\maketitle

\section{Introduction}
\label{sec:intro}

The LHC searches for jets + $\ETslash$ events,  as well as searches for
heavy Higgs bosons $H/A$ and the rare decay $B_s \to \mu^+ \mu^-$, are
placing ever-stronger constraints on supersymmetric models, in
particular those in which R-parity is conserved and the lightest
neutralino $\neu{1}$ is the stable lightest supersymmetric particle and
provides the cosmological dark matter. There have been a number of
analyses of the constraints imposed by ATLAS and CMS using 
$\sim 1$/fb at 7 TeV~\cite{mc7,mc75,post-LHC} 
and other data on the parameter spaces of simplified versions of the
minimal supersymmetric extension of the Standard Model (MSSM)~\cite{HK},
particularly the constrained MSSM (CMSSM)~\cite{cmssm1} in which the
soft supersymmetry-breaking parameters $m_0$, $m_{1/2}$ and $A_0$ are
universal at the GUT scale, and some also of the LHC
constraints on models in which the soft supersymmetry-breaking
contributions to the masses of the Higgs multiplets are
non-universal but equal to each other, the NUHM1 models~\cite{nuhm1}.
There have also been analyses of the CMSSM and NUHM1 of the prospective implications of a Higgs boson weighing
$\sim 125 \gev$~\cite{mc75,125-other,ATLAS5-other,Fittino5,BayesFits5}, and 
analyses within the CMSSM and
NUHM1~\cite{ATLAS5-other,Fittino5,BayesFits5,LMNW5} that incorporate
LHC searches for jets + $\ETslash$ events with $\sim 5$/fb of LHC
data at 7 TeV~\cite{ATLAS5,CMS5}. 
Here we extend our previous frequentist analyses of the CMSSM and
NUHM1~\cite{mc1,mc2,mc3,mc35,mc4,mc5,mc6,mc7,mc75} to include
the ATLAS jets + $\ETslash$ results with $\sim 5$/fb~\cite{ATLAS5} of data
as well as $\Mh \sim 125 \gev$ and recent data on \bmm\ from the ATLAS~\cite{ATLASbmm},
CDF~\cite{CDFbmm}, 
CMS~\cite{CMSbmm} and LHCb~\cite{LHCbbmm} Collaborations~\cite{LHCbmm}, and the recent CMS
exclusion of $H/A \to \mu^+ \mu^-$~\cite{CMSHA}.
We also implement the constraint on the spin-independent dark matter scattering cross
section imposed by 225 live days of XENON100 data~\cite{newXENON100}.
We give a critical discussion of the extent to which the CMSSM and NUHM1
can be said to be disfavoured by the available data from the LHC and elsewhere.

\medskip
The results of experimental searches for jets + $\ETslash$ events are typically
presented within the framework of the CMSSM for some fixed $A_0$ and
$\tb$. The applicability of these analyses to other $A_0$ and $\tb$ values,
as well as to constraining the NUHM1, requires some study and
justification. Within the CMSSM, one must verify that the constraints used are
indeed independent of $A_0$ and $\tb$, as often stated in the
experimental papers on searches for  jets + $\ETslash$ events. Similarly, 
within the NUHM1 one must verify whether, for any specific set of
values of $m_0$, $m_{1/2}$, $A_0$ and $\tb$, the sensitivities of ATLAS
and CMS searches for jets + $\ETslash$ events might depend on the degree 
of non-universality in the NUHM1. A second issue arising in the NUHM1 is that the range of
$m_0$ that is consistent with the $\neu{1}$ LSP requirement
depends on the degree of non-universality. In the CMSSM, for any given value of
$m_{1/2}$ there is a lower limit on $m_0$ that is, in general, violated
in the NUHM1. Since the LHC experiments generally quote
results only for the region of the $(m_0, m_{1/2})$ plane allowed within
the CMSSM, a dedicated study is needed to estimate the correct
extrapolation of LHC results outside this region. 

In order to explore the sensitivity to more general CMSSM and NUHM
parameters of the published LHC searches for supersymmetry within
specific CMSSM models, we use a version of the \delphes~\cite{Delphes}
generic simulation package with a `card' that emulates 
the performance of the ATLAS detector. As described below, in the
specific case of the 7 TeV ATLAS 5/fb jets + $\ETslash$ analysis,
\delphes\ reproduces reasonably accurately the quoted 95\%~CL limits on
CMSSM models with $A_0 = 0$ and $\tb = 10$, as well as other confidence
levels in the CMSSM $(m_0, m_{1/2})$ plane. These 
are well described by scaling of event numbers $\propto
1/{\cal M}^4$, where ${\cal M} \equiv \sqrt{m_0^2 + m_{1/2}^2}$, as
assumed in~\cite{mc7}. We also use the \delphes\ simulation to
confirm that the ATLAS 5/fb jets + $\ETslash$ constraint in the $(m_0,
m_{1/2})$ plane of the CMSSM is relatively insensitive to $\tb$ and $A_0$.

Having validated the application of the \delphes\ simulation
to the CMSSM, we then study whether the results of~\cite{ATLAS5} are
sensitive to the degree of Higgs non-universality in the NUHM1. Although
the supersymmetric decay cascades to which this analysis is
most sensitive are independent of the details of the supersymmetric
Higgs sector, they do in principle have some sensitivity to the Higgs
mixing parameter~$\mu$, which is in general different in the NUHM1
from its value in the CMSSM. We find that in practice the jets +
$\ETslash$ constraint 
is essentially independent of the degree of non-universality in the NUHM1. 
We then use \delphes\ to estimate the
sensitivity of the ATLAS analysis to the region with $m_0$ smaller than
is allowed in the CMSSM, finding that the ATLAS jets + $\ETslash$
constraint is essentially independent of $m_0$ in this range.

Following these validations of our implementation of \delphes\ in the
NUHM1, we use the Markov Chain Monte Carlo (MCMC) technique to sample
the CMSSM and NUHM1 parameter 
spaces with $\sim 95 \times 10^6$ and $\sim 221 \times 10^6$ points,
respectively. Given the constraints from $\Mh$ and the \lhco\ data, we find that the \atlasfive\ constraint by itself has
a relatively limited impact on the 68 and 95\% CL regions in the $(m_0, m_{1/2})$ planes of the CMSSM
and NUHM1, whereas the new \bmm\ constraint has  larger effects in both these and the $(\MA, \tb)$
planes, and the new XENON100 constraint also
impacts significantly both the $(m_0, m_{1/2})$ and $(\MA, \tb)$ planes of the CMSSM and NUHM1. 
After the inclusion of these constraints, we find that in the favoured
regions in the CMSSM and NUHM1 in $m_0$, $m_{1/2}$ and $\tb$ the global $\chi^2$ function
varies little, giving less relevance to the best-fit values of these parameters.
Nevertheless, we note that the best-fit values of $m_0$ and $m_{1/2}$ are decreased in
the CMSSM, and the best-fit values of $\tb$ are decreased in both the CMSSM and the NUHM1.
Within the CMSSM, the global $\chi^2$ increases from
31.0 to 32.8 and the $p$-value is correspondingly reduced from 
12\% to 8.5\%. 
Within the NUHM1, the value of $\tb$ rises to a value similar
to that in the CMSSM, while the total $\chi^2$ rises from 28.9 to 31.3,
reducing the 
$p$-value from 15\% to 9.1\%. The perilously low $p$-values of the CMSSM
and NUHM1 reflect the growth in tension between LHC constraints and
low-energy measurements such as \gmt~\cite{newBNL,Jegerlehner}.


\section{Analysis Procedure}

We follow closely the procedure described in~\cite{mc7}.
As in~\cite{mc7}, the sampling is based on the Metropolis-Hastings
algorithm, with a multi-dimensional Gaussian distribution as proposal density.
The width of this distribution is adjusted during the sampling, so as to
keep the MCMC acceptance rate between 20\% and 40\% in order to ensure
efficient sampling. We emphasize that we do not make use of the sampling
density to infer the underlying probability distribution. 
Rather, we use the MCMC method to construct a global likelihood function that receives contributions
from the usual full set of electroweak precision observables 
including \gmt~\cite{newBNL}, measurements of \bsg, \btn, \bmm, B-meson mixing and other
flavour-physics observables~\cite{LHCbmm,hfag}, the cosmological
$\neu{1}$ density and the XENON100 direct search 
for dark matter scattering~\cite{XE100}, as well the LHC searches for the Higgs boson
and supersymmetric signals.

Our treatment of the non-LHC constraints is similar to that
in~\cite{mc7}, with the exceptions that we use updated values of
$\MW = 80.385 \pm 0.015 \gev$~\cite{lepewwg} and 
$\mt = 173.2 \pm 0.9 \gev$~\cite{mt1732}. 
Our treatment of \bmm\ is based on a compilation of the recent
measurements by ATLAS~\cite{ATLASbmm}, CDF~\cite{CDFbmm},
CMS~\cite{CMSbmm} and LHCb~\cite{LHCbbmm,LHCbmm}.   
As already mentioned, we incorporate the public results of the search for
jets + $\ETslash$ events without leptons using $\sim 5$/fb of LHC data at 7 TeV
analyzed by ATLAS~\cite{ATLAS5}, which
has greater sensitivities to the models
discussed here than do searches with leptons and $b$-quarks in the final
state as well as searches with less integrated luminosity.
We also include the CMS constraint on
the heavier MSSM Higgs bosons, $H/A$~\cite{CMSHA}, which has greater
sensitivity than the corresponding ATLAS
search~\cite{ATLASHA}. 

Concerning the Higgs boson, the situation has evolved rapidly since our previous analysis~\cite{mc75}
and the data reported earlier this year~\cite{ATLASSMHiggs,CMSSMHiggs}, with
the recently-announced discovery by the CMS~\cite{CMS47} and ATLAS Collaborations~\cite{ATLAS47},
consistent with evidence from the CDF and D0 Collaborations~\cite{TEVH},
of a new particle that is a very strong candidate to be a Higgs boson resembling that in the Standard Model.
In the absence of an official combination of
these results, we model them by assuming, as in~\cite{mc75}, that
\begin{equation}
\Mh = 125 \pm 1 ~({\rm exp.}) \pm 1.5 ~({\rm theo.}) \gev~,
\label{Mh125}
\end{equation} 
and we incorporate this new constraint using the same `afterburner' approach 
as in~\cite{mc75}.

The contributions of these observables to the global likelihood function
are calculated within the {\tt MasterCode} framework~\cite{mcweb}. This
incorporates a code for the electroweak observables based
on~\cite{Svenetal} as well as the {\tt SoftSUSY}~\cite{Allanach:2001kg},
{\tt FeynHiggs}~\cite{FeynHiggs},  {\tt SuFla}~\cite{SuFla}, and 
{\tt MicrOMEGAs}~\cite{MicroMegas} codes, interfaced via the SUSY Les
Houches Accord~\cite{SLHA}. We use the {\tt SuperIso}~\cite{SuperIso}
and {\tt SSARD}~\cite{SSARD} codes as comparisons for B-physics
observables~%
\footnote{We do not include the isospin asymmetry in 
$B \to K^{(*)} \mu^+ \mu-$ decays~\cite{isoasymm} 
or the measurement of BR($B \to D^{(*)} \tau \nu$)~\cite{BABAR} in our
analysis, in view of the present experimental and theoretical
(long-distance) uncertainties.} 
and for the cold dark matter density, using also~{\tt DarkSUSY}~\cite{DarkSusy} as a comparison for the latter.
We use {\tt SSARD} for the spin-independent dark matter scattering cross section, comparing with
{\tt MicrOMEGAs} and {\tt DarkSUSY}.


\section{Validation and Extension of Jets + \boldmath{$\ETslash$} Constraints} 

ATLAS and CMS results on SUSY searches for jets + $\ETslash$
are typically shown as 95\% CL exclusion bounds in the CMSSM $(m_0,
m_{1/2})$ plane for fixed $\tb$ and  $A_0$, e.g., $\tb = 10$ and  $A_0 =
0$~\cite{ATLAS5,CMSsusy}. We first validate for these values of $\tb$
and  $A_0$ our implementation of the ATLAS constraint  using {\tt
  PYTHIA}~\cite{PYTHIA} and the generic {\tt Delphes}~\cite{Delphes}
simulation code with an ATLAS detector `card'. Next we extend it to
other values of the CMSSM parameters, and then to the NUHM1~%
\footnote{
We have made a similar validation analysis for the CMS $\alpha_T$ search
for jets + $\ETslash$ events using $\sim 1$/fb of data at
7~TeV~\cite{CMSsusy}. We do not discuss this validation in detail, as it
does not contribute to the  the likelihood function analyzed here, but
it does validate {\it a posteriori} our previous
treatments~\cite{mc7,mc75} of the CMS $\alpha_T$ analysis. It also
indicates that the CMS sensitivity with 5/fb of data at 7 TeV~\cite{CMS5} is
similar to the \atlasfive\ data discussed here.}.

\subsection*{\it Validation in the $(m_0, m_{1/2})$ plane}

As a first step in the validation procedure, we used 
{\tt SoftSUSY}~\cite{Allanach:2001kg} to generate the spectra for
various points in the $(m_0, m_{1/2})$ plane of the CMSSM  for $\tb =10$
and $A_0 = 0$. We then used {\tt PYTHIA}~\cite{PYTHIA} to generate
50,000 events for each of these points,  which were subsequently passed
through the {\tt Delphes}~\cite{Delphes} simulation code using the ATLAS
detector `card'.  We then used information from~\cite{ATLAS5} to
calculate the efficiencies for observing events that  would survive the
cuts used in the ATLAS jets + $\ETslash$ analysis using 5/fb of
data at 7 TeV. Taking into account the observed number of events,  together
with the information in Table~3 of~\cite{ATLAS5}, we then calculated the
confidence levels of these points, using the same prescription as
ATLAS. We recall that the ATLAS analysis combined searches using several
different channels $A, A^\prime, B, C, D$ and $E$, which, depending on
the channel, may have `tight', `medium' and `loose' event
selections. Following the procedure used in the ATLAS analysis, for each
set of CMSSM parameters we used the limit set by the most sensitive of
these various channel selections. 

\begin{figure*}[htb!]
\resizebox{8cm}{!}{\includegraphics{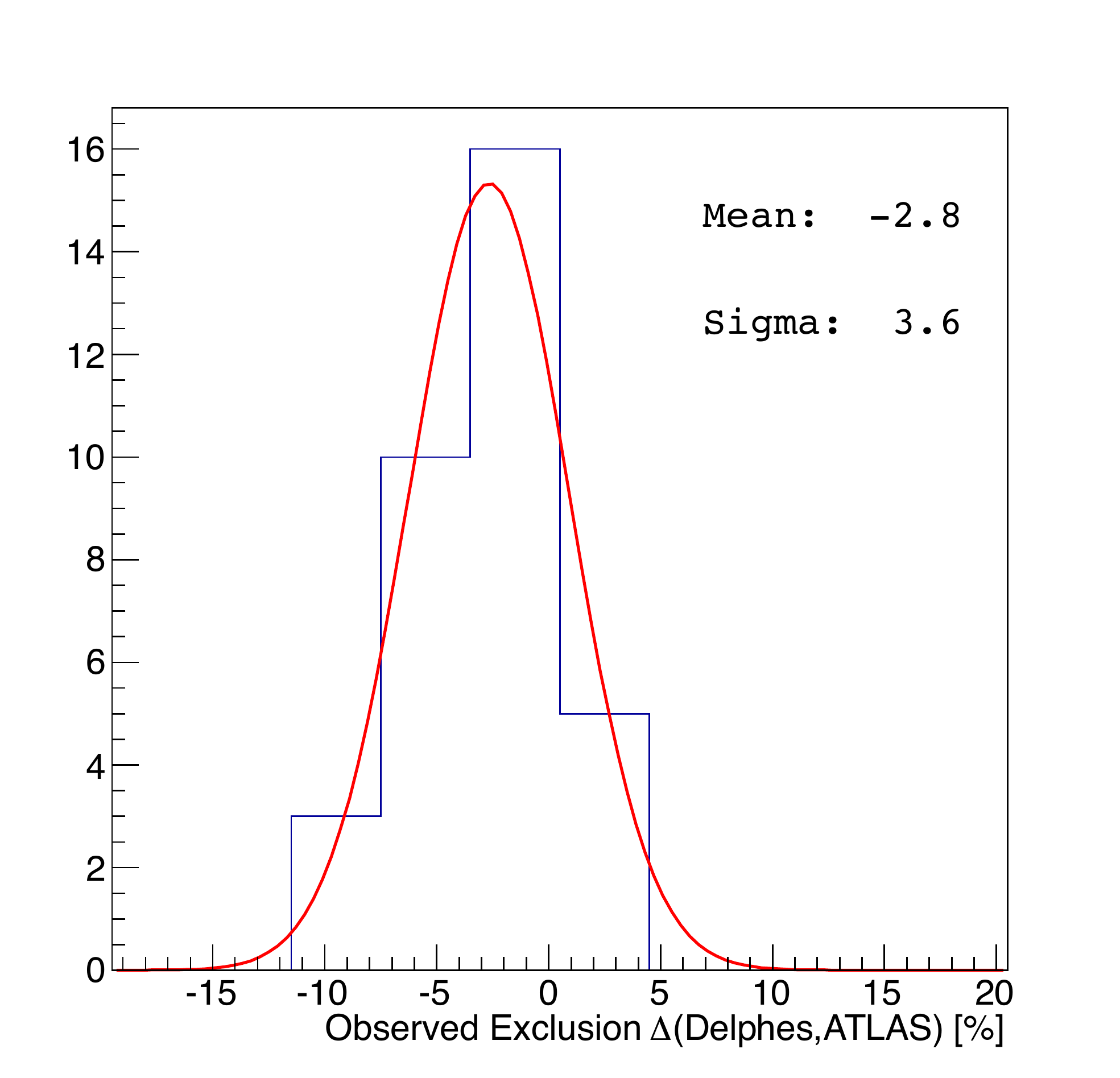}}
\resizebox{8cm}{!}{\includegraphics{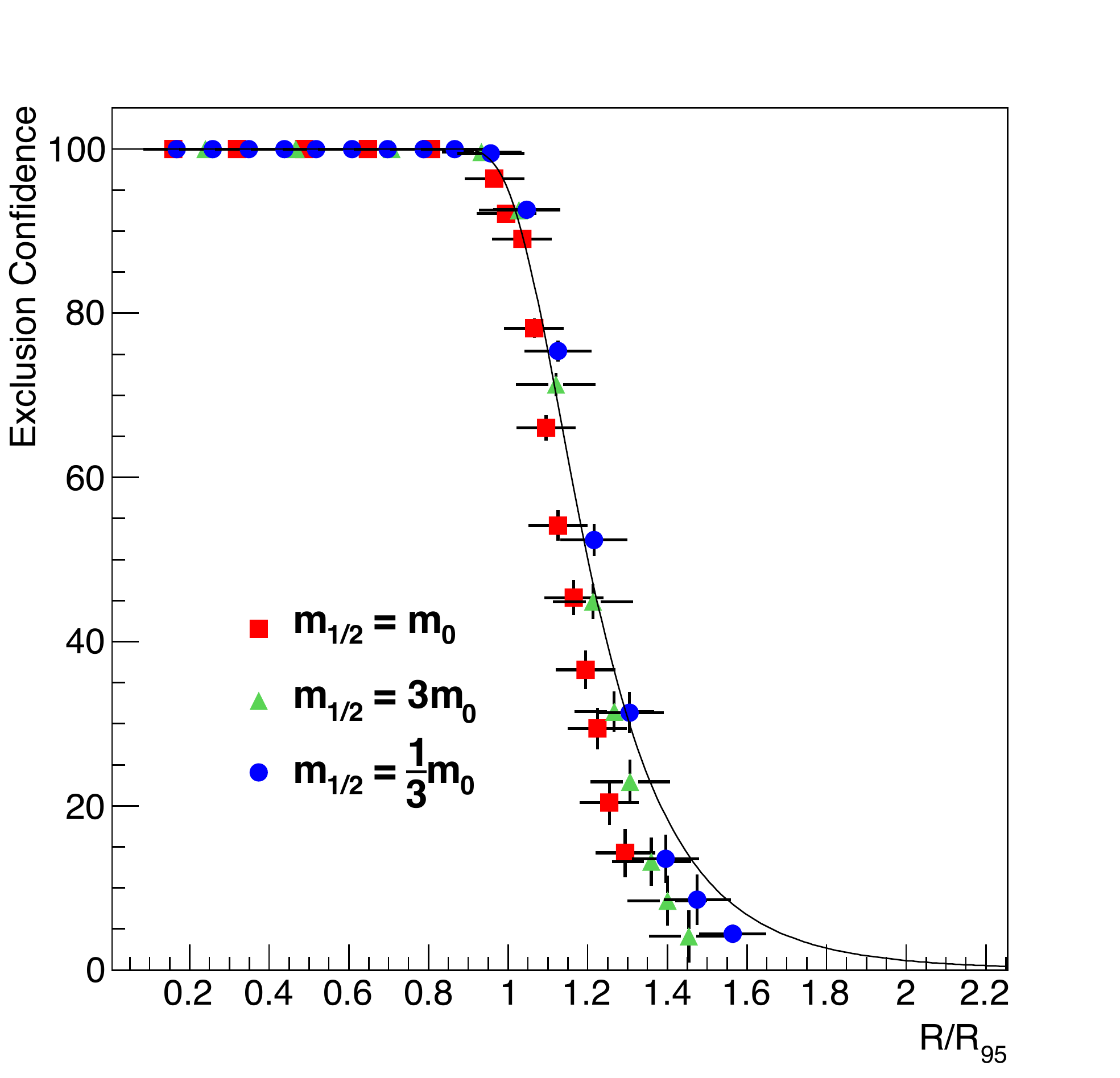}}

\caption{\it Left panel: A histogram of the differences between the
  confidence levels (CLs) we calculate using {\tt PYTHIA} and {\tt
    Delphes} from the values quoted by ATLAS for 34 points along the
  95\% exclusion contour in the $(m_0, m_{1/2})$ plane of the CMSSM with
  $\tb = 10$ and $A_0 = 0$. The mean difference is CL$_{\tt
    Delphes}$ - CL$_{\rm ATLAS} = (- 2.8 \pm 3.6)$\%. Right panel: 
  The points are the confidence levels for a selection of
  CMSSM points with $\tb = 10$ and $A_0 = 0$ and various values of $R
  \equiv {\cal M}/{\cal M}_0$, where ${\cal M} \equiv \sqrt{m_0^2 +
    m_{1/2}^2}$ and ${\cal M}_0$ is the value of ${\cal M}$ at the point
  on the LHC jets + $\ETslash$ 95\% exclusion line with the same ratio
  $m_0/m_{1/2}$. The red squares (blue circles, green triangles)
  were chosen along rays in the $(m_0, m_{1/2})$ plane with 
  $m_0/m_{1/2} = 1/3, 1, 3$, respectively.
  These points compare well with the solid line, which is the CL
  calculated assuming that the number   of signal events surviving the
  LHC cuts scales as $1/{\cal M}^4$. 
}
\label{fig:scaling}
\end{figure*}

The left panel of Fig.~\ref{fig:scaling} displays how closely this work flow 
reproduced the ATLAS CL for 34 test points along its 95\% CL exclusion
line, sampling the range $m_0 \in (200, 3500) \gev$ in $100 \gev$ steps.
Our results indicate a mean difference
CL$_{\tt Delphes}$ - CL$_{\rm ATLAS} = (- 2.8 \pm 3.6)$\%, so we conclude
that our simulation using {\tt PYTHIA} and {\tt Delphes} reproduces
quite accurately the 95\% CL exclusion contour in the  
$(m_0, m_{1/2})$ plane for $\tb = 10$ and $A_0 = 0$ reported by ATLAS.

Since it is not computationally feasible to use {\tt PYTHIA} and
\delphes\ in this way for every point generated in our MCMC analysis, we
make use of a simple analytic approximation to their results in order to
extrapolate their results to other regions of 
the $(m_0, m_{1/2})$ plane. The right panel of
Fig.~\ref{fig:scaling} compares the CL
values estimated using this \delphes\ implementation, for a number of CMSSM points with $\tb = 10$, $A_0 = 0$ and varying $m_0$
  and $m_{1/2}$, with the CL calculated
assuming, as 
in~\cite{mc7}, that the numbers of events for different points in the
CMSSM $(m_0, m_{1/2})$ plane scale as $1/{\cal M}^4$ (solid line). We see that the
agreement is quite good along rays in the $(m_0, m_{1/2})$ plane with
$m_{1/2}/m_0 = 1/3$ (blue circles), 1 (red squares) and 3 (green
triangles), indicating that this scaling law is an adequate
approximation to the true sensitivity of the ATLAS 5/fb jets +
$\ETslash$ analysis.


\subsection*{\it Extension to other $\tb$ and $A_0$ values}

\begin{figure*}[htb!]
\resizebox{8cm}{!}{\includegraphics{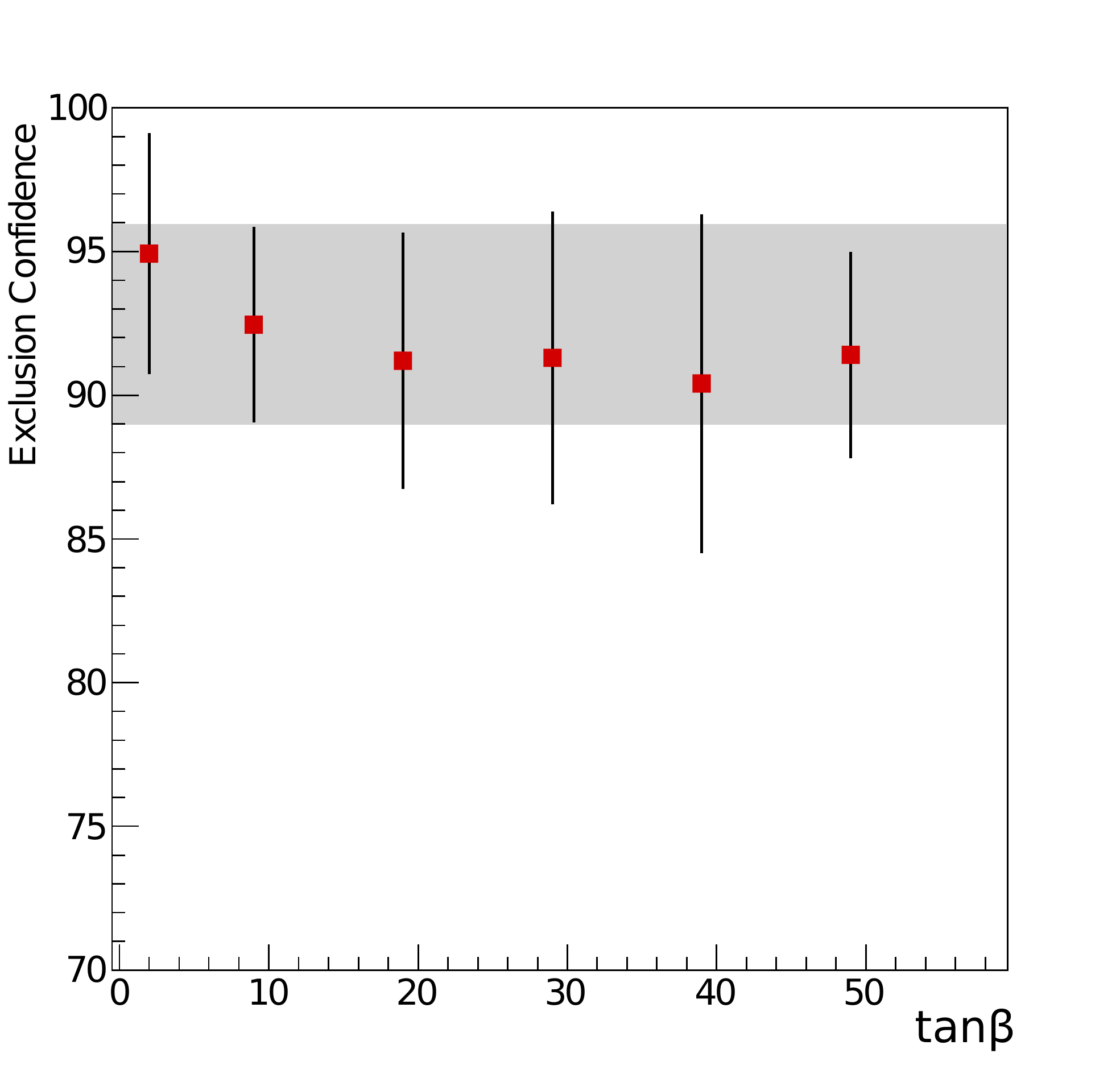}}
\resizebox{8cm}{!}{\includegraphics{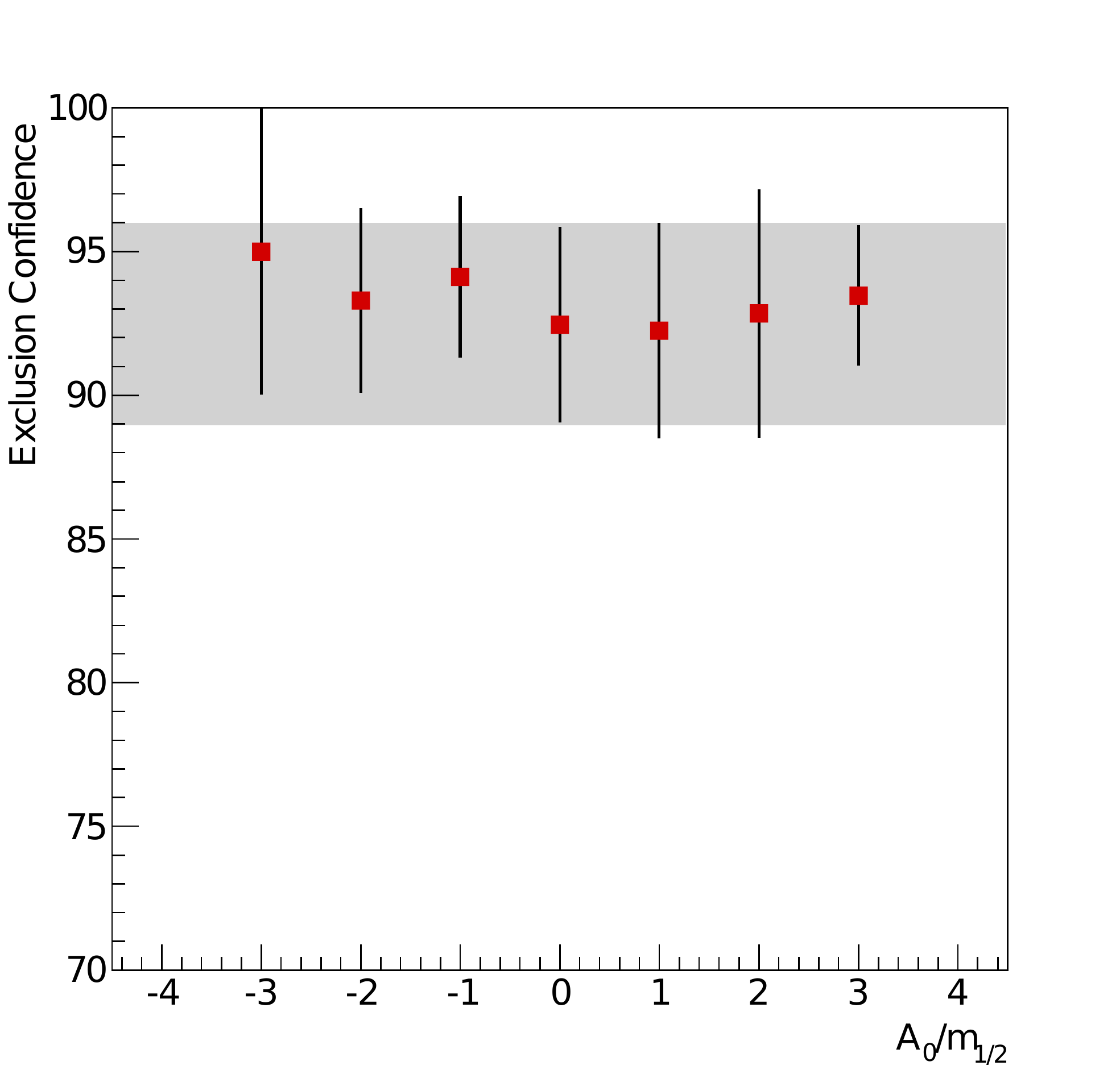}}
\caption{\it Calculations using \delphes\ of the average and spread in
  exclusion confidence levels of 34 CMSSM points selected to have the
  same $(m_0, m_{1/2})$ values as points on the 95\% ATLAS 5/fb jets +
  $\ETslash$ exclusion line  for $\tb = 10$ and $A_0 = 0$, but with
  different values of $\tb$ (left panel) and $A_0$ (right panel). The CL
  is almost independent of these parameters, with the error bar always
  embracing 95\%. The  band represents the $\pm 1 \sigma$ range of
  the fit shown in the left panel of Fig.~\ref{fig:scaling}. 
} 
\label{fig:tbA0}
\end{figure*}

The next step is to use our \delphes\ implementation to evaluate the
sensitivity of the ATLAS analysis to other values of $\tb$ and $A_0$. To
this end, for each of the 34 points along the 95\% CL contour in the
$(m_0, m_{1/2})$ plane for $\tb = 10$ and $A_0 = 0$ mentioned
previously, we used {\tt PYTHIA} and \delphes\ to estimate the
confidence levels of these points as the other parameters are varied, as
shown in Fig.~\ref{fig:tbA0}. The left panel exhibits the variation of
the confidence level with $\tb$ for fixed $A_0 = 0$. We see that the CL
is quite stable, with its central value ranging between 95 and 90\% and
its 68\% CL range lying between 99\% and 84\% and always including the
95\% CL found by ATLAS for $\tb = 10$ and $A_0 = 0$~%
\footnote{The different sizes of the error bars, here and in subsequent
validation plots, are due to {\tt PYTHIA} failing for varying numbers
of points.}%
. The right panel of Fig.~\ref{fig:tbA0} displays a similar analysis for
varying $A_0/m_0 \in [-3, 3]$ with $\tb$ fixed at 10. We see again that
the CL is quite stable, with central values between 95 and 92\% and the
68\% CL ranges lying between 100\% and 88\% and again always including
the 95\% CL found by ATLAS for $\tb = 10$ and $A_0 = 0$. These results
are consistent with the statement by ATLAS~\cite{ATLAS5} that their
$(m_0, m_{1/2})$ exclusion contours are largely independent of $\tb$ and
$A_0$, as we have assumed previously~\cite{mc7}, and also provide
consistency checks on our \delphes\ implementation. 

\subsection*{\it Extensions to Non-Universal Higgs Masses}

Following these validations, as a third step we now use our \delphes\
implementation to extend the interpretation of the ATLAS 5/fb jets +
$\ETslash$ analysis to models with non-universal Higgs masses. We
recall that in the CMSSM, the electroweak vacuum conditions can be used
to fix the Higgs mixing parameter $\mu$ and $\MA$ for any given choice
of the parameters $m_0, m_{1/2}, A_0$ and $\tb$. Introducing the single
extra degree of freedom $m^2_{H_u} /m_0^2 = m^2_{H_d} /m_0^2$ in the
NUHM1, one can treat either $\mu$ or $\MA$ as a free parameter, whereas
these quantities may be treated as two free parameters in the NUHM2, in
which $m^2_{H_u} /m_0^2$ and $m^2_{H_d} /m_0^2$ are treated as two
independent parameters~\cite{nuhm2}. The effects of varying $\mu$ and $\MA$ on the
gluino mass and the spectra of squarks (the main sparticles expected to
be produced at the LHC) are of secondary importance~%
\footnote{Except
that varying $\mu$ affects the masses and mixings of third-generation
squarks, but the dominant LHC jets + $\ETslash$ searches are 
less sensitive to these.}%
. However, they may also affect the branching fractions of the cascade
decays to which the LHC experiments are sensitive, and so could in
principle have some effect on the sensitivity of the ATLAS 5/fb jets +
$\ETslash$ search in the $(m_0, m_{1/2})$ plane. 

\begin{figure*}[htb!]
\resizebox{8cm}{!}{\includegraphics{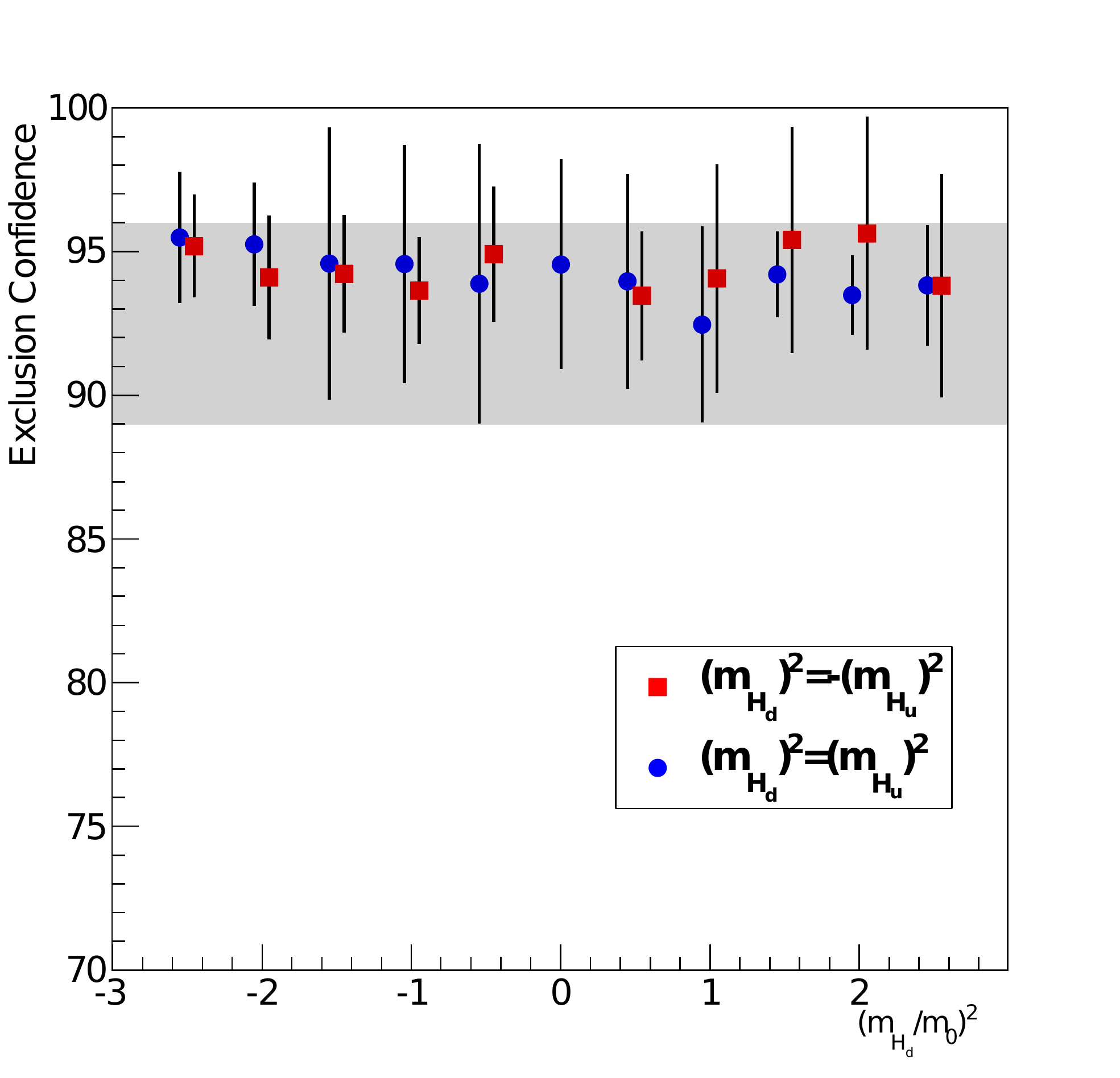}}
\resizebox{8cm}{!}{\includegraphics{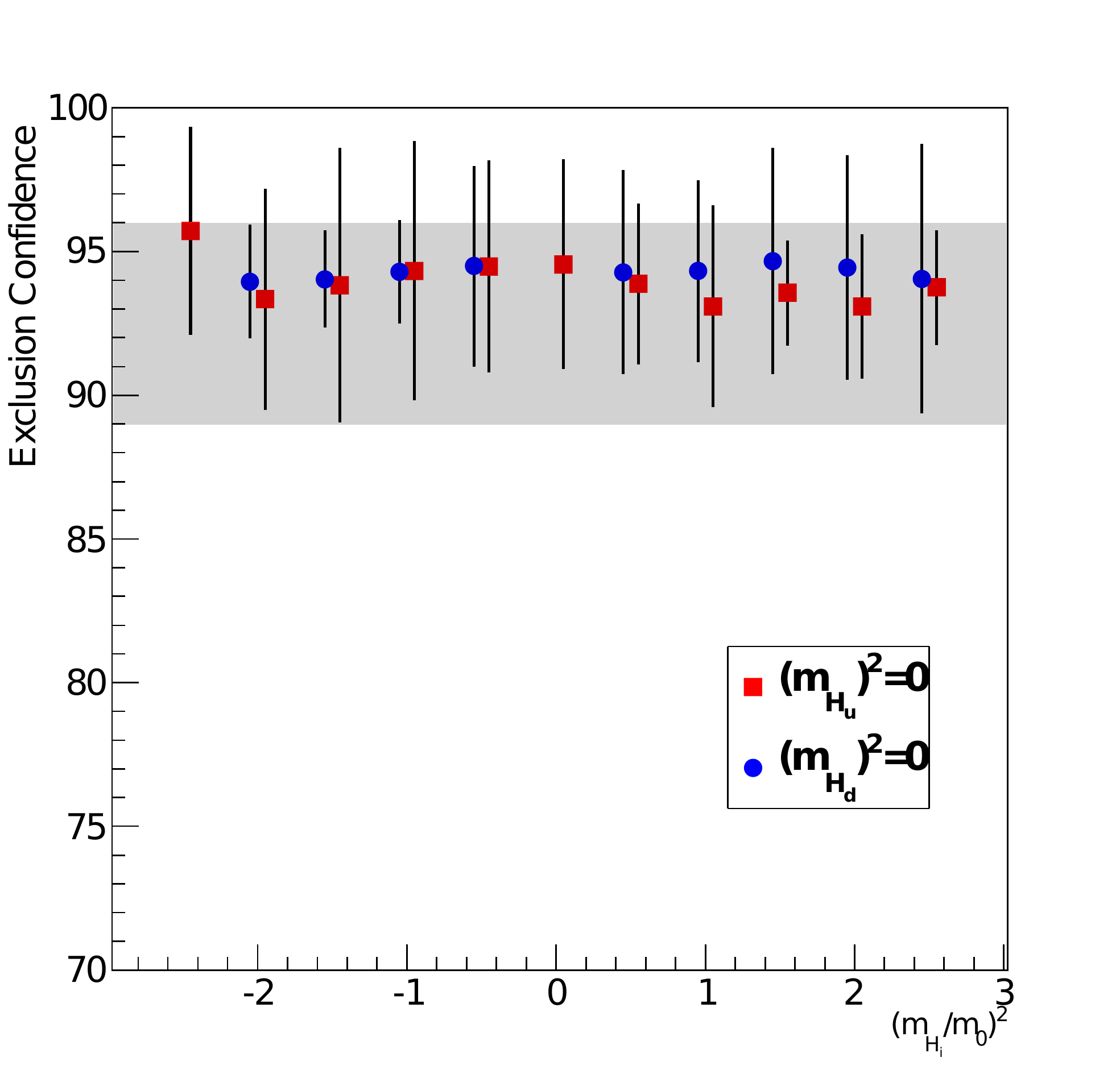}}
\caption{\it Calculations using \delphes\ of the average and spread in
  exclusion confidence levels of 34 CMSSM points selected to have the
  same $(m_0, m_{1/2})$ values as points on the 95\% LHC jets +
  $\ETslash$ exclusion line  for $\tb = 10$ and $A_0 = 0$, 
  as in Fig.~\ref{fig:tbA0}, but (left panel) varying 
  $m^2_{H_d} /m_0^2 = m^2_{H_u} /m_0^2$ in the NUHM1 (red squares)   and
  $m^2_{H_d} /m_0^2 = - m^2_{H_u} /m_0^2$ in the NUHM2 (blue circles),
  and (right panel) varying $m^2_{H_d}$ for fixed $m^2_{H_u} = 0$ (red
  squares) and varying $m^2_{H_u}$ for fixed $m^2_{H_d} = 0$ (blue
  circles) in the NUHM2. In both panels, the  bands represent the
 $\pm 1 \sigma$ range of the fit shown in the left panel of
  Fig.~\protect\ref{fig:scaling}.
  }
\label{fig:NUHM}
\end{figure*}

For this reason, we have taken the same 34 points along the 95\% CL
contour in the $(m_0, m_{1/2})$ for $\tb = 10$ and $A_0 = 0$ studied
above, and used {\tt PYTHIA} and \delphes\ to estimate how the
confidence levels of these points vary as the Higgs
non-universality parameters $m^2_{H_u} /m_0^2$ and $m^2_{H_d} /m_0^2$
are varied. The red squares in the left panel of Fig.~\ref{fig:NUHM}
show how the CL depends on the common non-universality parameter
$m^2_{H_u} /m_0^2 = m^2_{H_d} /m_0^2$ in the NUHM1. As before, we see
that the CL does not depart significantly from the CMSSM value of
95\%. Also shown in the left panel of Fig.~\ref{fig:NUHM} (as blue
circles) are {\tt PYTHIA} and \delphes\ estimates of the confidence
levels of points with the same values of $m_0, m_{1/2}, \tb = 10$ and
$A_0 = 0$ but with Higgs non-universality parameters of opposite signs
for the two Higgs doublets: $m^2_{H_u} /m_0^2 = - m^2_{H_d} /m_0^2$. The
right panel of Fig.~\ref{fig:NUHM} shows how the CL in the NUHM2 varies
as a function of $m^2_{H_d}$ for fixed $m^2_{H_u} = 0$ (red squares) and
as a function of $m^2_{H_u}$ for fixed $m^2_{H_d} = 0$ (blue
circles). We again see that the CL is similar to the 95\% found in the
CMSSM for the same values of $m_0, m_{1/2}, \tb = 10$ and $A_0 = 0$. 

\subsection*{\it Extension to $m_0^2 < 0$}

\begin{figure*}[htb!]
\begin{center}
\resizebox{8cm}{!}{\includegraphics{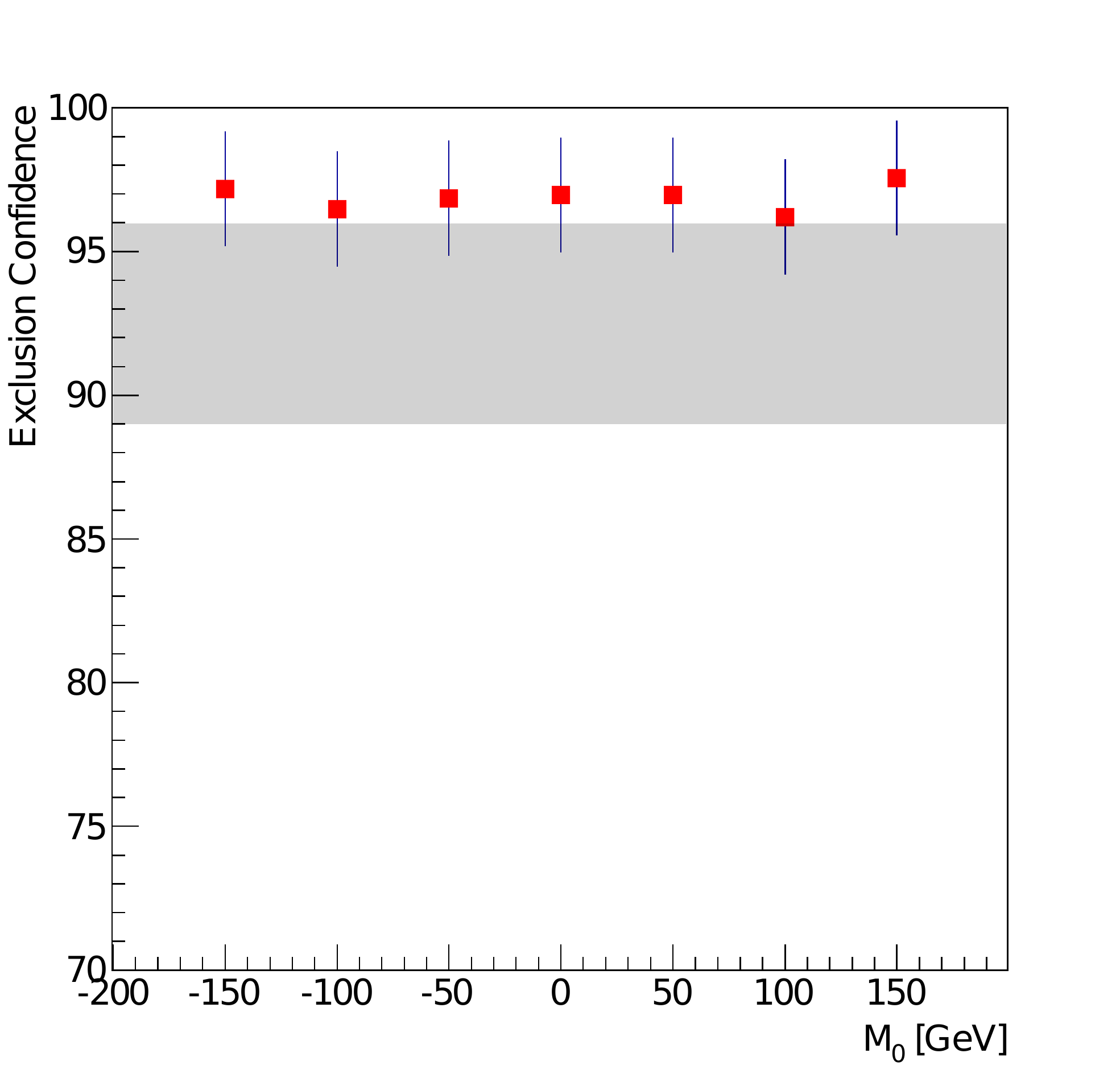}}
\end{center}
\caption{\it Calculations using \delphes\ of the exclusion confidence levels of NUHM1 
points selected to have $\tb = 10, A_0 = 0, m_{1/2} = (630 \pm 25) \gev$ and varying
  $M_0 = sign(m_0^2) \times \sqrt{|m_0^2|} \in (-150, + 150) \gev$. The  band represents the $\pm 1 \sigma$ 
  range of the fit shown in the left panel of
  Fig.~\protect\ref{fig:scaling}.
}
\label{fig:negm0}
\end{figure*}

To complete the validation for the NUHM1, and in preparation for
future analyses in the NUHM2, we have
also investigated the applicability of the \atlasfive\ results to models
with smaller values of $m_0^2$ than in the CMSSM, including negative
values.  We have evaluated the CL along a line of points with  
$\tb = 10, A_0 = 0, m_{1/2} = 630 \gev$ and 
$- (150 \gev)^2 < m_0^2 < + (150 \gev)^2$, i.e., 
$M_0 \equiv {\rm sign}(m_0^2) \times \sqrt{|m_0^2|} \in (- 150, + 150) \gev$,
as shown in Fig.~\ref{fig:negm0}. 
We see that the CL remains constant, within
errors, over this range of $M_0$, and is quite consistent with the
95\% CL reported for the \atlasfive\ analysis at $m_0 = 100 \gev$.

\section{Combination of \bmm\ Constraints}

Since our previous analysis~\cite{mc75}, new constraints on \bmm\ have
been published by ATLAS~\cite{ATLASbmm}, CDF~\cite{CDFbmm},
CMS~\cite{CMSbmm} and LHCb~\cite{LHCbbmm} Collaborations~\cite{LHCbmm}. Their
measurements are to be compared with the theoretical (TH) value in the
SM, which we take to be $\bmm_{\rm SM, TH} = (3.2 \pm 0.2)
\times 10^{-9}$~\cite{bmmtheo}. 
This comparison is not direct, because experiments
measure the time-averaged (TA) branching ratio, which differs from the
TH value because of the difference between the lifetimes of the heavier and lighter $B_s$
mesons~\cite{Fleischer1,Fleischer2}. In general, $\bmm_{\rm TH}
= [(1 - y_s^2)/(1 + {\cal A}_{\Delta \Gamma} y_s)] \times  \bmm_{\rm TA}$,
where ${\cal A}_{\Delta \Gamma} = +1$ in the SM and 
$y_s = 0.088 \pm 0.014$. Thus we compare the measurements to the SM
prediction for the time-averaged branching ratio: $\bmm_{\rm SM, TA} =
\bmm_{\rm SM, TH}/(1 - y_s) = (3.5 \pm 0.2) \times 10^{-9}$~\cite{Fleischer2}. 

\begin{figure*}[htb!]
\resizebox{8cm}{!}{\includegraphics{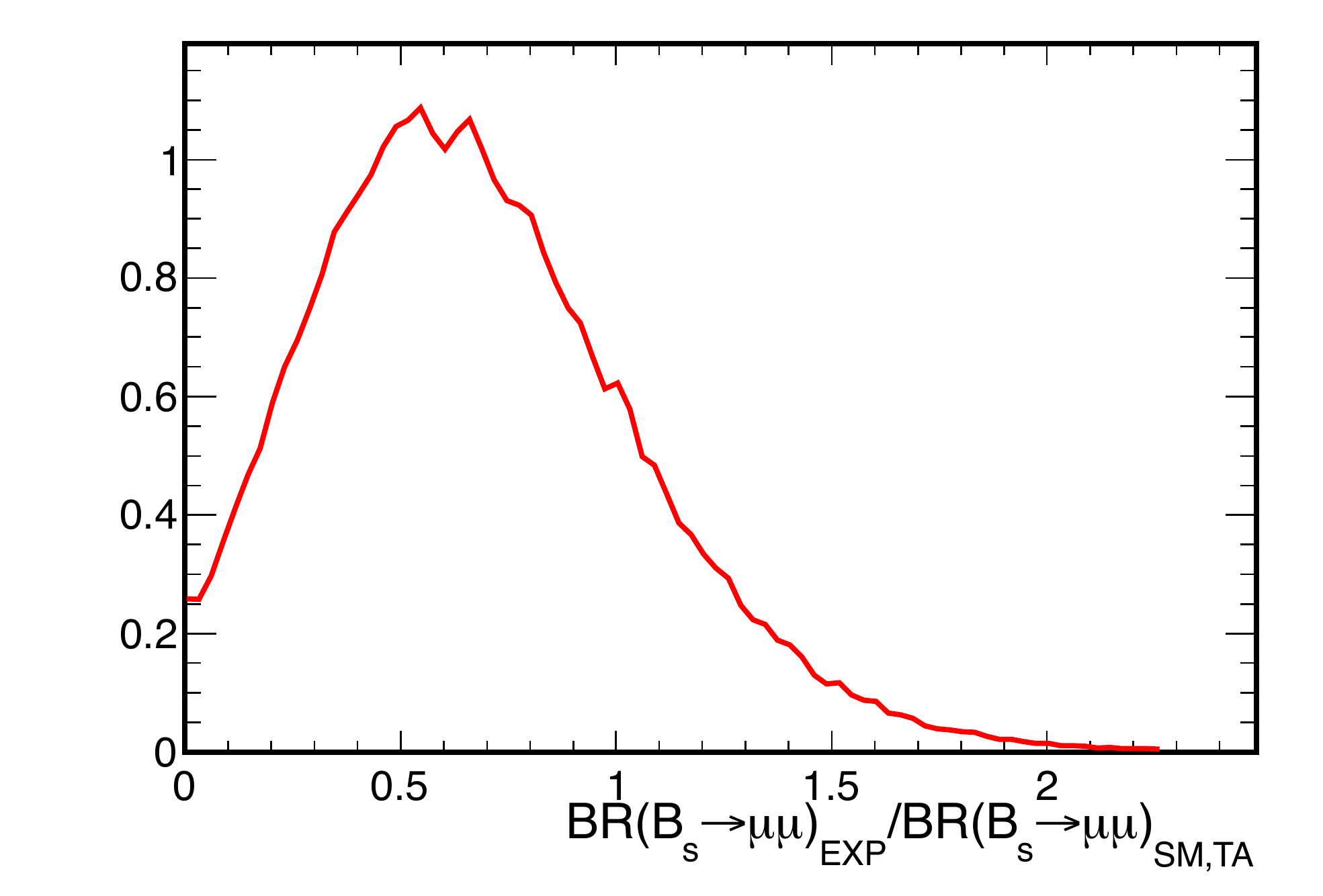}}
\resizebox{8cm}{!}{\includegraphics{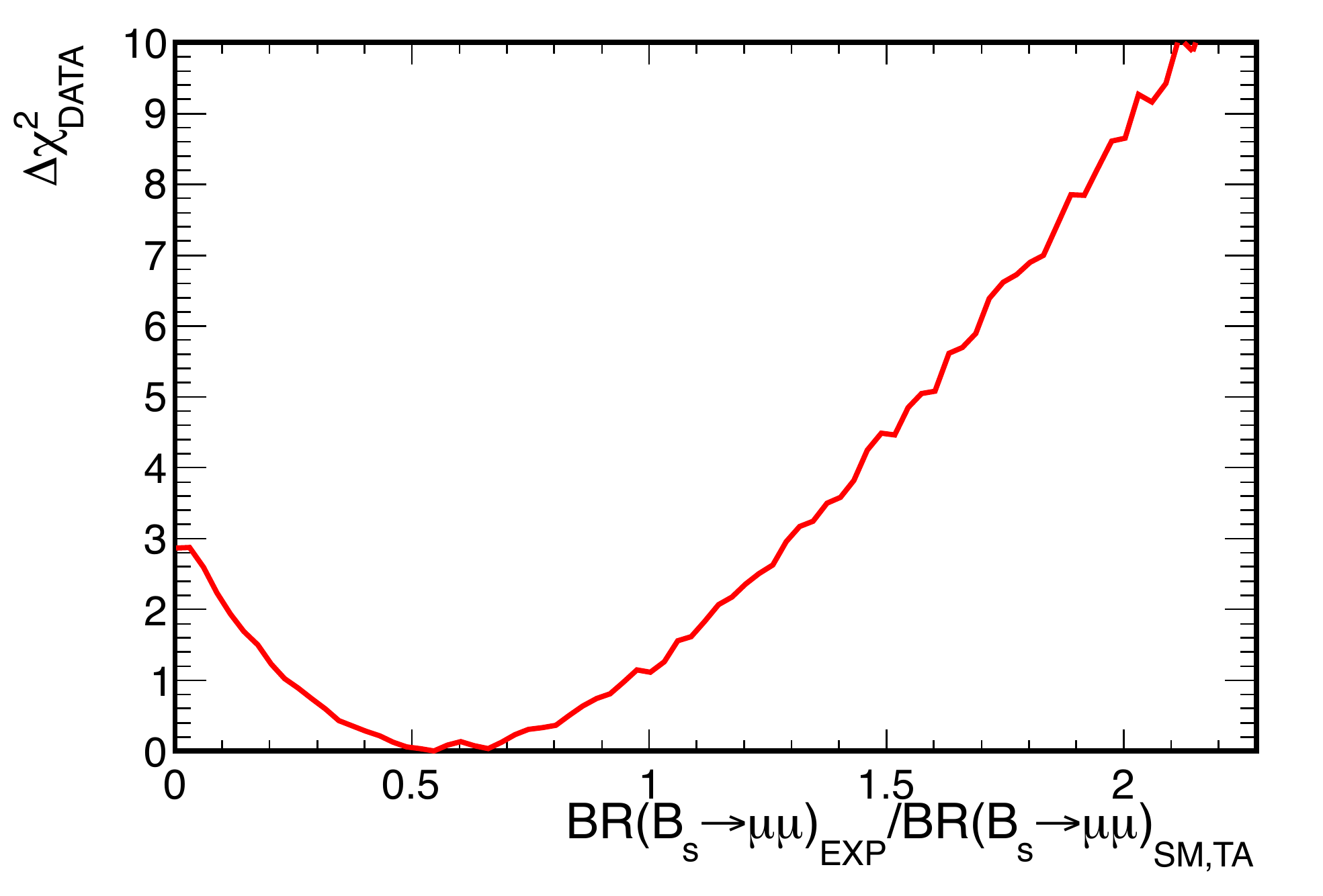}}
\caption{\it The probability distribution (left panel) and the
  log-likelihood function (right panel) found in an unofficial
  combination of the constraints on \bmm\ from ATLAS, CDF, CMS and  
LHCb~\protect\cite{ATLASbmm,CDFbmm,CMSbmm,LHCbbmm,LHCbmm}, normalized to 
$\bmm_{\rm SM, TA} = \bmm_{\rm SM, TH}/(1 - y_s) 
= (3.5 \pm 0.2) \times 10^{-9}$, as described in the text.}
\label{fig:Bmumu}
\end{figure*}

We use for this comparison an unofficial combination of the ATLAS, CDF,
CMS and  LHCb data~\cite{ATLASbmm,CDFbmm,CMSbmm,LHCbbmm,LHCbmm}, 
from which we compute a $\chi^2$ penalty for each possible
(calculated) value of \bmm.
In Fig.~\ref{fig:Bmumu} we show the probability distribution (left) and
the log-likelihood function for the ratio
$\bmm/\bmm_{\rm SM, TA}$, folding together the
theoretical and experimental errors. The log-likelihood
function is used as the $\Delta \chi^2$ penalty for the values of
this observable as they are calculated in the CMSSM and NUHM1~%
\footnote{We note in passing that one expects  
${\cal A}_{\Delta \Gamma} = +1$ also in these models.}%
.

\section{The Constraint on Spin-Independent Dark Matter Scattering from 225 Live Days of XENON100 Data}

The XENON100 Collaboration has recently published the result of a search for spin-independent dark matter 
scattering based on 225 live days of data~\cite{newXENON100}. They report two events in a fiducial region 
where $1.0 \pm 0.2$ background events were expected, i.e., no evidence for any excess, quote a 90\% CL 
upper limit of $\ssi\ < 2.0 \times 10^{-45}$~cm$^2$ for a dark matter particle mass of 55~GeV, 
and provide a 90\% CL exclusion line for other masses.

Following the procedure used in~\cite{mc6} for the previous XENON100 constraint, 
we use a CL$_s$ calculator, now implemented in our {\tt Delphes} framework, 
to model the contribution to the global $\chi^2$ function of this new XENON100 constraint 
from the number of signal events as a Gaussian with mean $\mu = 1.0$ and standard deviation 
$\sigma = 2.7$ events. This corresponds to the mean (2 - 1 = 1 event) given in~\cite{newXENON100} 
and the upper limit of 5.1 signal events at the 90\% CL given by the CL$_s$ calculator, 
which we identify with the published XENON100 90\% exclusion line.  
For models predicting values of \ssi\ above or below the 90\% CL line in~\cite{newXENON100}, 
we rescale linearly the number of events and estimate the $\Delta \chi^2$ contribution of 
XENON100 using the Gaussian model described above.

We use {\tt SSARD}~\cite{SSARD} to calculate \ssi, assuming a nominal value of the $\pi$-N $\sigma$ term
$\Sigma_{\pi N} = 50$~MeV, and assigning an asymmetric error $^{+ 14}_{-7}$~MeV.
This accounts for the fact that $\Sigma_{\pi N} = \sigma_0 \sim 36$~MeV if
$\langle N | {\bar s} s | N \rangle = 0$, while allowing for significantly larger values of
$\Sigma_{\pi N}$ as suggested by some analyses~\cite{Pavan}. 

The XENON100 analysis
assumes that the local dark matter density is $0.3 \gev$/cm$^3$, and makes
supplementary assumptions on the dark matter velocity distribution. We have 
explored the potential implications of the uncertainties in these quantities by
convoluting with our above-mentioned model of the XENON100 constraint on \ssi\ a fractional uncertainty 
of a factor $\sqrt{2}^{\pm 1}$ at the 68\% CL.  We find that it has negligible impact
on the results shown below, which are for the nominal dark matter distribution
parameters assumed by XENON100.


\section{The CMSSM and NUHM1 after \boldmath{\atlasf}, \boldmath{\bmm} and XENON100}

As already mentioned, in addition to the \delphes\ implementation of the
\atlasfive\ constraint, the implementation of the combined
\bmm\ constraints described above, and the new CMS constraint on
$H/A \to \tau^+\tau^-$~\cite{CMSHA}, we include in our global fit the
constraint $\Mh = 125 \pm 1.0 \pm 1.5 \gev$  and the other
constraints included in~\cite{mc75}, with updated values of 
the $W$~boson mass~\cite{lepewwg} and the top quark 
mass~\cite{mt1732}, see above.

\subsection*{\it The $(m_0, m_{1/2})$ planes}

\begin{figure*}[htb!]
\resizebox{8.5cm}{!}{\includegraphics{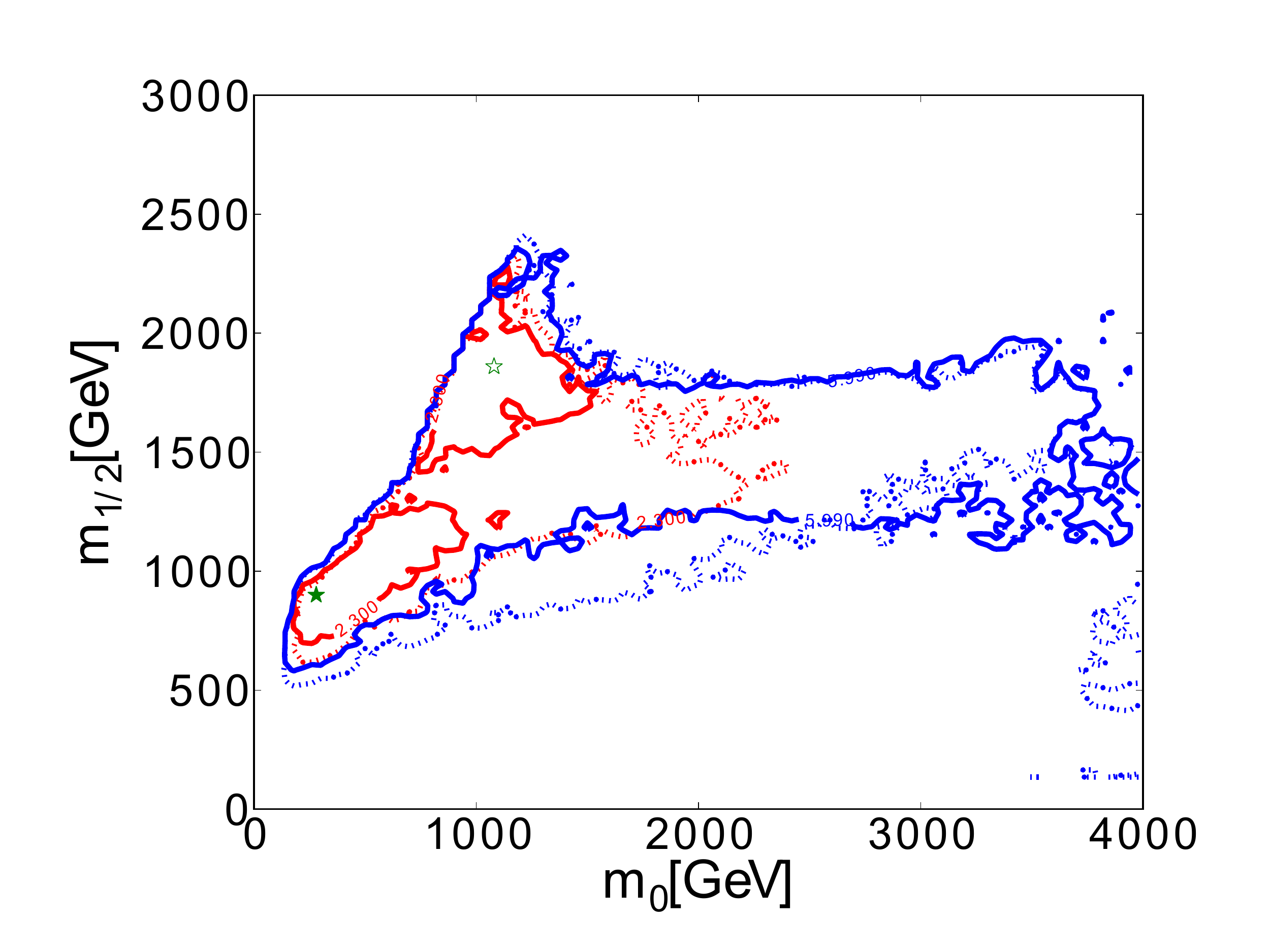}}
\resizebox{8.5cm}{!}{\includegraphics{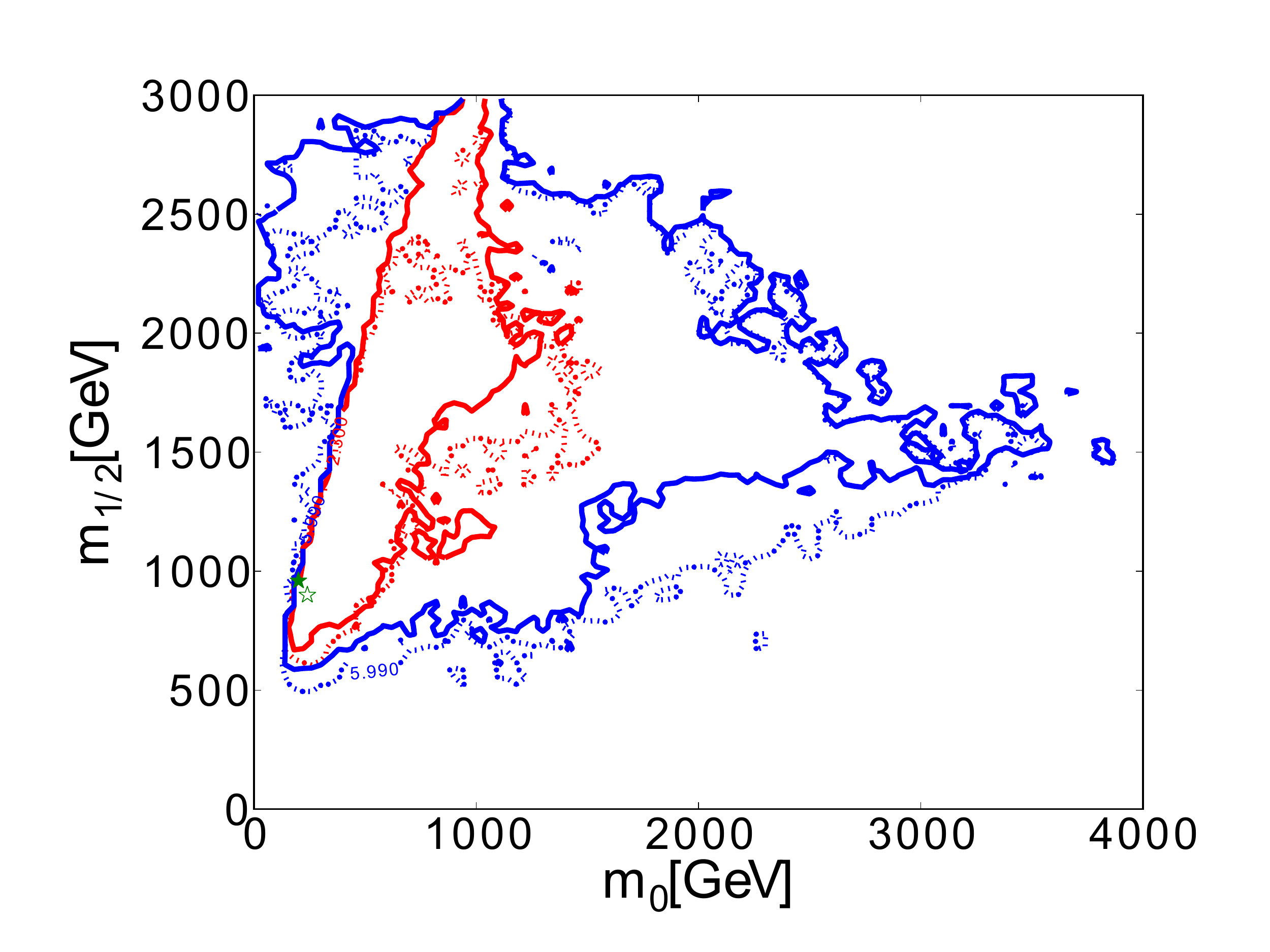}}
\vspace{-1cm}
\caption{\it The $(m_0, m_{1/2})$ planes in the CMSSM (left panel) and
  the NUHM1 (right panel) including the \atlasfive\ constraint~\cite{ATLAS5}, a
  combination of the ATLAS~\cite{ATLASbmm}, CDF~\cite{CDFbmm}, CMS~\cite{CMSbmm} and LHCb~\cite{LHCbbmm} constraints on 
  \bmm~\cite{LHCbmm} and the recent XENON100 result~\cite{newXENON100},
  assuming $\Mh = 125 \pm 1 ~({\rm exp.}) \pm 1.5 ~({\rm theo.})
  \gev$. The   results of the current fits are indicated by solid lines
  and filled   stars, and previous fits based on $\sim 1$/fb of LHC data
  are indicated by dashed lines and open stars.    The blue lines denote
  68\%~CL  contours, and the red lines denote 95\%~CL contours.} 
\label{fig:m0m12}
\end{figure*}

Fig.~\ref{fig:m0m12} displays the $(m_0, m_{1/2})$ planes for the CMSSM
(left panel) and the NUHM1 (right panel). The results of the new global
fits are shown as solid lines and filled stars, and previous fits based
on $\sim 1$/fb of LHC data are indicated by dashed lines and open stars.
The blue and red lines denote levels $\Delta \chi^2 = 2.30$ and
5.99, which we interpret as the 68\% and 95\%~CLs.

We note the existence in the CMSSM of a grouping of points in the focus-point region at large $m_0$ and small $m_{1/2}$
that were previously allowed at the 95\% CL but are now disallowed~\footnote{This could also be inferred from
Figs.~7 and 9 of the paper by JE and KAO in~\cite{125-other}.}. This exclusion is due to the new XENON100
constraint~\cite{newXENON100}. The main effect of the \atlasfive\ constraint is to shave off a slice of points
at low $m_{1/2}$ and moderate $m_0$. We also note the appearance in the NUHM1 sample of a grouping of points at large
$m_{1/2}$ that have smaller $m_0$ than is possible in the CMSSM, and are allowed at the 95\% CL. The relic density in this region is brought into
the WMAP range by multiple neutralino and chargino coannihilation processes. Points are found in this region using each of the codes
{\tt MicrOMEGAs}~\cite{MicroMegas}, {\tt DarkSUSY}~\cite{DarkSusy}, {\tt SuperIso}~\cite{SuperIso} and {\tt SSARD}~\cite{SSARD}, 
though they yield somewhat different values of the relic density. As mentioned previously, here we use {\tt MicrOMEGAs} as our default.

The values of $\chi^2$ in the most-favoured regions of the CMSSM and NUHM1 are relatively
  shallow functions of $m_0$, $m_{1/2}$ and $\tb$, so that little importance
  can be attributed to the positions of the best-fit points or to their movements relative to those in~\cite{mc75}.
Nevertheless, we note that, as summarized in Table~\ref{tab:bestfits}, the nominal best-fit point in the CMSSM moves to smaller $m_0$ and $m_{1/2}$,
  whereas that in the NUHM1 moves little in this plane.
The best-fit points have $\chi^2 = 32.8$ in the CMSSM and
31.3 in the NUHM1, and
the $p$-values of these models are reduced to 8.5 and 9.1\%, respectively.

We see in Fig.~\ref{fig:m0m12} that the 95\% CL region in the CMSSM extends 
to $m_0 \sim 4000 \gev$ and $m_{1/2} \sim 2400 \gev$. The corresponding region
in the NUHM1 extends to smaller $m_0$ but larger $m_{1/2}$. We also note that the CMSSM fit features two
disconnected 68\%~CL `islands', the one at lower $m_0$ and $m_{1/2}$
corresponding to the stau coannihilation region, and that at larger $m_0$
and $m_{1/2}$ corresponding to the rapid-annihilation funnel region. We
note, however, that our sampling also includes a few intermediate points
that also have $\Delta \chi^2 < 2.30$, reflecting the relative flatness of
the global $\chi^2$ function along the band between $(m_0, m_{1/2}) \sim
(300, 900) \gev$ and $\sim (1100, 1900) \gev$. Typically, these points
sit in rapid-annihilation funnels, and it is possible that a more
complete sampling might reveal a lower `isthmus'  connecting these
islands, as happens in the NUHM1.

We report in Table~\ref{tab:bestfits} the values of
$m_0, m_{1/2}, A_0$ and $\tb$ for the best fits in both the CMSSM `islands',
as well as for the best fit in the NUHM1, which is at relatively low masses, and
a local minimum of the NUHM1 $\chi^2$ function at high masses.
We note that the best fit in the low-mass CMSSM `island' is similar to the best fit in the NUHM1,
and that both have smaller $\tb$ than in the previous best fits to the
LHC$_{\rm 1/fb}$ data set in these models. As we discuss later, this effect is due to the new \bmm\ constraint. On the other hand,
the best fit in the high-mass CMSSM  `island' is more similar to the previous best fit to the
LHC$_{\rm 1/fb}$ data set. In general terms, we see that the area of the NUHM1 $(m_0,
m_{1/2})$ plane that is allowed at the 68 and 95\%~CL is larger than
in the $(m_0, m_{1/2})$ plane of the CMSSM. This reflects the fact
that the dependence of the global $\chi^2$ function on these variables
is weaker in the NUHM1 than in the CMSSM, reflecting the additional
freedom offered by the degree of Higgs-mass non-universality.  

In particular, the 68\%~CL region in the NUHM1 is connected, and extends to larger
$m_0$ than in the CMSSM. This extension reflects the fact that in the
NUHM1 a rapid-annihilation funnel region may appear at smaller $m_{1/2}$ and $\tb$
than in the CMSSM for the same value of $m_0$, thanks to the variation
in $\MA$ that is possible in the NUHM1, allowing $\MA \sim 2 \mneu{1}$
in different regions of the $(m_0, m_{1/2})$ plane.  It is this freedom
in the location of the rapid-annihilation funnel, a characteristic of
the NUHM1, that is largely responsible for the expansion in the region
of the $(m_0, m_{1/2})$ plane that is favoured at the 68\%~CL in the
NUHM1, compared with the CMSSM~\footnote{The relic density may also be
  brought into the WMAP range because the $\neu{1}$ acquires a
  relatively large higgsino component, another possibility made possible
  by the variation in $\mu$ that is possible in the NUHM1.}. 

\begin{table*}[!tbh!]
\renewcommand{\arraystretch}{1.5}
\begin{center}
\begin{tabular}{|c|c||c|c|c|c|c|c|} \hline
Model & Data set & Minimum & Prob- & $m_0$ & $m_{1/2}$ & $A_0$ & $\tb$ \\
  &    & $\chi^2$/d.o.f.& ability & (GeV) & (GeV) & (GeV) & \\ 
\hline \hline
CMSSM & LHC$_{\rm 1/fb}$     
    & 31.0/23 & {12\%} & $1120$ & $1870$ 
    & $1220$ & $46$\\
&ATLAS$_{\rm 5/fb}$ (low)
    & 32.8/23 & {8.5\%} & $300$ & $910$ 
    & $1320$ & $16$ \\
&ATLAS$_{\rm 5/fb}$ (high)    & 33.0/23 & {8.0\%} & $1070$ & $1890$     & $1020$ & $45$ \\
\hline
NUHM1 & LHC$_{\rm 1/fb}$     
    & 28.9/22 & 15\% & $270$ & $920$ 
    & $1730$ & $27$ \\
& ATLAS$_{\rm 5/fb}$ (low)
    & 31.3/22 & 9.1\% & $240$ & $970$ 
    & $1860$ & $16$ \\
& ATLAS$_{\rm 5/fb}$ (high) & 31.8/22 & 8.1\% & $1010$ & $2810$ & $2080$ & $39$ \\
\hline
\end{tabular}
\caption{\it The best-fit points found in global CMSSM and NUHM1 fits
  using the \atlasfive\ constraint~\cite{ATLAS5}, the combination of the
  ATLAS~\cite{ATLASbmm}, CDF~\cite{CDFbmm}, CMS~\cite{CMSbmm} and
  LHCb~\cite{LHCbbmm} \bmm~\cite{LHCbmm} constraints and the updated values of $\MW$
  and   $\mt$, compared with those found previously in global fits based
  on   the LHC$_{\rm 1/fb}$ data set. In both cases, we include a 
 measurement of $\Mh = 125 \pm 1.0 \pm 1.5 \gev$ and the new
  XENON100 constraint~\cite{newXENON100}. In the case of the CMSSM, we list the parameters of the best-fit
  points in both the low- and high-mass `islands' in Fig.~\protect\ref{fig:m0m12}, and we quote results
  for a high-mass NUHM1 point as well as the low-mass best-fit point in this model. We note that the
  overall likelihood function is quite flat in bot the CMSSM and the NUHM1, so that the precise
  locations of the best-fit points are not very significant, and we do not quote uncertainties.
  For completeness, we note that in the best NUHM1
  fit   $m_H^2 \equiv m_{H_u}^2 = m_{H_d}^2 = - 6.5 \times 10^6 \gev^2$,
  compared with $-5.5 \times 10^6 \gev^2$ previously.
 }
\label{tab:bestfits}
\end{center}
\end{table*}

Table~\ref{tab:chi2} summarizes some interesting contributions to
the global $\chi^2$ function, namely those contributing $\Delta \chi^2 >
1$, \atlasfive\ and the new global combination of \bmm\ measurements. We
present their contributions at the best-fit point in the CMSSM (which
has relatively high mass parameters), at the local best-fit point in the
low-mass `island' in Fig.~\ref{fig:m0m12}, at the best-fit point in
the NUHM1, and at a high-mass point in the NUHM1. We see that \bsg\ favours lower masses in the
NUHM1, whereas \btn\ applies an essentially constant $\Delta \chi^2$
penalty~\footnote{We note that the Belle Collaboration has recently reported a new
measurement of \btn\ that is in better agreement with the Standard Model and the classes of
supersymmetric models discussed here~\cite{Bellebtn}.}. None of the preferred models gives a good fit to \gmt,
since they reproduce approximately the SM value, though
lower masses are somewhat preferred, particularly in the NUHM1. The lower-mass fits worsen
the fit to $\MW$ and that in the CMSSM (NUHM1) worsens (improves) the fit to $A_\ell({\rm SLD})$, but there are only
small changes in $ \sigma_{\rm had}^0$. 
We note that the \atlasfive\ constraint favours higher masses, whereas the
new \bmm\ constraint favours the low-mass region in the CMSSM and (marginally)
in the NUHM1, and all the models fit XENON100 equally well.

\begin{table*}[htb!]
\renewcommand{\arraystretch}{1.2}
\begin{center}
\small{
\begin{tabular}{|c||c|c|c|c|} \hline
Observable & $\Delta \chi^2$ & $\Delta \chi^2$ & $\Delta \chi^2$ & $\Delta \chi^2$ \\
& CMSSM (high) & CMSSM (low) & NUHM1 (high) & NUHM1 (low) \\
\hline \hline
Global & 33.0 & 32.8 & 31.8 & 31.3 \\
\hline \hline
BR$_{\rm b \to s \gamma}^{\rm EXP/SM}$ & 1.15 & 1.19 & 0.94 & 0.18 \\
\hline
BR$_{\rm B \to \tau\nu}^{\rm EXP/SM}$ & 1.10 & 1.03 & 1.04 & 1.08 \\
\hline
$ a_{\mu}^{\rm EXP} - a_{\mu}^{\rm SM}$ & 9.69 & 8.48 & 10.47 & 7.82 \\
\hline
$\MW$ [GeV] & 0.10 & 1.50 & 0.24 & 1.54 \\
\hline
$ R_\ell$ & 0.95 & 1.09 & 1.09& 1.12 \\ 
\hline
$ A_{\rm fb}({b})$ & 8.16 & 6.64 & 5.68 & 6.43 \\ 
\hline
$ A_\ell({\rm SLD})$ & 2.49 & 3.51 & 4.36 & 3.68 \\ 
\hline
$ \sigma_{\rm had}^0$ & 2.58 & 2.50 & 2.55 & 2.50 \\ 
\hline \hline
\atlasfive & 0.09 & 1.73 & 0.02 & 1.18 \\
\hline
\bmm\ & 2.52 & 1.22 & 1.59 & 1.70 \\ 
\hline
XENON100 & 0.13 & 0.12 & 0.14 & 0.13 \\ 
\hline
\end{tabular}
\caption{\it Summary of the contributions of the most important
  observables to the global $\chi^2$ function at the best-fit high- and low-mass points in the
  CMSSM and NUHM1 (those with $\Delta \chi^2 > 1$) , and of the 
  main updated observables \atlasfive, \bmm\ and XENON100.}
  \label{tab:chi2}}  
\end{center}

\end{table*}

\subsection*{\it The $(\tb, m_{1/2})$ planes}

\begin{figure*}[htb!]
\resizebox{8.5cm}{!}{\includegraphics{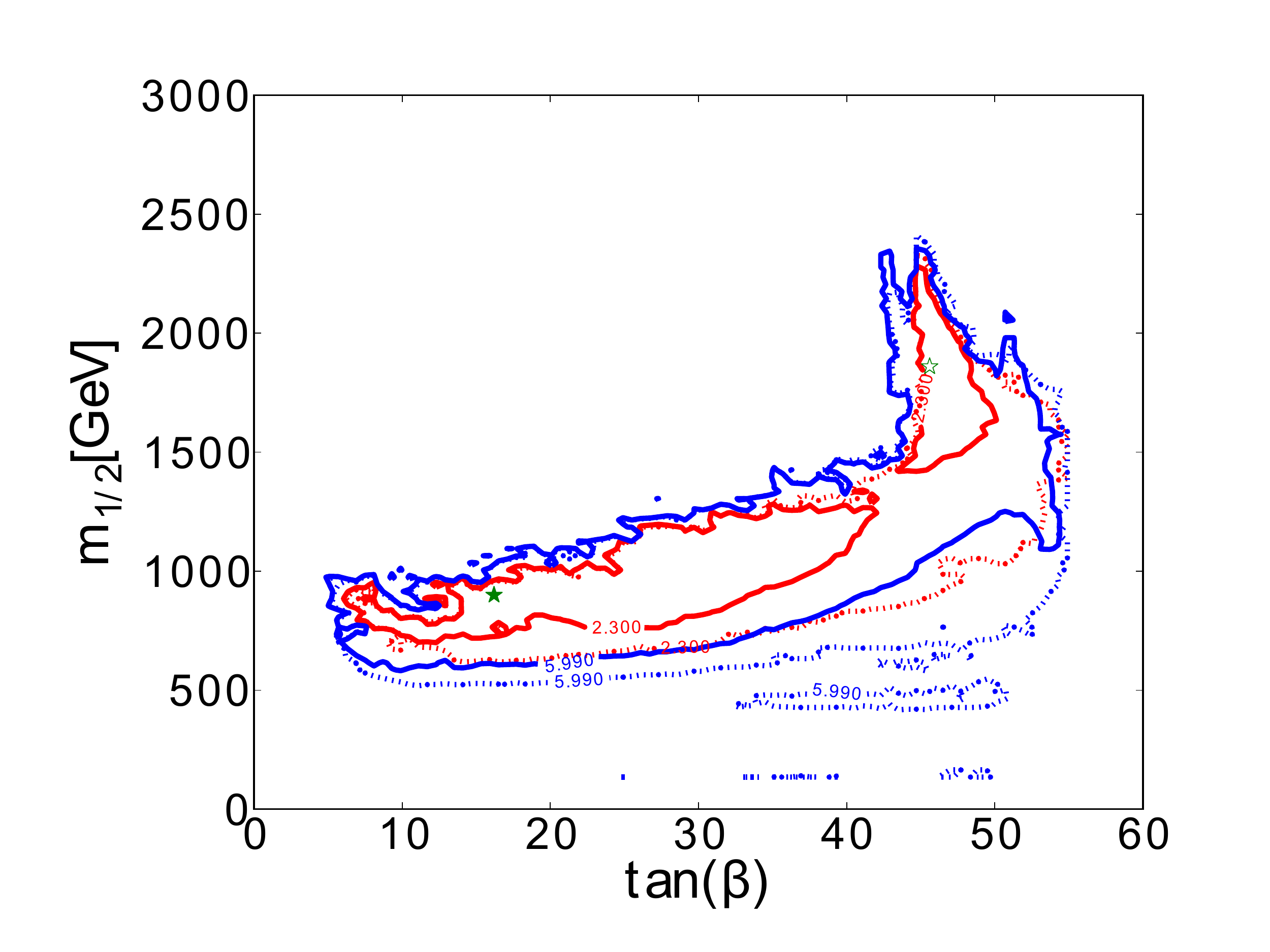}}
\resizebox{8.5cm}{!}{\includegraphics{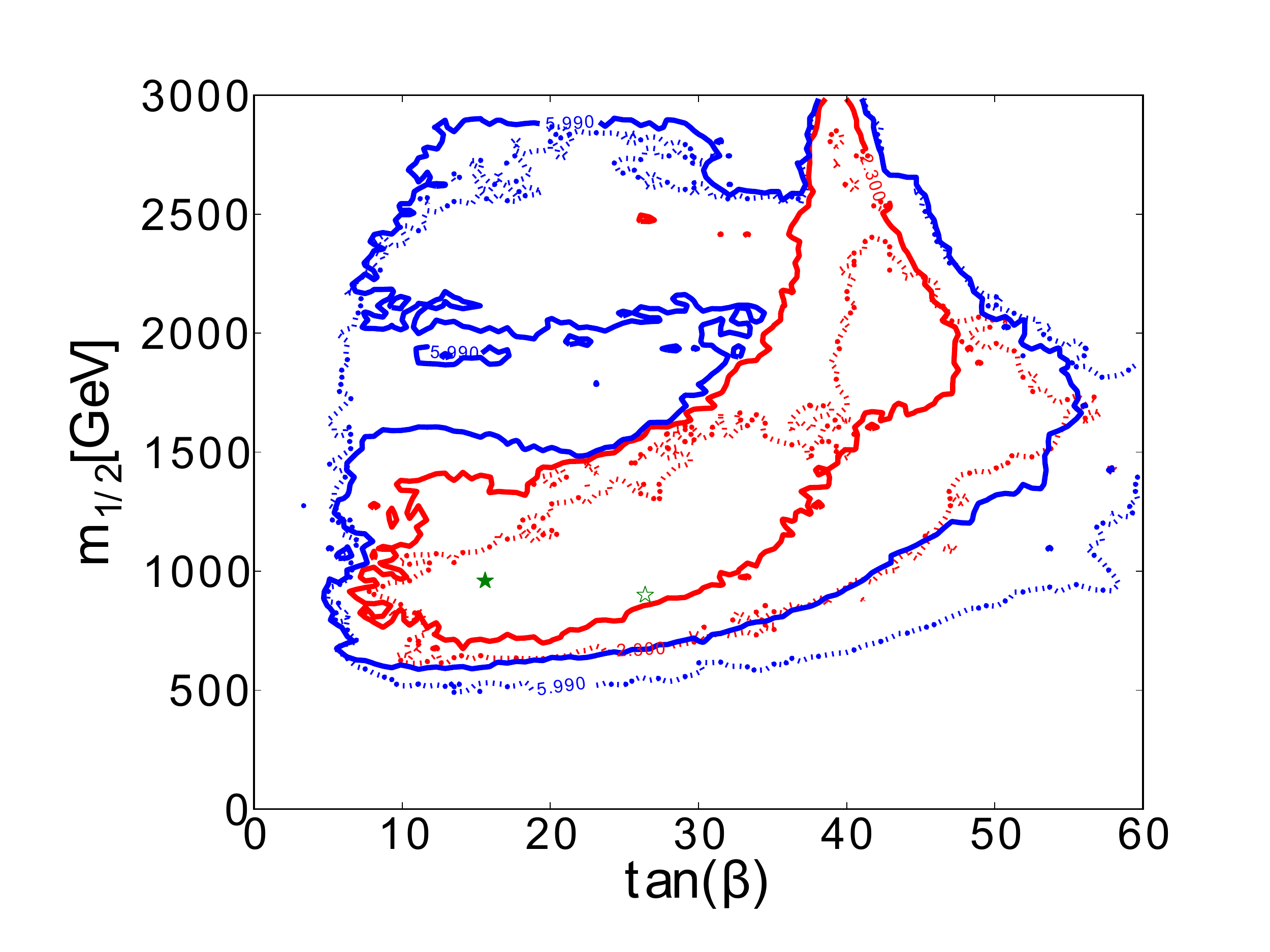}}
\vspace{-1cm}
\caption{\it The $(\tb, m_{1/2})$ planes in the CMSSM (left panel) and
  the NUHM1 (right panel) including the \atlasfive\ constraint~\cite{ATLAS5}, a
  combination of the ATLAS~\cite{ATLASbmm}, CDF~\cite{CDFbmm}, CMS~\cite{CMSbmm} and LHCb~\cite{LHCbbmm} constraints 
  on \bmm~\cite{LHCbmm}. In both cases, we include a 
  measurement of $\Mh = 125 \pm 1.0 \pm 1.5 \gev$ and the new
  XENON100 constraint~\cite{newXENON100}. The   results of the current fits are indicated by solid lines
  and filled   stars, and previous fits based on $\sim 1$/fb of LHC data
  are indicated by dashed lines and open stars.    The blue lines denote
  68\%~CL  contours, and the red lines denote 95\%~CL contours.} 
\label{fig:tbm12}
\end{figure*}

As can be seen in the left panel of Fig.~\ref{fig:tbm12}, where the
$(\tb, m_{1/2})$ plane in the CMSSM is shown, the two
`islands' visible in the $(m_0, m_{1/2})$ plane of the CMSSM (left panel
of Fig.~\ref{fig:m0m12}) correspond to different ranges of $\tb$. Large
values of $\tb \sim 45$ are favoured only at large $m_{1/2} > 1500
\gev$, and values of $\tb$ as low as $\sim 10$ come within the $\Delta
\chi^2 = 2.30$ (68\%~CL) region for $m_{1/2} \sim 800 \gev$, consistent with
their association with funnel and stau coannihilation regions respectively. We see again in the right panel of
Fig.~\ref{fig:tbm12} that the region of the $(\tb, m_{1/2})$ plane that is
allowed in the NUHM1 at the 68\%~CL is connected, and there is no strong separation between
the low- and high-mass regions. As in Fig.~\ref{fig:m0m12}, the region
of the $(\tb, m_{1/2})$ plane allowed at the 95\%~CL is also
enlarged in the NUHM1, compared to the CMSSM. The region at large $m_{1/2}$ and relatively small $\tb$
corresponds to the grouping of points at low $m_0$ and large $m_{1/2}$ already noted in the
right panel of Fig.~\ref{fig:m0m12}.

\subsection*{\it The $(\MA, \tb)$ planes}

\begin{figure*}[htb!]
\resizebox{8.5cm}{!}{\includegraphics{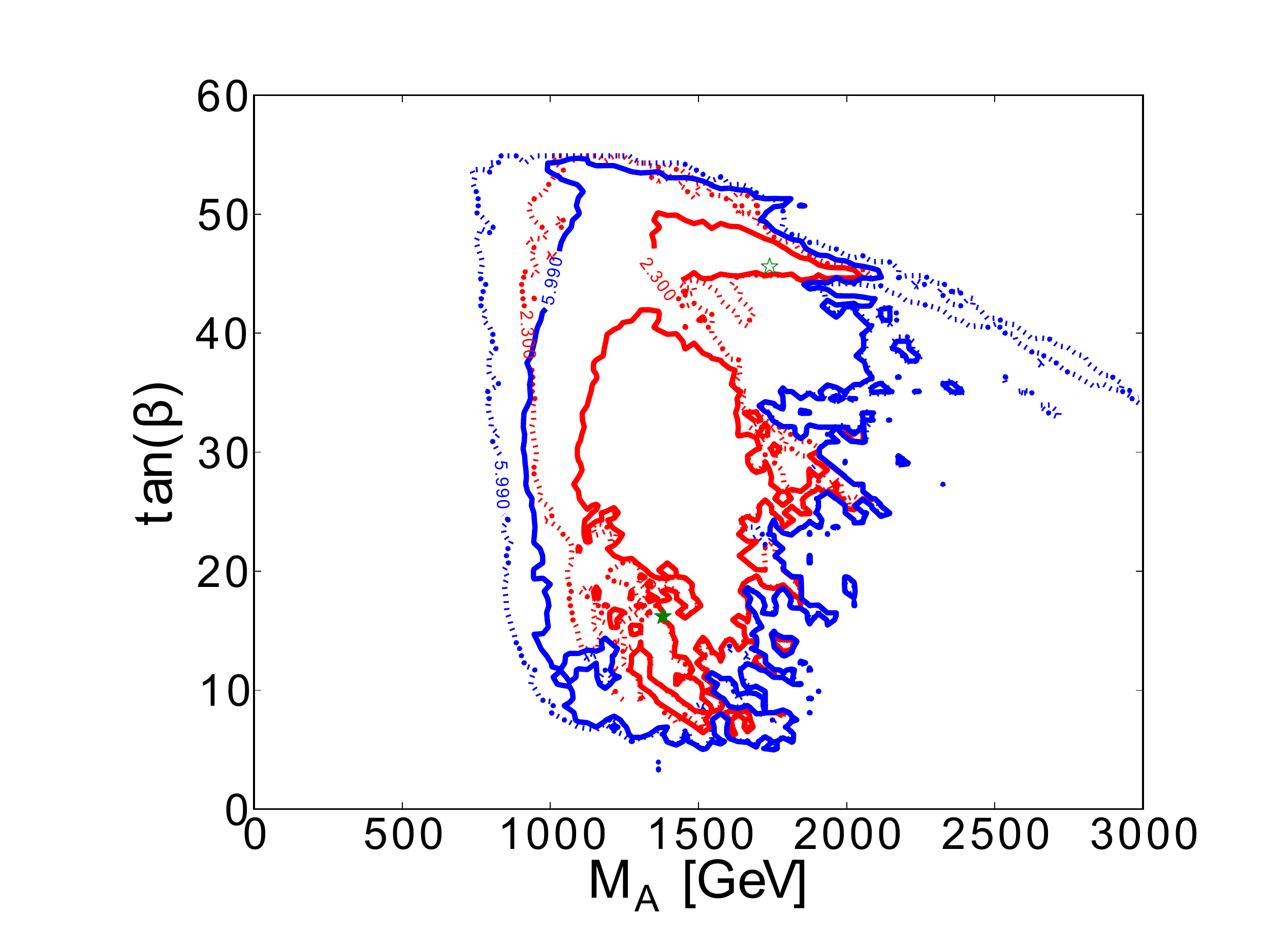}}
\resizebox{8.5cm}{!}{\includegraphics{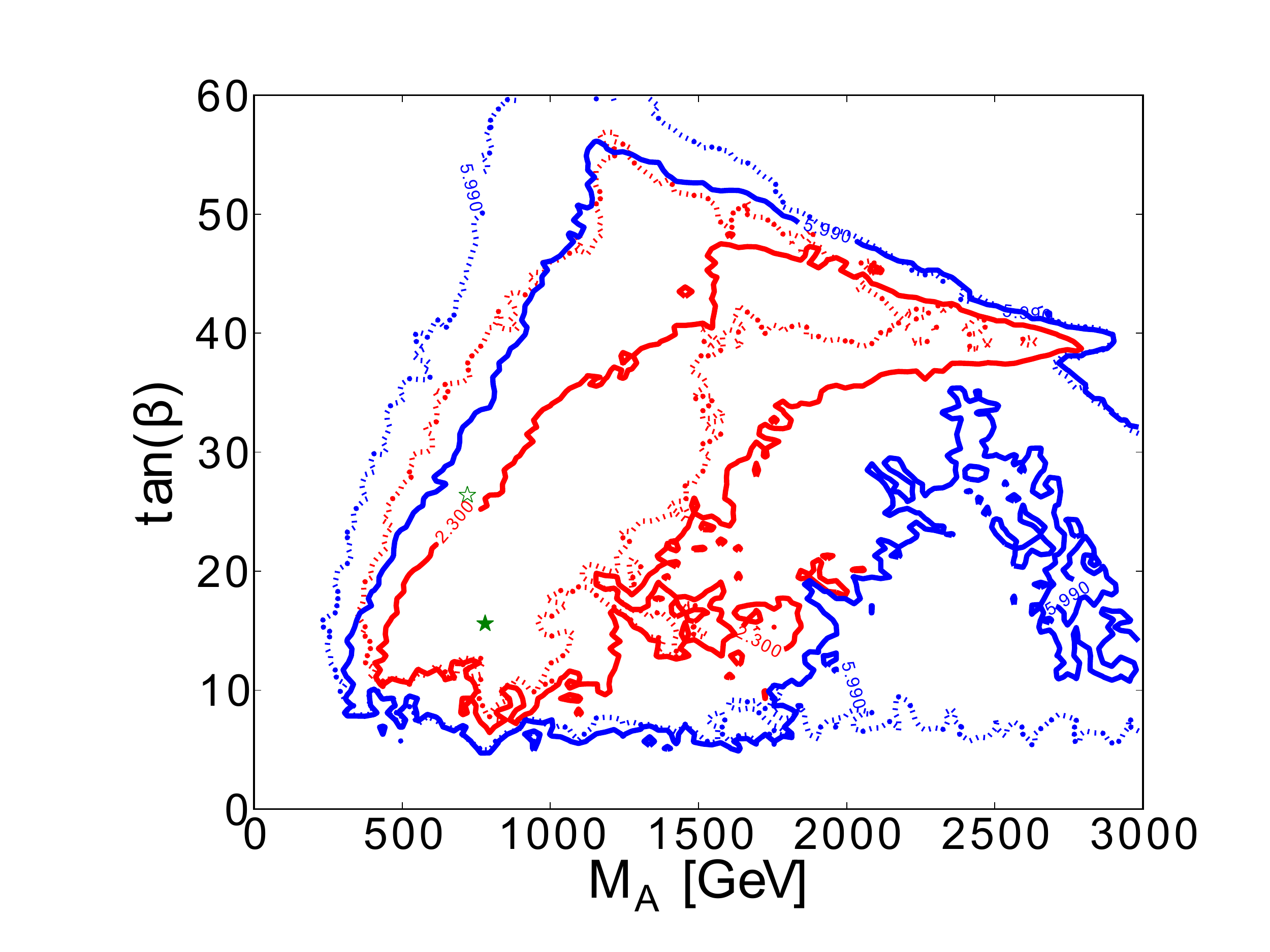}}
\vspace{-1cm}
\caption{\it The $(\MA, \tb)$ planes in the CMSSM (left panel) and the
  NUHM1 (right panel) including the \atlasfive\ constraint, a
  combination of the ATLAS, CDF, CMS and LHCb constraints on \bmm. In both cases, we include a 
  measurement of $\Mh = 125 \pm 1.0 \pm 1.5 \gev$ and the new
  XENON100 constraint~\cite{newXENON100}. The   results of the current fits are indicated by solid lines
  and filled   stars, and previous fits based on $\sim 1$/fb of LHC data
  are indicated by dashed lines and open stars.    The blue lines denote
  68\%~CL  contours, and the red lines denote 95\%~CL contours.} 
\label{fig:MAtb}
\end{figure*}

We see in the left panel of Fig.~\ref{fig:MAtb}, where the 
$(\MA, \tb)$ plane in the CMSSM is shown, that relatively large
values of $\MA$ are favoured in this model, $\MA > 1200 \gev$
at the 68\%~CL
and $> 1000 \gev$ at the 95\%~CL. As is seen by comparing the solid contours
corresponding to $\Delta \chi^2$ contours in this fit with the dotted lines
corresponding to the analysis in~\cite{mc75}, we see substantial retreats at
small $\MA$ and large $\tb$ that are due to the updated \bmm\ constraint.
In the NUHM1, smaller values
of $\MA$ are allowed at low $\tb$, as low as $\sim 500 \gev$ at the 68\%
CL and $\sim 350 \gev$ at the 95\%~CL. On the other hand, a substantial
region of the NUHM1 $(\MA, \tb)$ plane at small $\MA$ and large $\tb$ is 
also excluded by the new \bmm\ constraint~\footnote{We see in both
panels of Fig.~\ref{fig:MAtb} `archipelagos' at large $\MA$ that might evolve
with more sampling into connected regions allowed at the 95\% CL.
In the NUHM1, this region is mainly populated by the points with small $m_0$
and large $m_{1/2}$ visible in the corresponding panel of Fig.~\ref{fig:m0m12}.}.

\section{Relative Impacts of \boldmath{\atlasf}, \boldmath{\bmm} and XENON100}

In this Section we analyze in more detail the different
impacts of the new \atlasfive, \bmm\ and XENON100 constraints on the CMSSM and NUHM1.

\begin{figure*}[htb!]
\begin{center}
\resizebox{7.5cm}{!}{\includegraphics{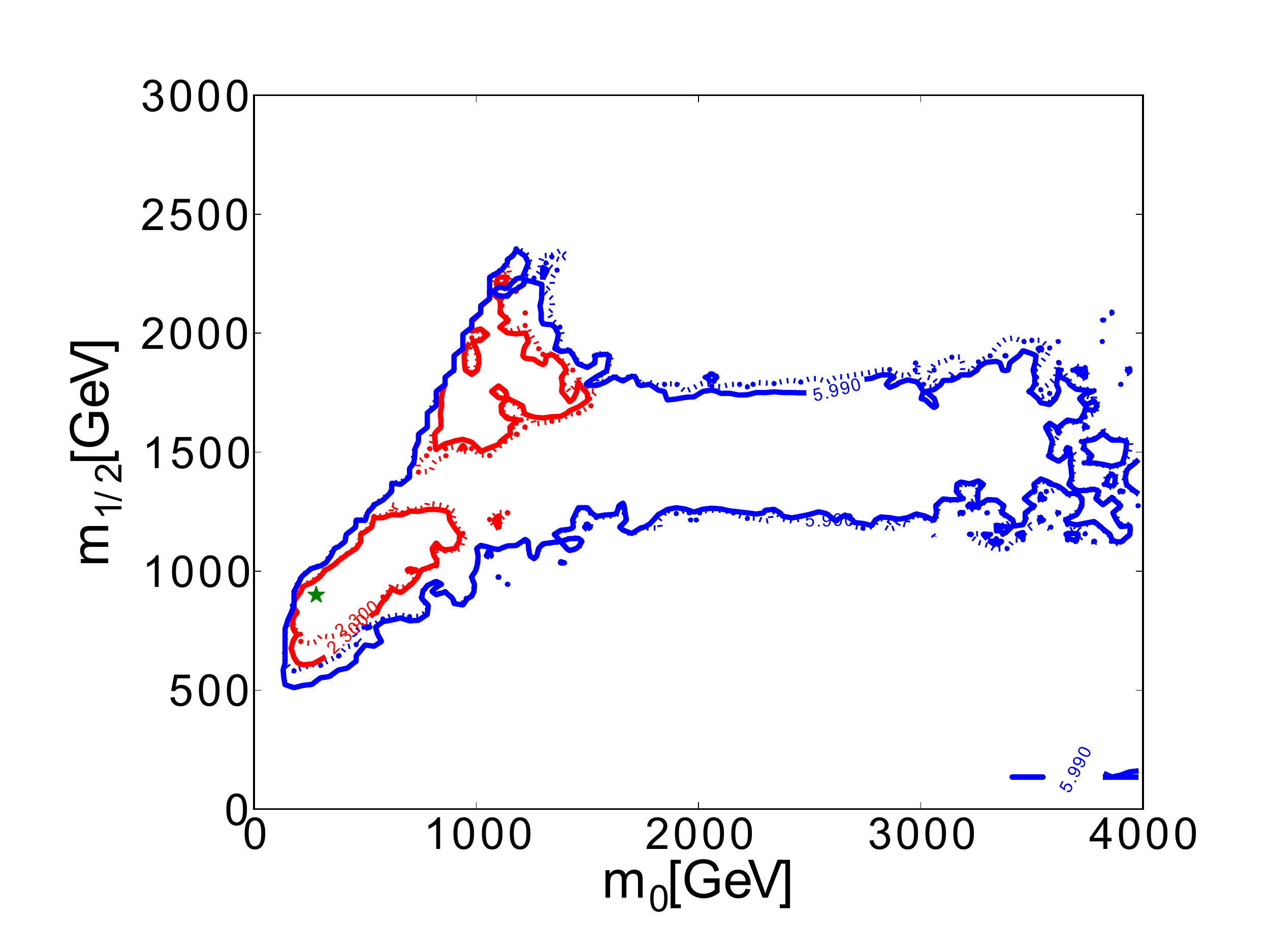}}
\resizebox{7.5cm}{!}{\includegraphics{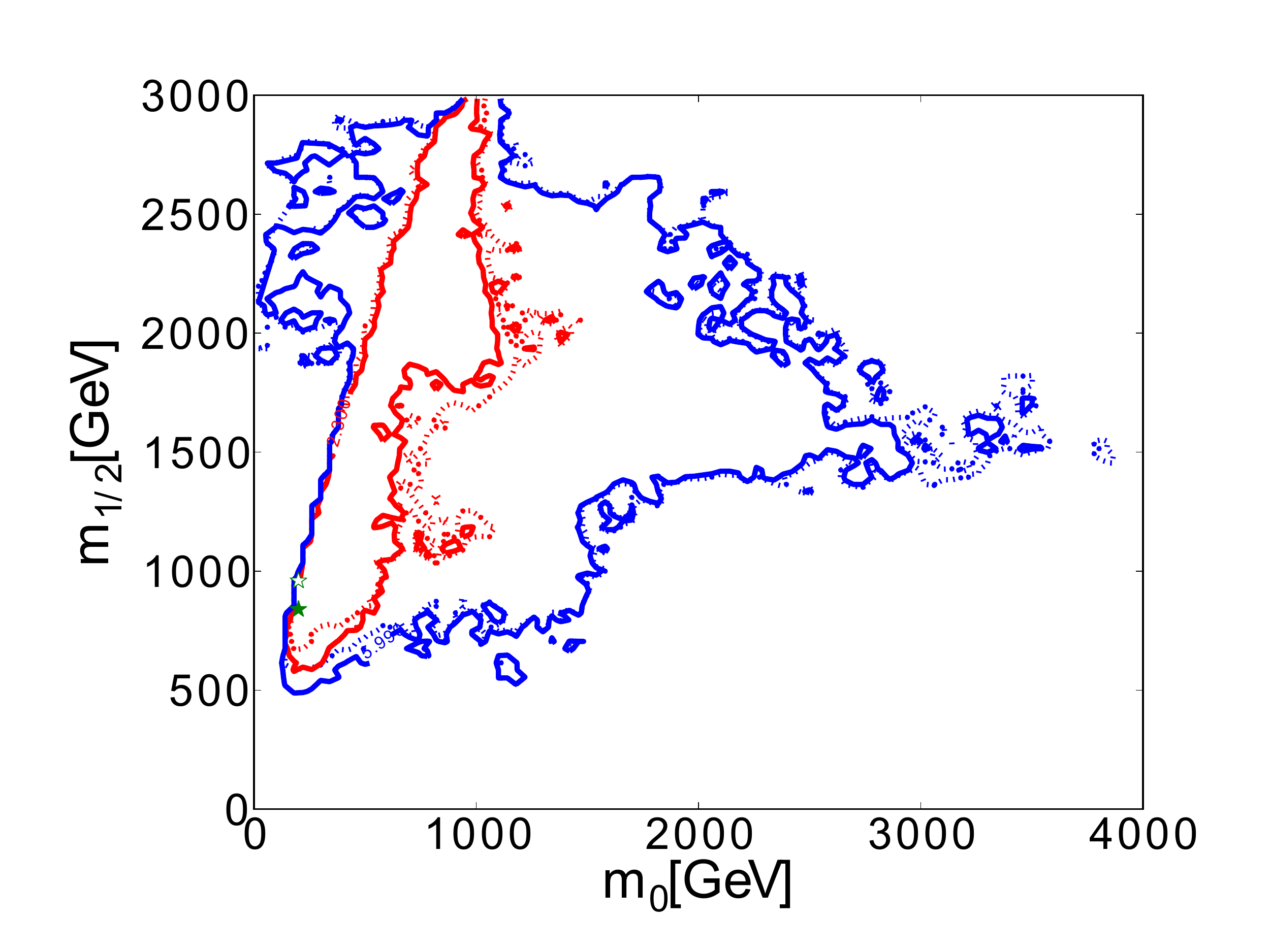}}
\resizebox{7.5cm}{!}{\includegraphics{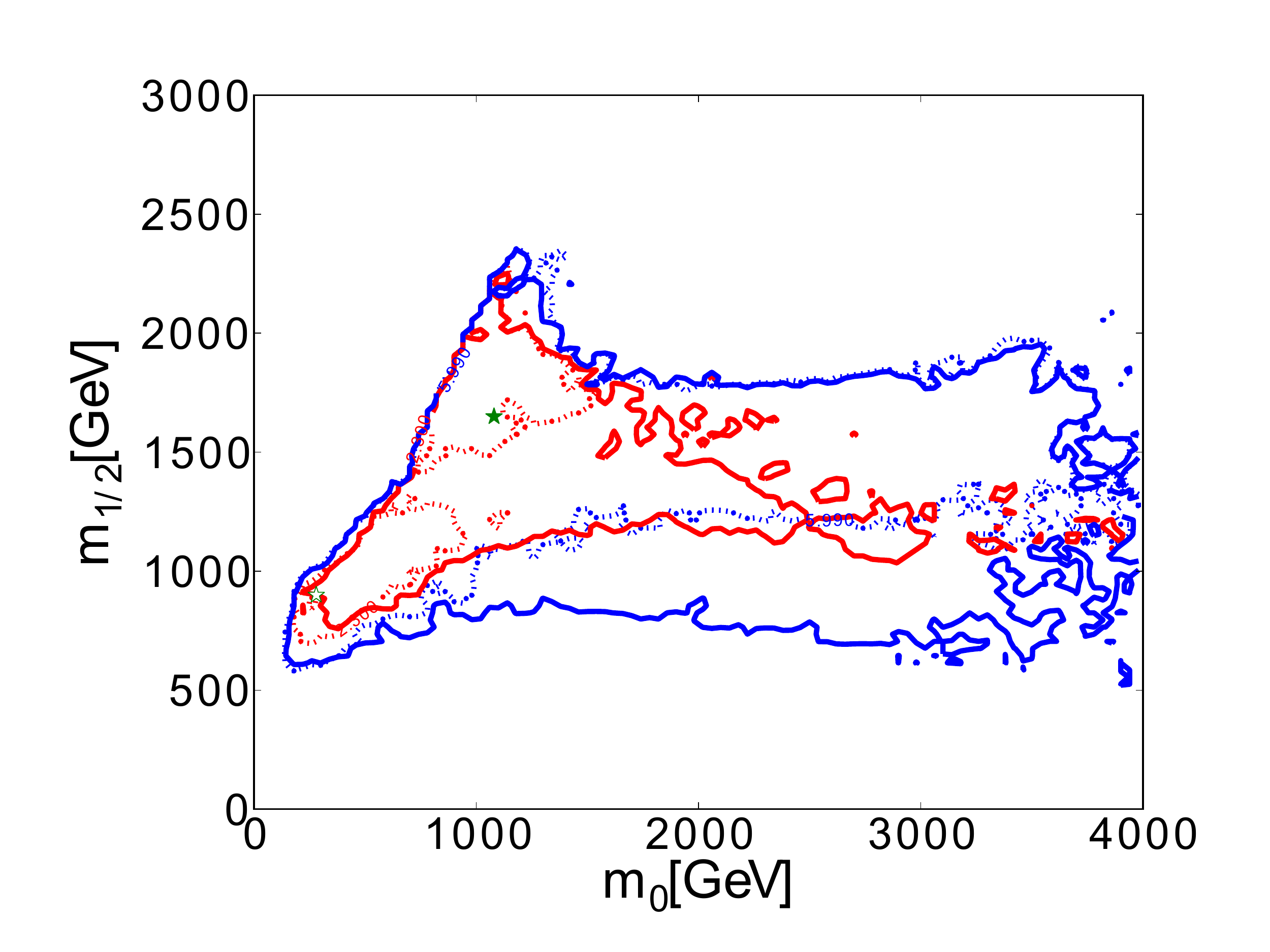}}
\resizebox{7.5cm}{!}{\includegraphics{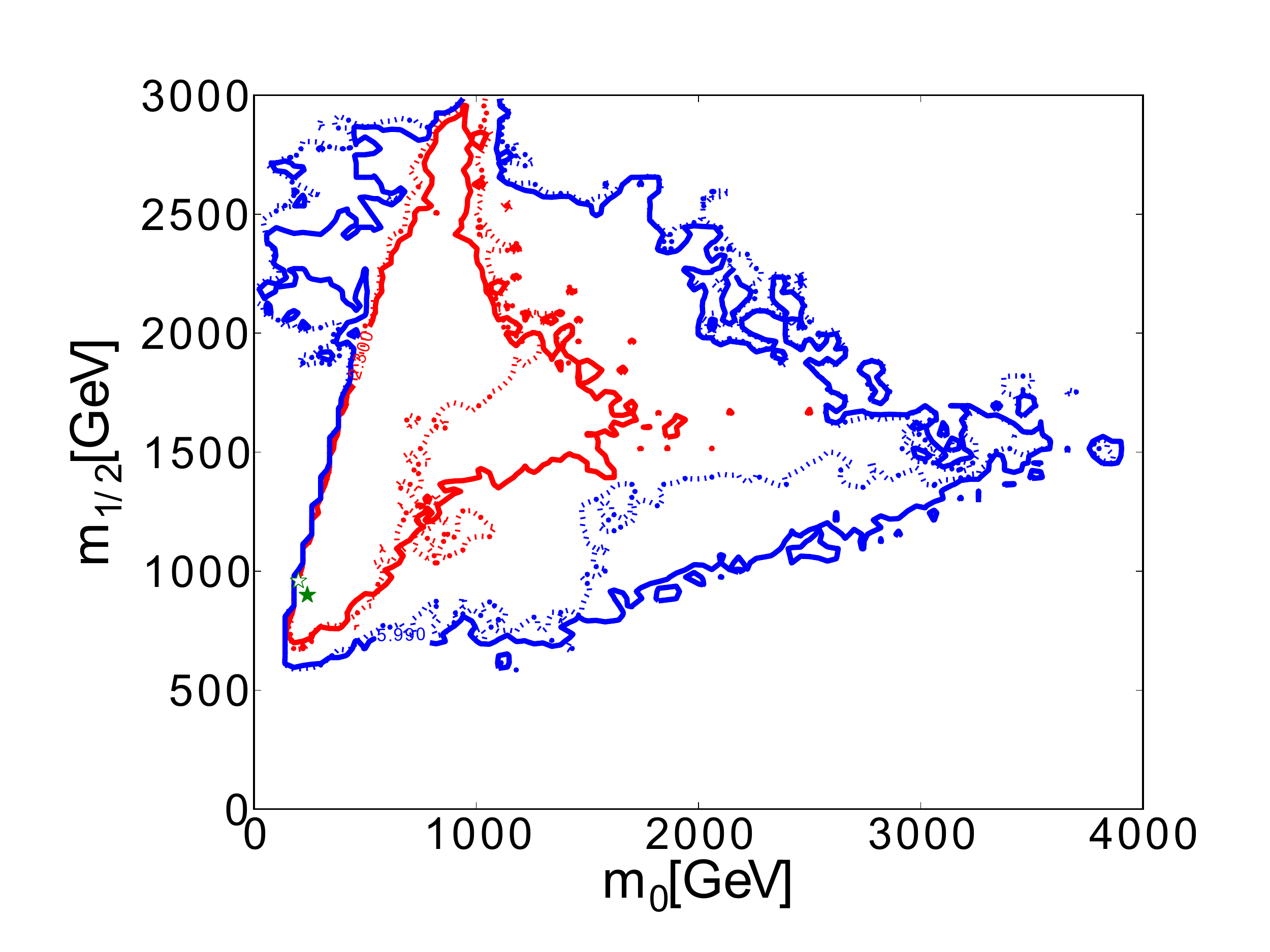}}
\resizebox{7.5cm}{!}{\includegraphics{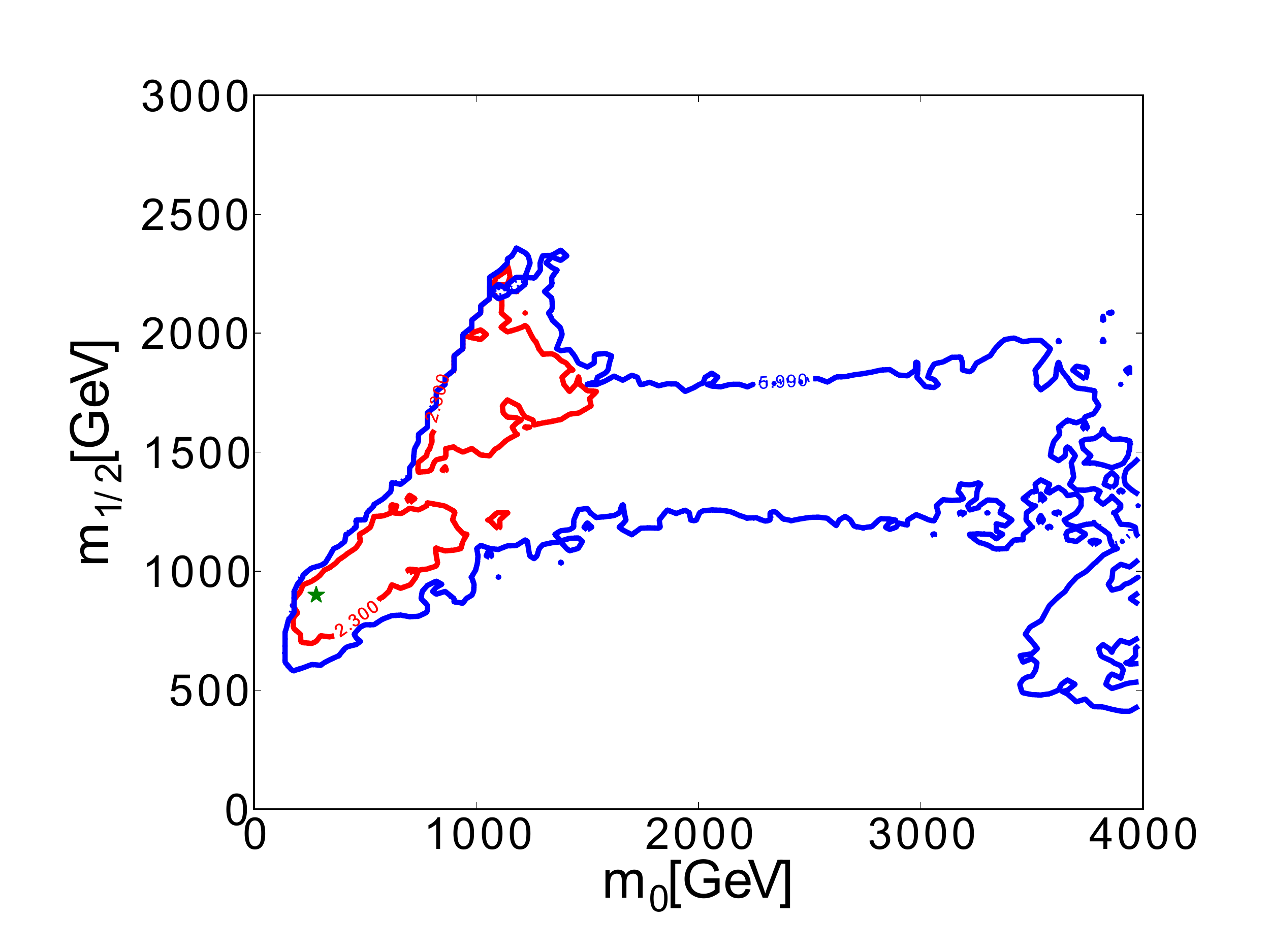}}
\resizebox{7.5cm}{!}{\includegraphics{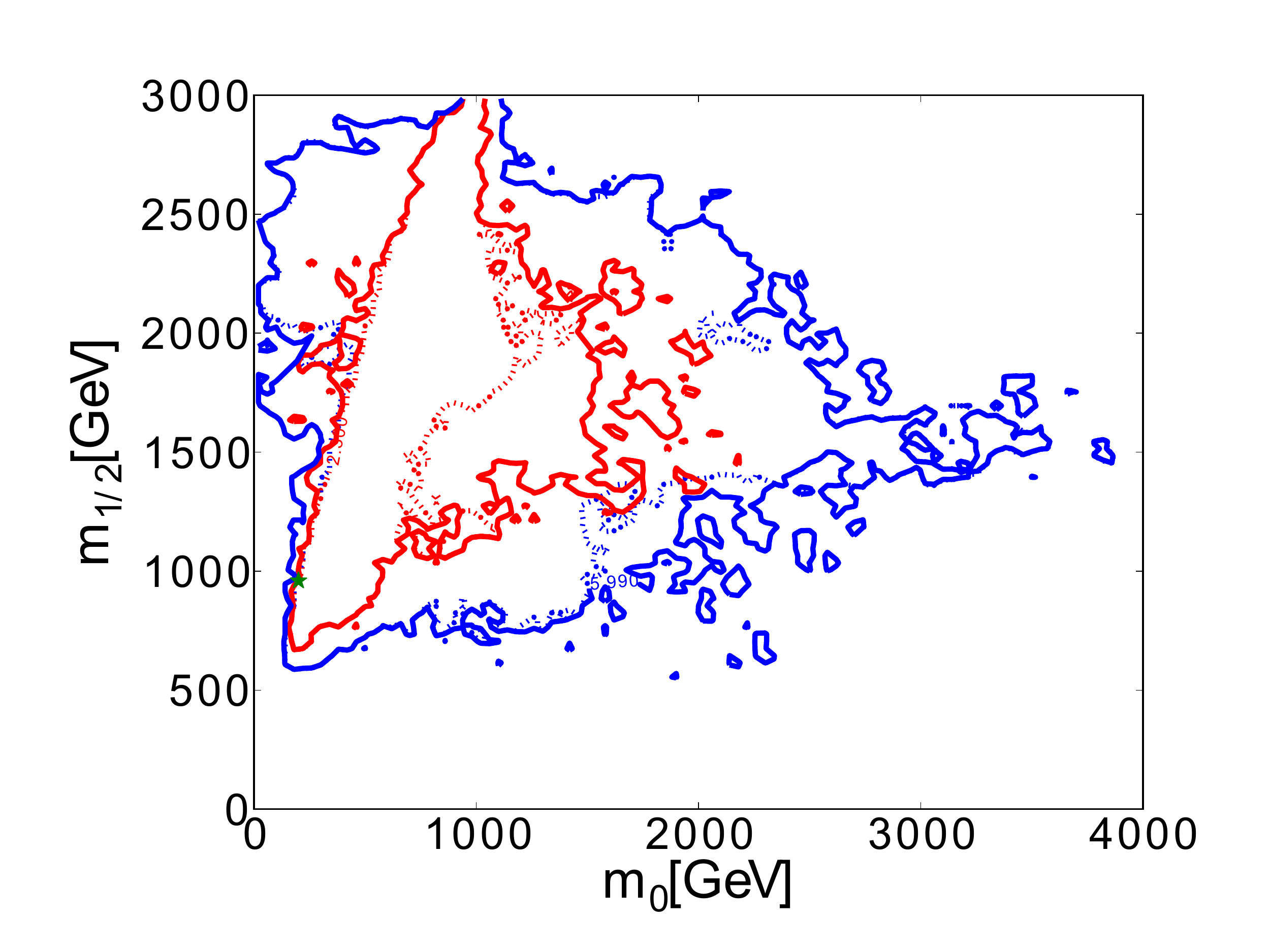}}
\end{center}
\vspace{-1cm}
\caption{\it The $(m_0, m_{1/2})$ planes in the CMSSM (left panels) and
  the NUHM1 (right panels) displaying the effects of dropping the
  \atlasfive\ constraint (top panels), the new \bmm\ constraint
  (middle panels) and the new XENON100 constraint (bottom panels). In each case, the full global fit is represented by an
  open green star and dashed blue and red lines for the 68 and 95\%~CL
  contours, whilst the fits to the incomplete data sets are represented
  by closed stars and solid contours.} 
\label{fig:comparem0m12}
\end{figure*}

Fig.~\ref{fig:comparem0m12} displays the impacts in the $(m_0, m_{1/2})$
planes of the CMSSM (left panels) and the NUHM1 (right panels) of
dropping the \atlasfive\ constraint (top panels), the new
\bmm\ constraint (middle panels), or the new XENON100 constraint (bottom panels). In each panel, we compare the 68 and
95\%~CL contours in the full global fit (dashed lines) with a fit
dropping one of the constraints (solid lines). The corresponding best fit points
are shown as open and closed stars.

In the CMSSM case, very little effect is seen when the \atlasfive\ constraint
is dropped. On the
other hand, when \bmm\ is dropped there are significant extensions
of the 68 and 95\% CL regions: the two 68\% CL `islands' at low $m_0$ merge and
extend to larger $m_0$, and the 95\% band extending to large $m_0$ broadens
to a much larger range of $m_{1/2}$. When the new XENON100 constraint is
dropped, the main effect is the appearance of the focus-point region at large $m_0$
and small $m_{1/2}$.

In the NUHM1 case, dropping the \atlasfive\ constraint again has little
effect on the contours, except at small $m_0$ and $m_{1/2}$.
Dropping the \bmm\ constraint has more effect on the 68 and
95\%~CL contours in the NUHM1.
The effect of dropping the new XENON100 constraint is
qualitatively similar to that of dropping the new \bmm\ constraint,
except that it allows a larger region at small $m_{1/2}$ and large $m_0$.

\begin{figure*}[htb!]
\begin{center}
\resizebox{7.5cm}{!}{\includegraphics{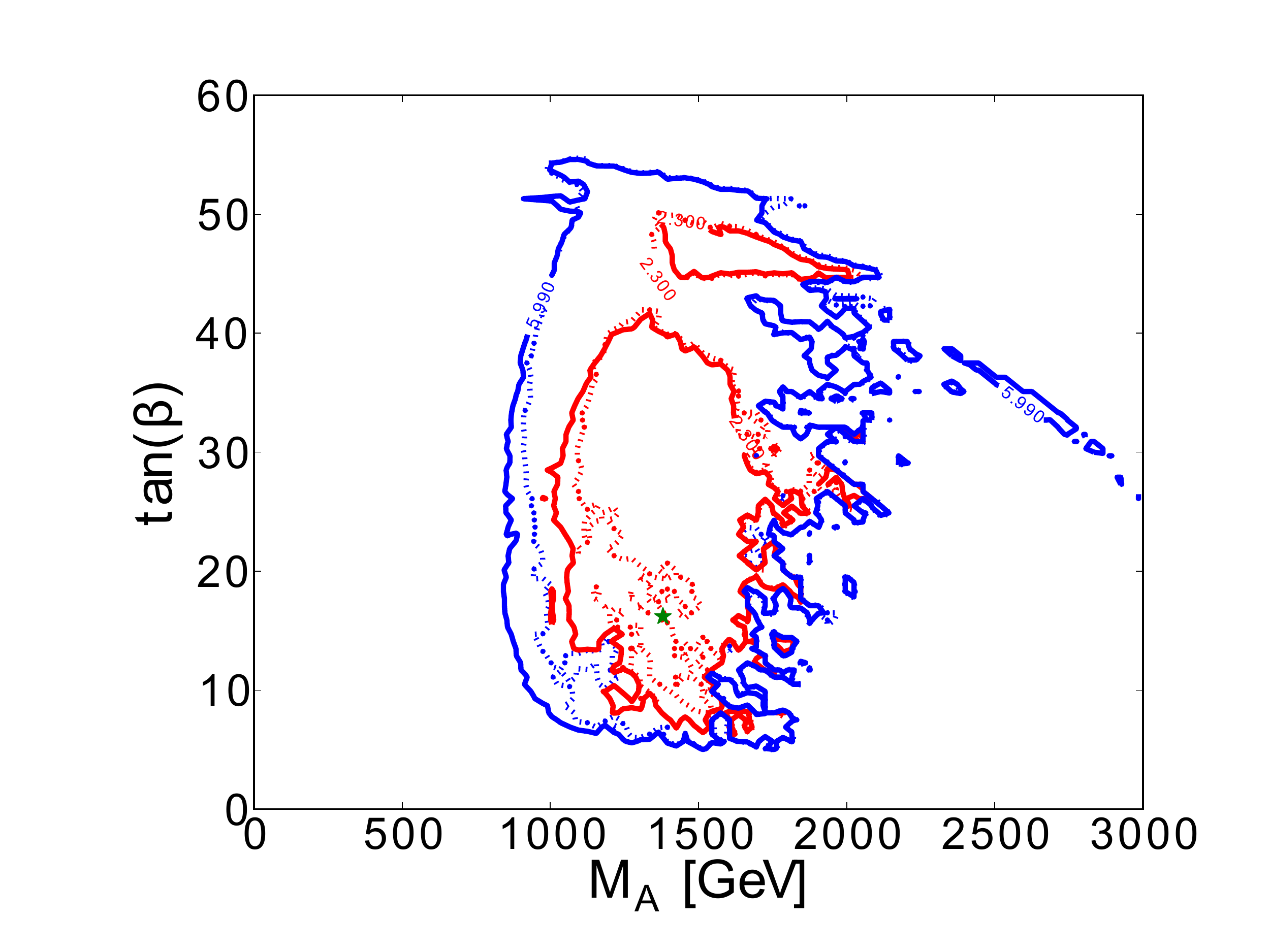}}
\resizebox{7.5cm}{!}{\includegraphics{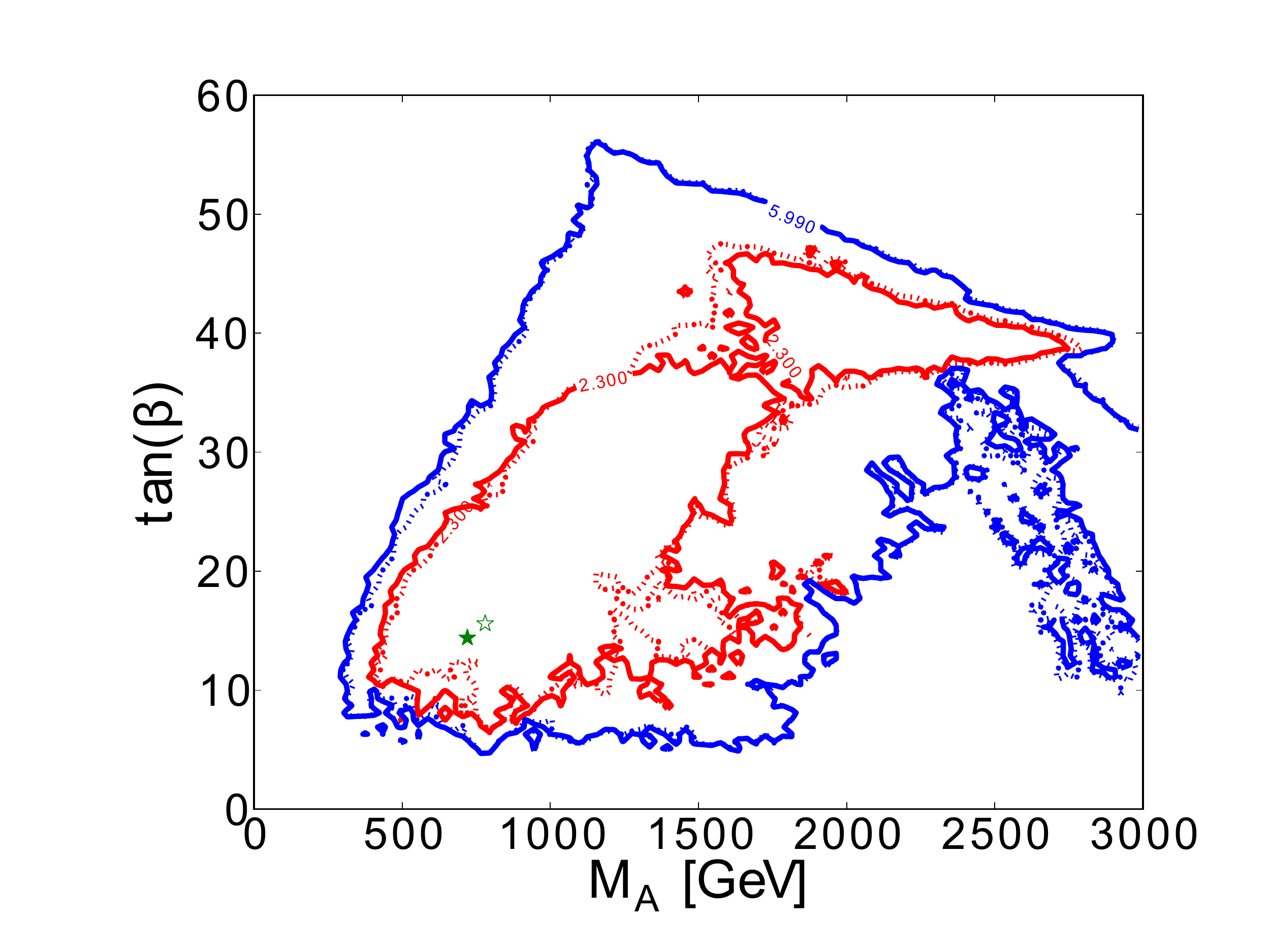}}
\resizebox{7.5cm}{!}{\includegraphics{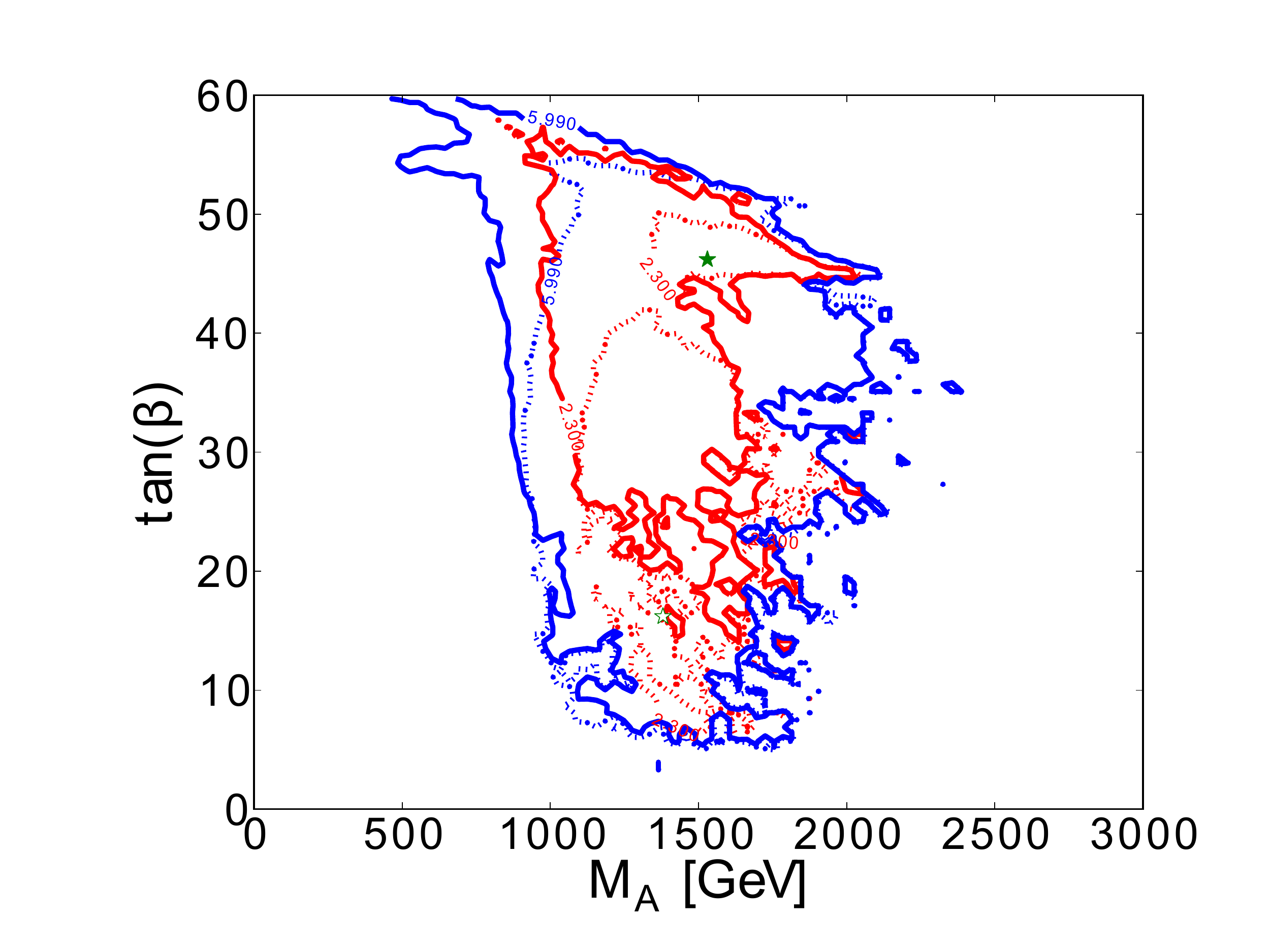}}
\resizebox{7.5cm}{!}{\includegraphics{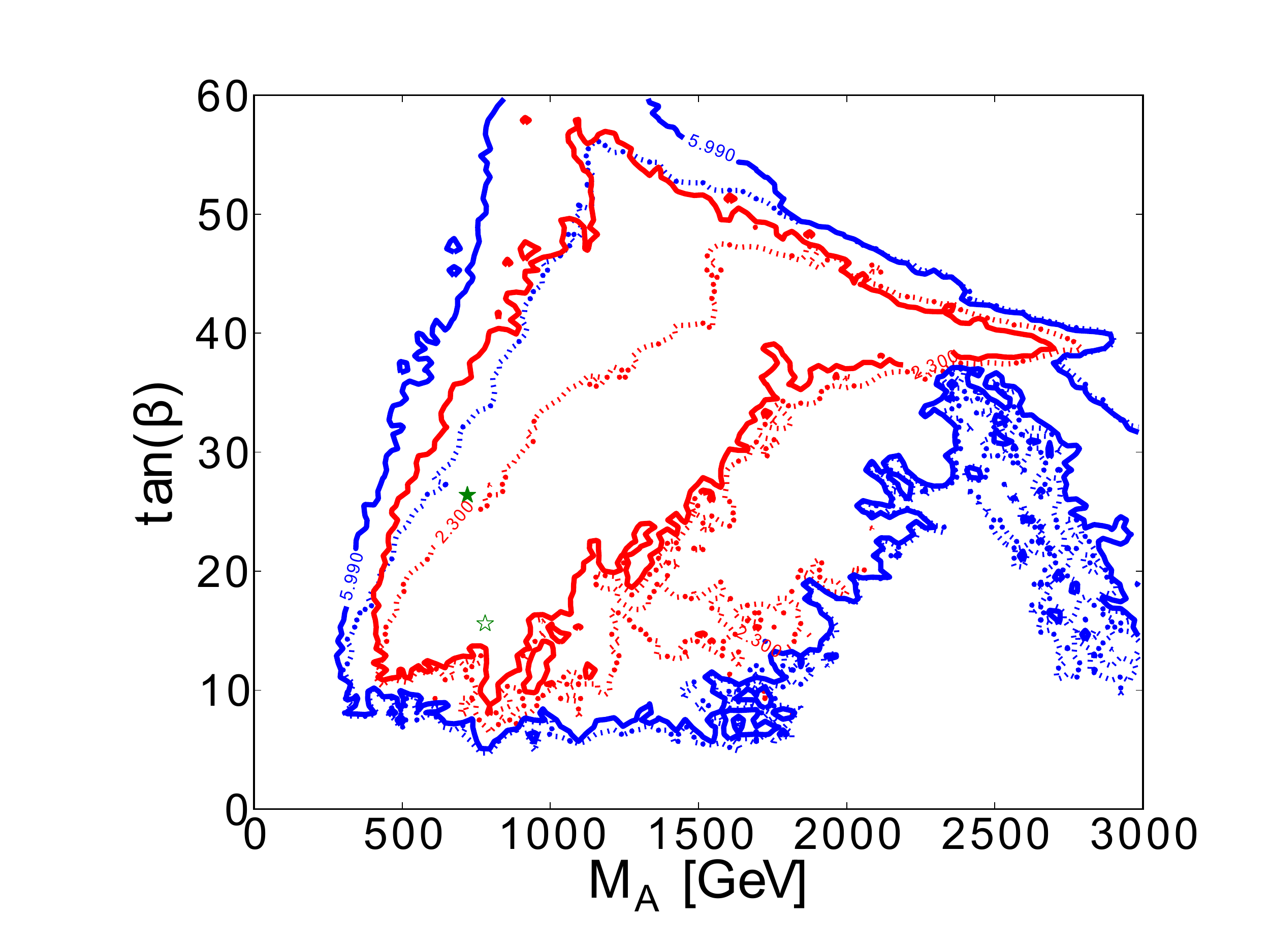}}
\resizebox{7.5cm}{!}{\includegraphics{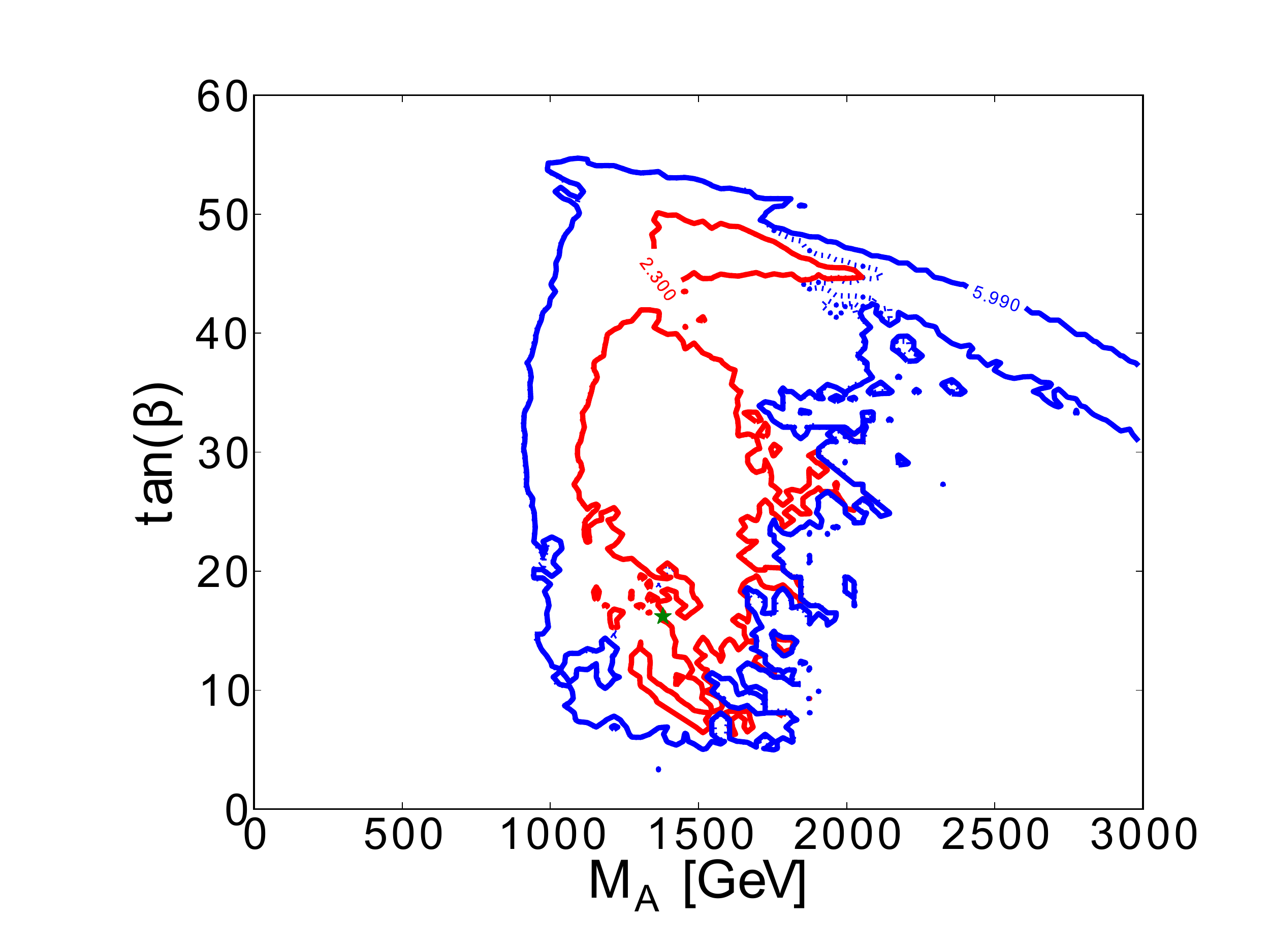}}
\resizebox{7.5cm}{!}{\includegraphics{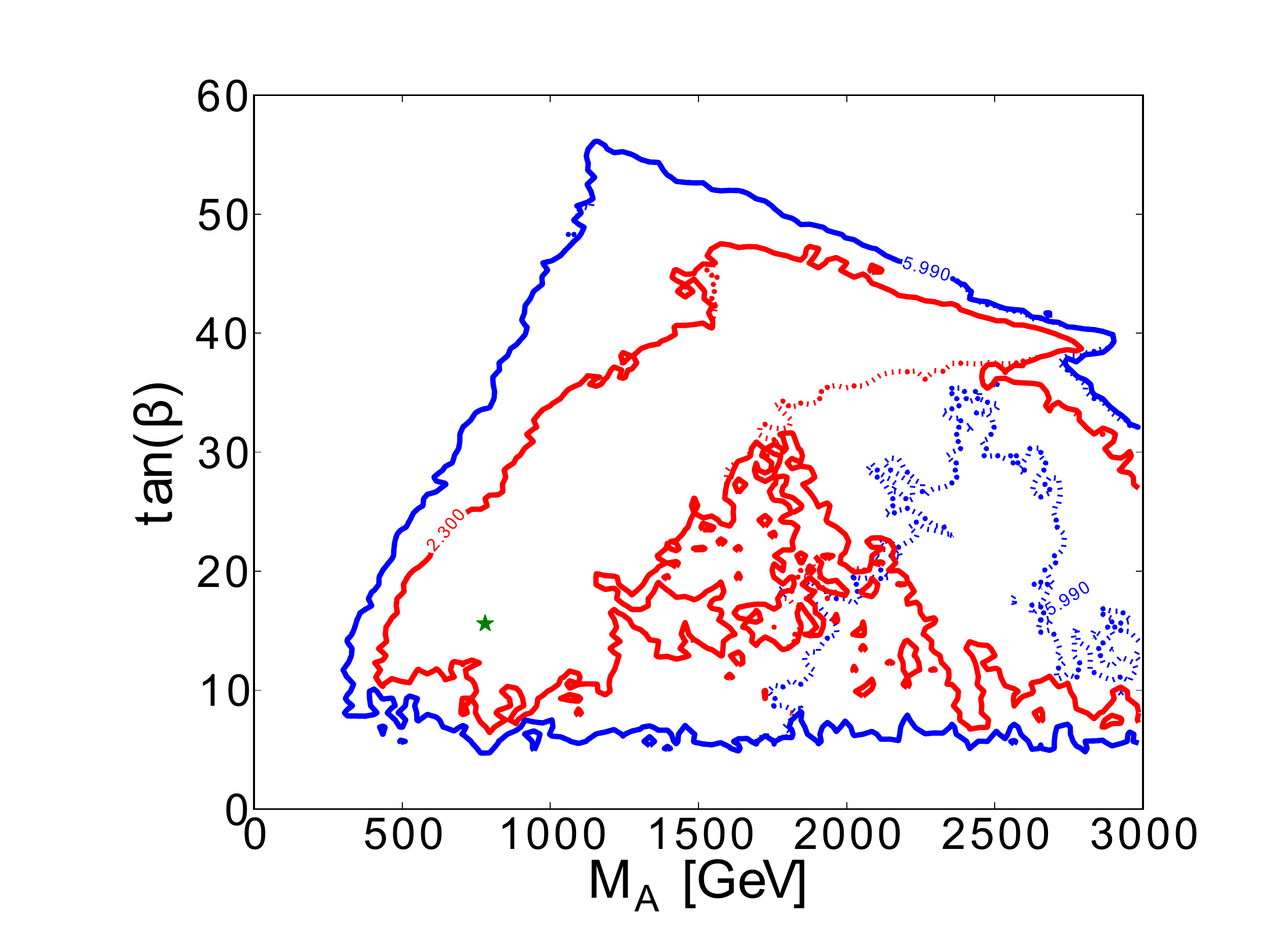}}
\end{center}
\vspace{-1cm}
\caption{\it The $(\MA, \tb)$ planes in the CMSSM (left panels) and the
  NUHM1 (right panels) displaying the effects of dropping the
  \atlasfive\ constraint (top panels), the new \bmm\ constraint
  (middle panels) or the new XENON100 constraint (bottom panels). In each case, the full global fit is represented by an
  open green star and dashed blue and red lines for the 68 and 95\%~CL
  contours, whilst the fits to the incomplete data sets are 
represented by closed stars and solid contours.}
\label{fig:compareMAtb}
\end{figure*}

Similarly, Fig.~\ref{fig:compareMAtb} displays the impacts in the 
$(\MA, \tb)$ planes of the CMSSM (left) panels and the NUHM1 (right
panels) of dropping the \atlasfive\ constraint (top panels), the new
\bmm\ constraint (middle panels) or the new XENON100 constraint (bottom panels). As in the $(m_0, m_{1/2})$ planes, we
see relatively little impact in either the CMSSM or the NUHM1 from dropping the
\atlasfive\ constraint, and a much bigger effect from dropping the new
\bmm\ constraint, particularly at small $\MA$ and large $\tb$. In both cases,
the \bmm\ constraint reduces drastically the value of $\tb$ at the best-fit point from $\sim 46 (27)$ in the CMSSM (NUHM1) to
$\sim 16$. On the other hand, most of the impact of dropping the new XENON100 constraint is to expand the
preferred region at large $\MA$ and $tb < 40$. 

We analyse the various $\chi^2$
contributions and best-fit points in more detail in 
Table~\ref{tab:NUHM1fits}. Here we display the principal changes in the
best-fit 
parameters from the new global fit (second column) when dropping the
ATLAS$_{\rm 5/fb}$ constraint (third column), dropping the new
\bmm\ constraint (fourth column), dropping the new XENON100 constraint (fifth column), and dropping all three
new constraints, i.e., the analysis
of~\cite{mc75} (last column). We see that it is \bmm\ that, among the new
constraints, has the most significant impact on the best-fit parameters of the
NUHM1, principally $\tb$.
In the lower rows we show the $\chi^2$ contributions of the direct
  SUSY searches, denoted as $\Delta \chi^2$ for $\ETslash$, \bmm\ and \gmt.
One can clearly observe that
the contribution of \gmt\ to the total $\chi^2$ function increases when
\bmm\ and the other new constraints are incorporated, demonstrating once again 
that the new data are pushing simple supersymmetry-breaking models such
as the NUHM1 (and CMSSM) out of the `sweet spot' where the experimental 
measurement of \gmt\ could easily be explained.

\begin{table*}[!tbh!]
\renewcommand{\arraystretch}{1.5}
\begin{center}
\begin{tabular}{|c||c|c|c|c|c|} \hline
Quantity & Full fit & w/o ATLAS$_{\rm 5/fb}$ & w/o $B_s \to \mu^+ \mu^-$ & w/o XENON100 & LHC$_{\rm 1/fb}$ fit \\
\hline \hline
Total $\chi^2$ & 31.3  & 30.1 & 28.7 & 31.3 & 28.9 \\
\hline
$m_0$ (GeV)  & 240 & 240 & 270 & 240 & 270 \\
$m_{1/2}$ (GeV) & 970 & 870 & 920 & 970 & 920 \\
$A_0$ (GeV) & 1860 & 1860 & 1730 & 1860 & 1730 \\
$\tb$ & 16 & 15 & 27 & 16 & 27 \\
$m_H^2$ (GeV$^2$) & $- 6.5 \times 10^6$ & $- 5.6 \times 10^6$ & $- 5.5 \times 10^6$ & $- 6.5 \times 10^6$ &$- 5.5 \times 10^6$ \\
\hline \hline
$\Delta \chi^2$ $\ETslash$ & 1.18 & 1.06$^*$ & 1.36 & 1.18 & 0.85$^*$ \\
$\Delta \chi^2$ $B_s \to \mu^+ \mu^-$ & 1.70 & 1.12 & 0.21$^*$ & 1.70 & 0.21$^*$ \\
$\Delta \chi^2$ XENON100 & 0.13 & 0.13 & 0.13 & 0.14$^*$ & 0.14$^*$ \\
$\Delta \chi^2$ \gmt\ & 7.82 & 7.41 & 5.99 & 7.82 & 5.99 \\
\hline
\end{tabular}
\caption{\it 
Principal changes in the best-fit parameters in the NUHM1 from the
  new global fit (second column) when dropping the ATLAS$_{\rm 5/fb}$
  constraint (third column), dropping the new \bmm\ constraint (fourth
  column), dropping the new XENON100 constraint (fifth column) and dropping
  all three new constraints, i.e., the analysis 
  of~\cite{mc75} (last column). The bottom
  four lines display the contributions to $\chi^2$ of the $\ETslash$,
  \bmm, XENON100 and \gmt\ constraints. The numbers marked by $^*$ are for the
  implementations of these constraints in~\protect\cite{mc75}.
 }
\label{tab:NUHM1fits}
\end{center}
\end{table*}


\section{Predictions for Physical Observables}

In this section we present one-dimensional
$\chi^2$ likelihood functions for slected physical observables, based on
the above surveys of the CMSSM and NUHM1 parameter spaces, comparing
results based on the \lhco\ and \lhcf\ data sets.

\subsection*{\it The Gluino Mass}

\begin{figure*}[htb!]
\resizebox{7.9cm}{!}{\includegraphics{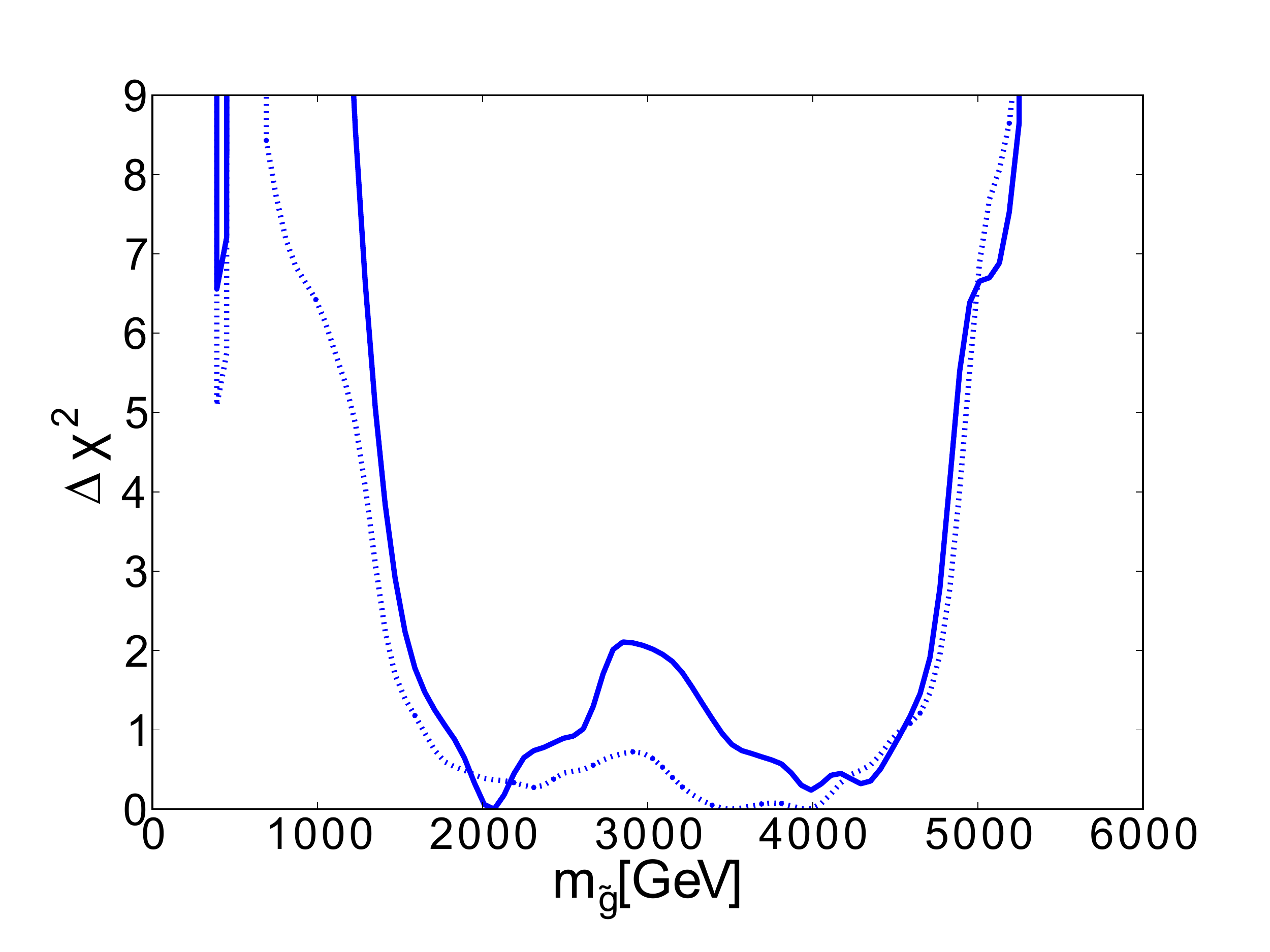}}
\resizebox{7.9cm}{!}{\includegraphics{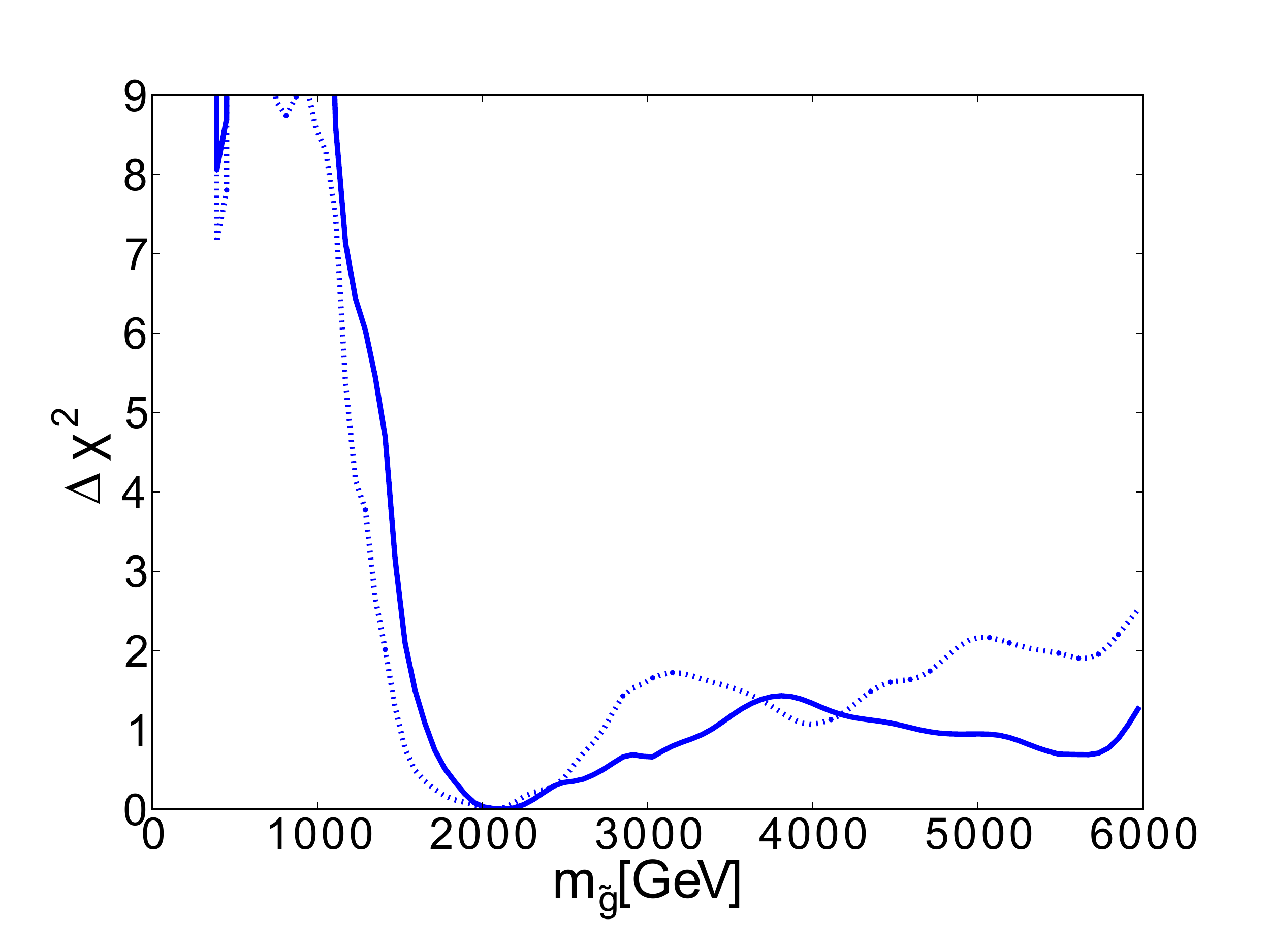}}
\caption{\it The $\chi^2$ likelihoods in the CMSSM (left panel) and
  NUHM1 (right panel) as functions of   $\mgl$ based on global fits to
  the \lhcf, \bmm\ and new XENON100 data set (solid lines), and to the \lhco\ data set (dashed
  lines).
}
\label{fig:mgl}
\end{figure*}

Fig.~\ref{fig:mgl} displays the $\chi^2$ functions for $\mgl$ in the
CMSSM (left panel) and the NUHM1 (right panel), for fits to the
\lhcf\ data including the new \bmm\ and XENON100 limits (solid lines) 
and to the \lhco\ data including the \bmm\ limits then available
(dashed lines). 
In the case of the CMSSM, we see a bimodal distribution, with local
minima at $\mgl \sim 2000 \gev$ and $\sim 4000 \gev$, corresponding to
the best fits in the low- and high-mass islands discussed
above, whose values of $\chi^2$ are almost indistinguishable. 
Intermediate values of $\mgl$ are disfavoured by $\Delta \chi^2 \lsim
2$, though the envelope of the $\chi^2$ function could only be reduced by
further sampling. In the case of the NUHM1, a much more weakly bimodal
distribution is visible.
In the CMSSM, values of $\mgl$
between $\sim 1500 \gev$ and $\sim 4800 \gev$ are allowed at the $\Delta
\chi^2 < 2$ level, and in the NUHM1 values over $6000 \gev$ are 
allowed at this level. Once again, the ability to move to large masses (in this case the 
gluino mass or $m_{1/2}$) can be traced to the freedom in the NUHM1 of
adjusting $\MA \sim 2 m_\chi$ so that the relic density is at the WMAP value.
At such large masses, there is essentially no further price paid by
$g_\mu - 2$ or other low energy observables as we approach the decoupling limit.

\subsection*{\it The Squark Mass}

\begin{figure*}[htb!]
\resizebox{7.9cm}{!}{\includegraphics{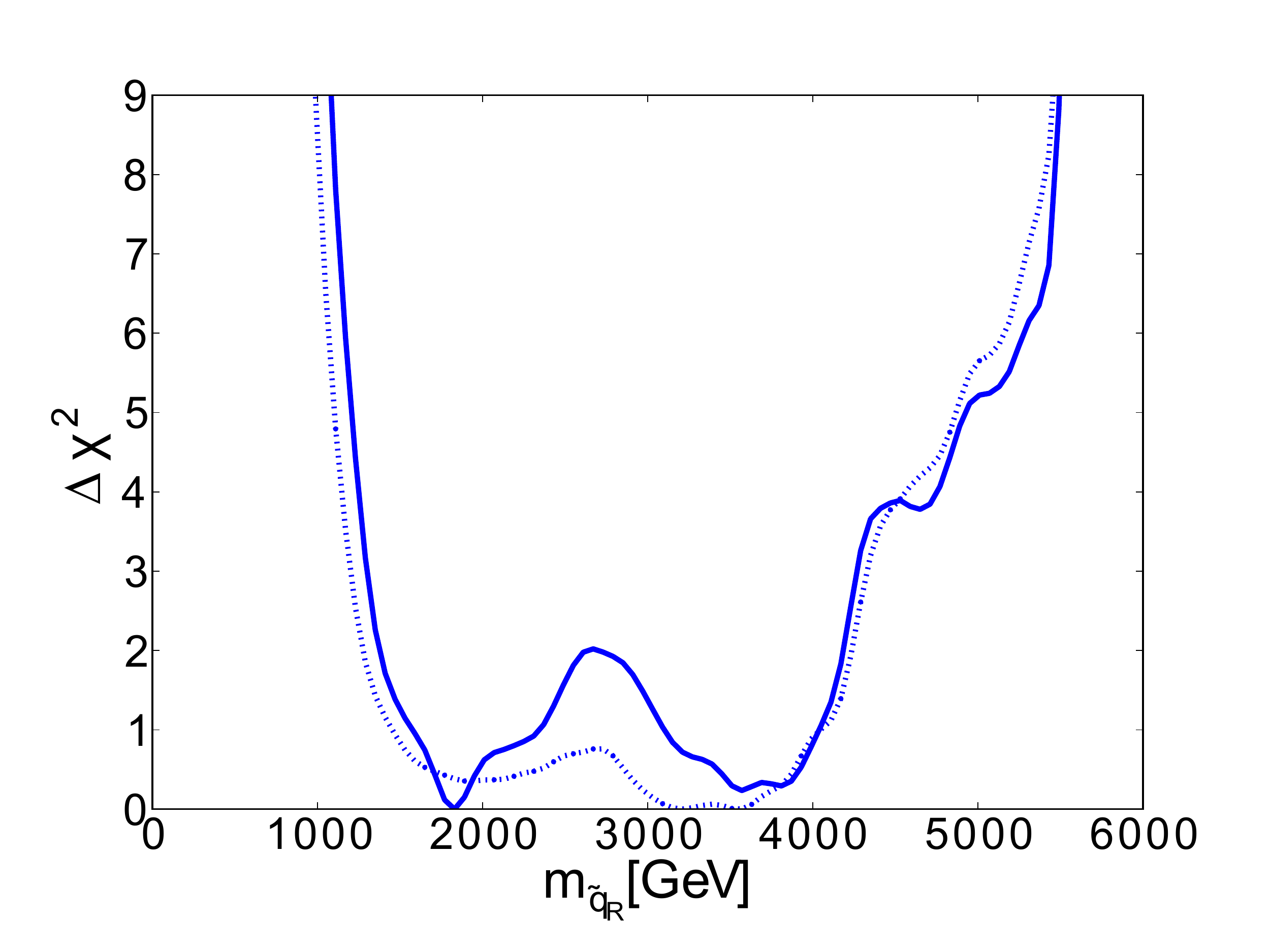}}
\resizebox{7.9cm}{!}{\includegraphics{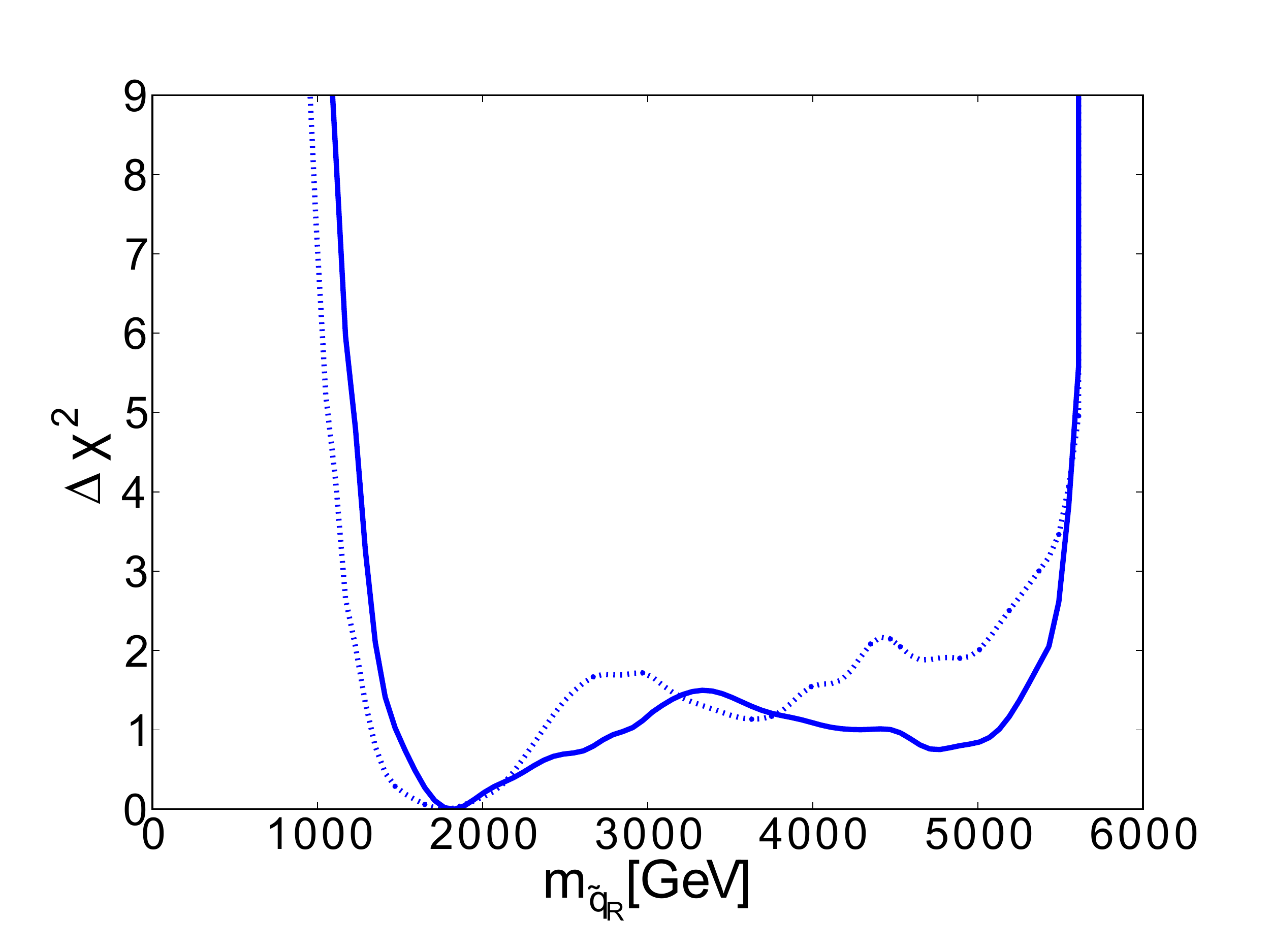}}
\caption{\it The $\chi^2$ likelihoods in the CMSSM (left panel) and NUHM1 (right panel) as functions of
  $\msq$ based on global fits to the \lhcf, \bmm\ and new XENON100 data set (solid lines), and
  to the \lhco\ data set (dashed lines).
}
\label{fig:msq}
\end{figure*}

Fig.~\ref{fig:msq} displays the $\chi^2$ functions for the average mass of the supersymmetric partners
of the five lightest right-handed quarks, $\msq$, in the CMSSM (left panel) and the
NUHM1 (right panel), for fits to the \lhcf, \bmm\ and new XENON100 data (solid lines) and the
\lhco\ data (dashed lines), as in Fig.~\ref{fig:mgl}. 
In the case of the CMSSM, we see local
minima of $\chi^2$ at $\msq \sim 1800$ and $3500 \gev$, separated again
by a range with $\Delta \chi^2 \lsim 2$. At this level of $\Delta \chi^2$, any value of $\msq$
between $\sim 1200 \gev$ and $\sim 4200 \gev$ is allowed. In the case of the NUHM1, there is
a similar structure, but it is less pronounced and the intermediate range has
$\Delta \chi^2 \sim 1$. Only squark mass values above $\sim 5.4 \tev$
receive a substantial $\chi^2$ penalty, which is due to large $m_0$ being disfavoured in this model.

\subsection*{\it The Stop Mass}

\begin{figure*}[htb!]
\resizebox{7.9cm}{!}{\includegraphics{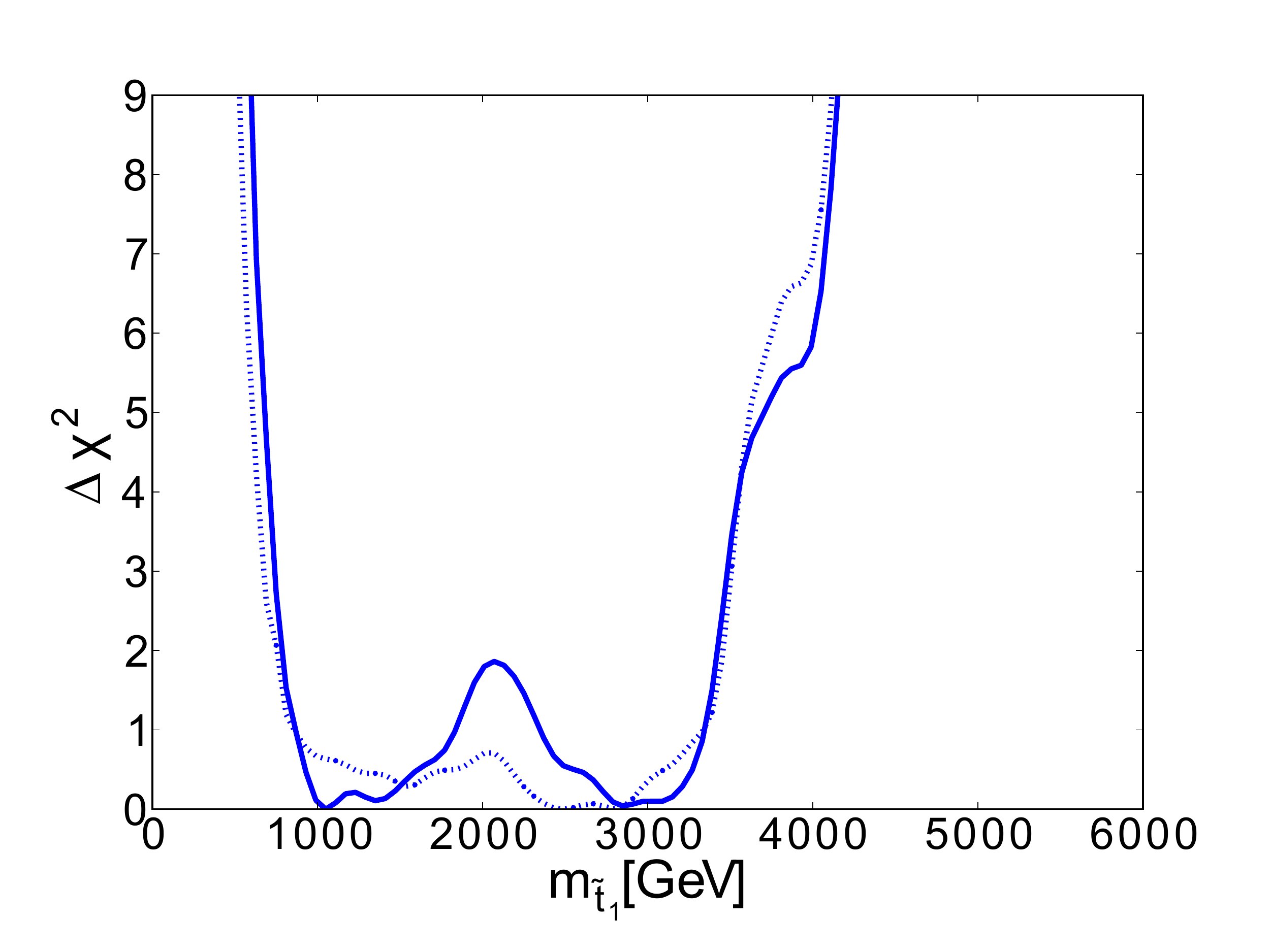}}
\resizebox{7.9cm}{!}{\includegraphics{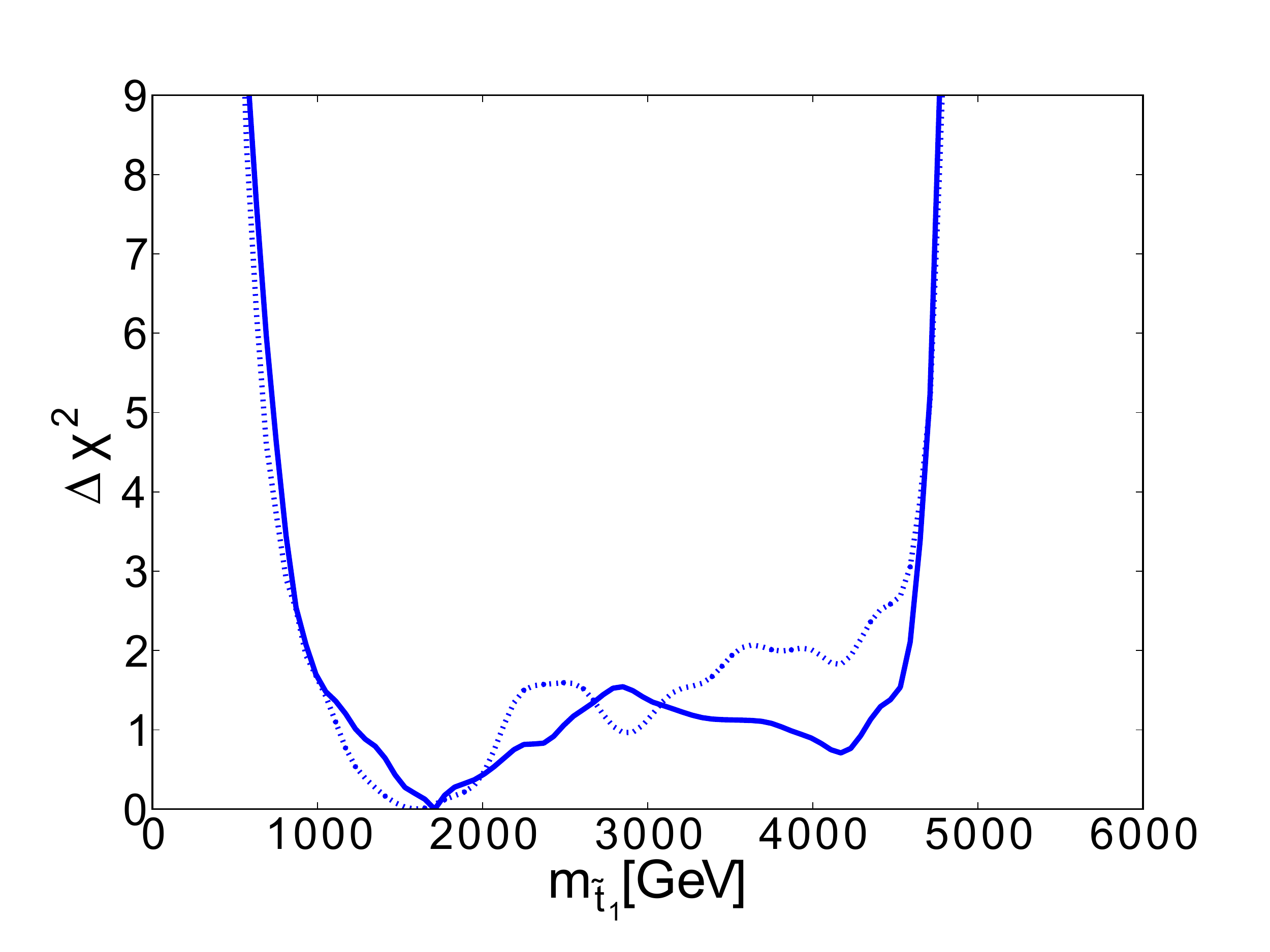}}
\caption{\it The $\chi^2$ likelihoods in the CMSSM (left panel) and
  NUHM1 (right panel) as functions of   $m_{\tilde t_1}$ based on global fits to
  the \lhcf, \bmm\ and new XENON100 data set (solid lines), and to the \lhco\ data set (dashed
  lines).
}
\label{fig:mstop}
\end{figure*}

Fig.~\ref{fig:mstop} displays the $\chi^2$ functions for the lighter stop squark mass $m_{\tilde t_1}$ in
the CMSSM (left panel) and the NUHM1 (right panel), for fits to the
\lhcf, \bmm\ and new XENON100 data set (solid lines) and the \lhco\ data (dashed lines), 
as in Fig.~\ref{fig:mgl}. We see that the CMSSM $\chi^2$ is minimized for $m_{\tilde t_1} \sim 1100 \gev$, 
with a secondary minimum for $m_{\tilde t_1} \sim 2900 \gev$,
whereas in the NUHM1 $\chi^2$ is minimized for $m_{\tilde t_1} \sim 1700 \gev$ with values above around 
$4.5 \tev$ being disfavoured. In both models, values of $m_{\tilde t_1} < \msq$ are expected.

\subsection*{\it The Stau Mass}

\begin{figure*}[htb!]
\resizebox{7.9cm}{!}{\includegraphics{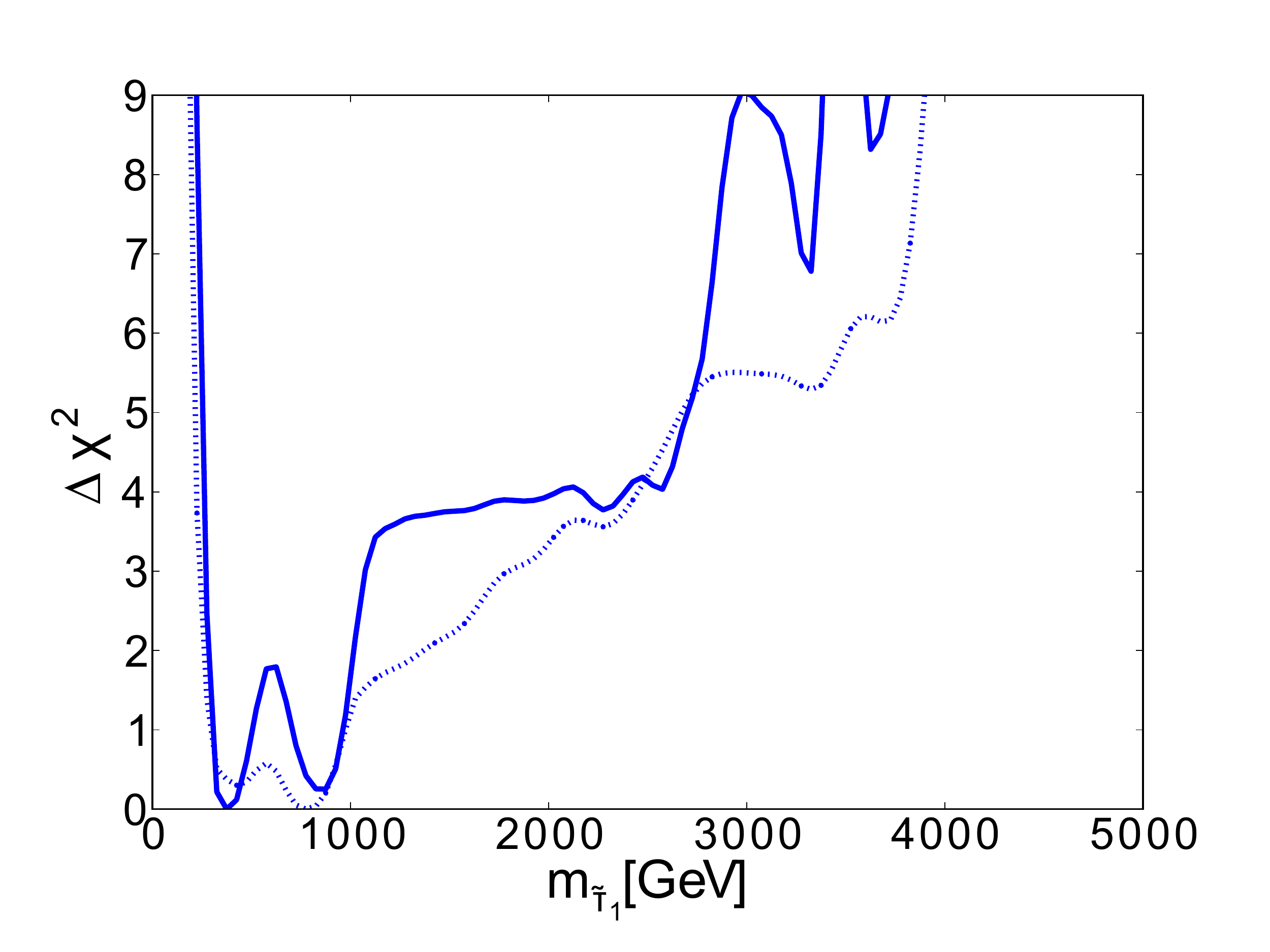}}
\resizebox{7.9cm}{!}{\includegraphics{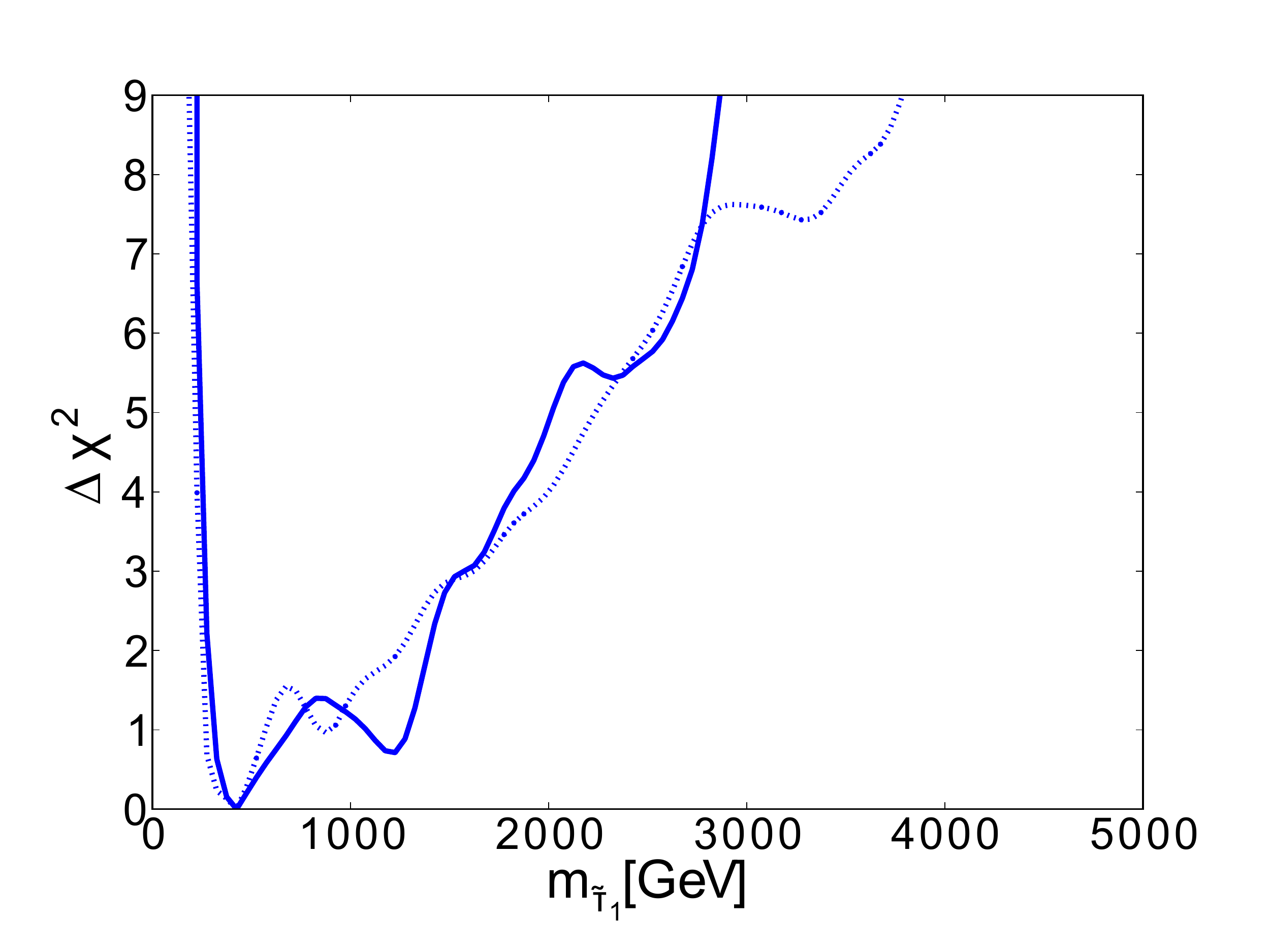}}
\caption{\it The $\chi^2$ likelihoods in the CMSSM (left panel) and
  NUHM1 (right panel) as functions of   $\mstaue$ based on global fits to
  the \lhcf, \bmm\ and new XENON100 data set (solid lines), and to the \lhco\ data set (dashed
  lines).
}
\label{fig:mstaue}
\end{figure*}

Likewise, Fig.~\ref{fig:mstaue} displays the $\chi^2$ functions for $\mstaue$ in
the CMSSM (left panel) and the NUHM1 (right panel), for fits to the
\lhcf, \bmm\ and new XENON100 data set (solid lines) and the \lhco\ data (dashed lines), 
as in Fig.~\ref{fig:mgl}. In this
case, in the CMSSM $\chi^2$ is minimized for $\mstaue \sim 400$ and
$\sim 800 \gev$, whereas in the NUHM1 $\chi^2$ is minimized for $\mstaue
\sim 700 \gev$ with an upper limit around 
$2.5 \tev$.

\subsection*{\it The Higgs Mass}

\begin{figure*}[htb!]
\resizebox{7.9cm}{!}{\includegraphics{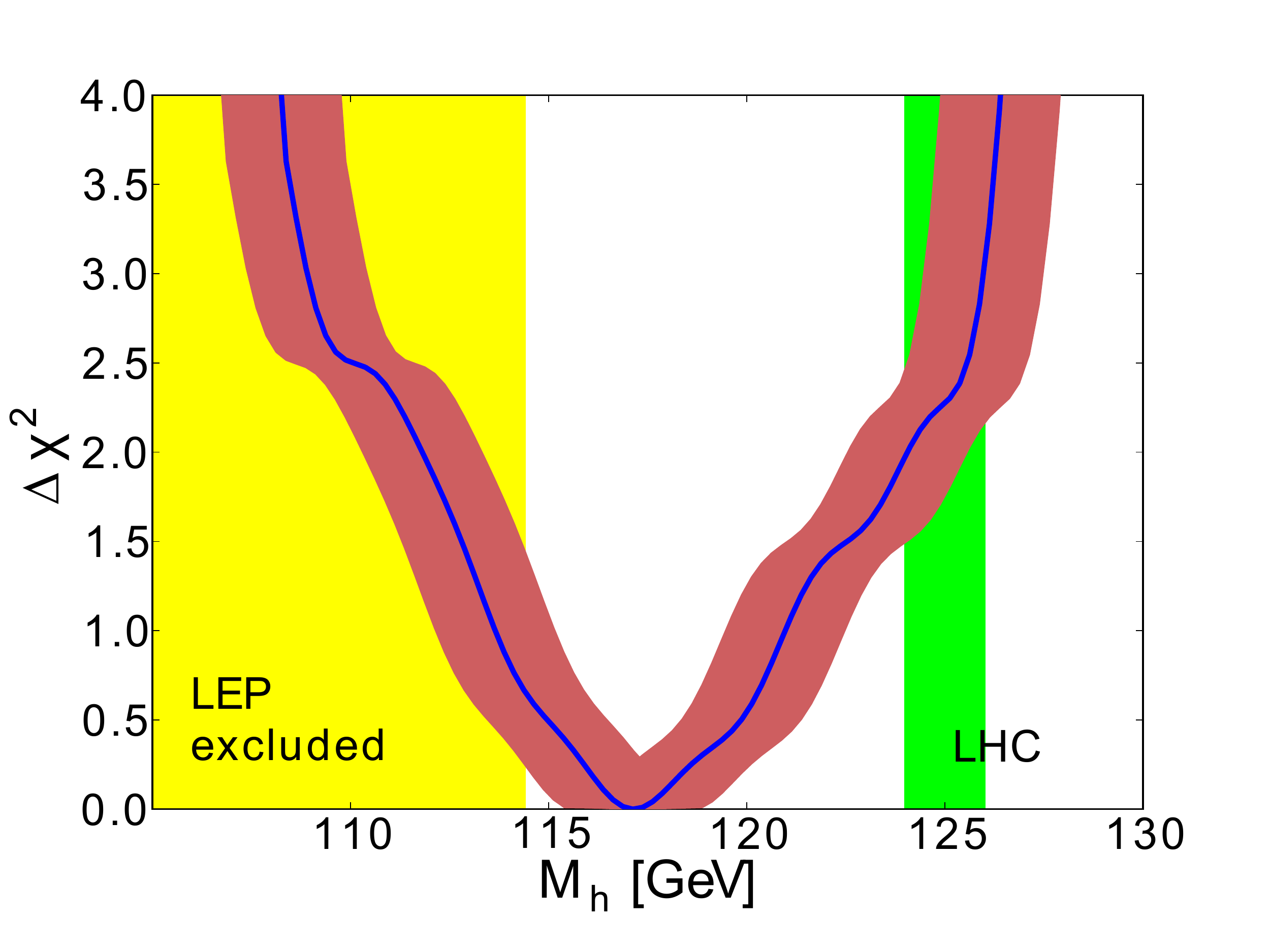}}
\resizebox{7.9cm}{!}{\includegraphics{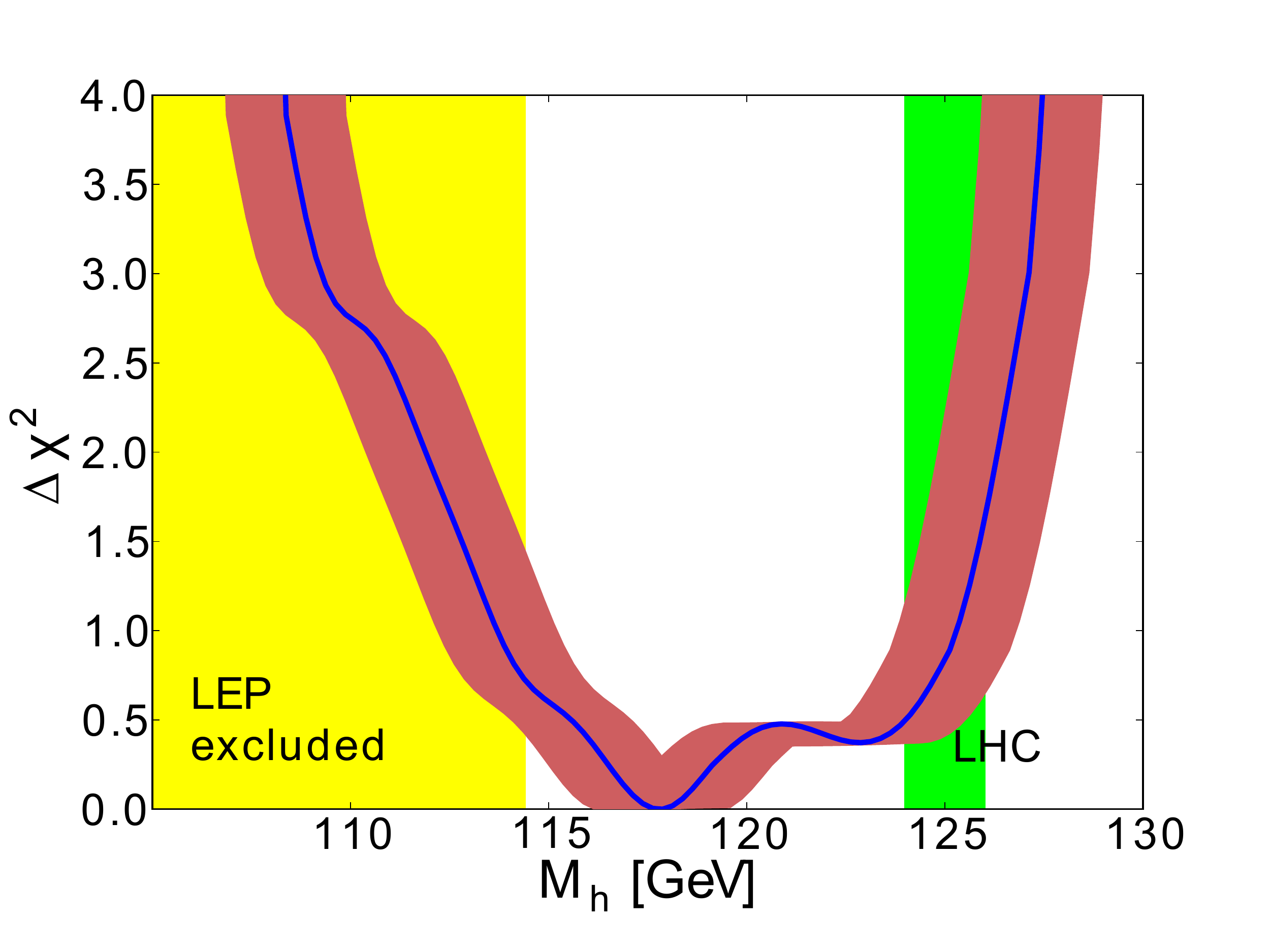}}
\caption{\it The one-dimensional $\Delta \chi^2$ functions for $\Mh$ in
the CMSSM (left) and the NUHM1 (right). The solid lines are for
fits including all the available data including \gmt\ but excluding the direct
LEP~\protect\cite{Barate:2003sz,Schael:2006cr},
Tevatron~\protect\cite{TeVH} and LHC~\protect\cite{CMS47,ATLAS47}  
constraints on $\Mh$, with a red band indicating the estimated theoretical
uncertainty in the calculation of $\Mh$
of $\sim 1.5 \gev$. The shaded ranges are those excluded by LEP searches
for a SM Higgs boson (yellow) and indicated by the LHC discovery (green).
} 
\label{fig:mh2}
\end{figure*}

In Fig.~\ref{fig:mh2} we show as solid lines the $\chi^2$ functions for
$\Mh$ in the CMSSM (left panel) and the NUHM1 (right panel), for   fits
to the \lhcf, \bmm\ and new XENON100 data set, but {\em omitting} all the direct
constraints   on the Higgs mass from the LHC, LEP and Tevatron. Thus,
Fig.~\ref{fig:mh2}   displays the predictions for $\Mh$ in the CMSSM and
NUHM1 on the basis of   the other constraints. As in   previous
publications (see~\cite{mc7,mc75,mc1} and references therein), the plots in
Fig.~\ref{fig:mh2}   include a red band with a width of $1.5 \gev$ to
represent the theoretical uncertainty. These plots show 
{\em directly} the extent to which the indirect prediction of $\Mh$ agrees
with the (provisional)   direct measurement. Also shown in
Fig.~\ref{fig:mh2} is the LEP exclusion for a SM Higgs boson (shaded yellow),  and the range indicated by the LHC
discovery (shaded green). We see that the LHC measurement of $\Mh$
is somewhat larger than would be indicated by our best fit based on other data,
but not disastrously so, with $\Delta \chi^2 \sim 1.5$ in the CMSSM and $\sim 0.5$ in the NUHM1.

\subsection*{\it \bmm\ }

\begin{figure*}[htb!]
\resizebox{7.9cm}{!}{\includegraphics{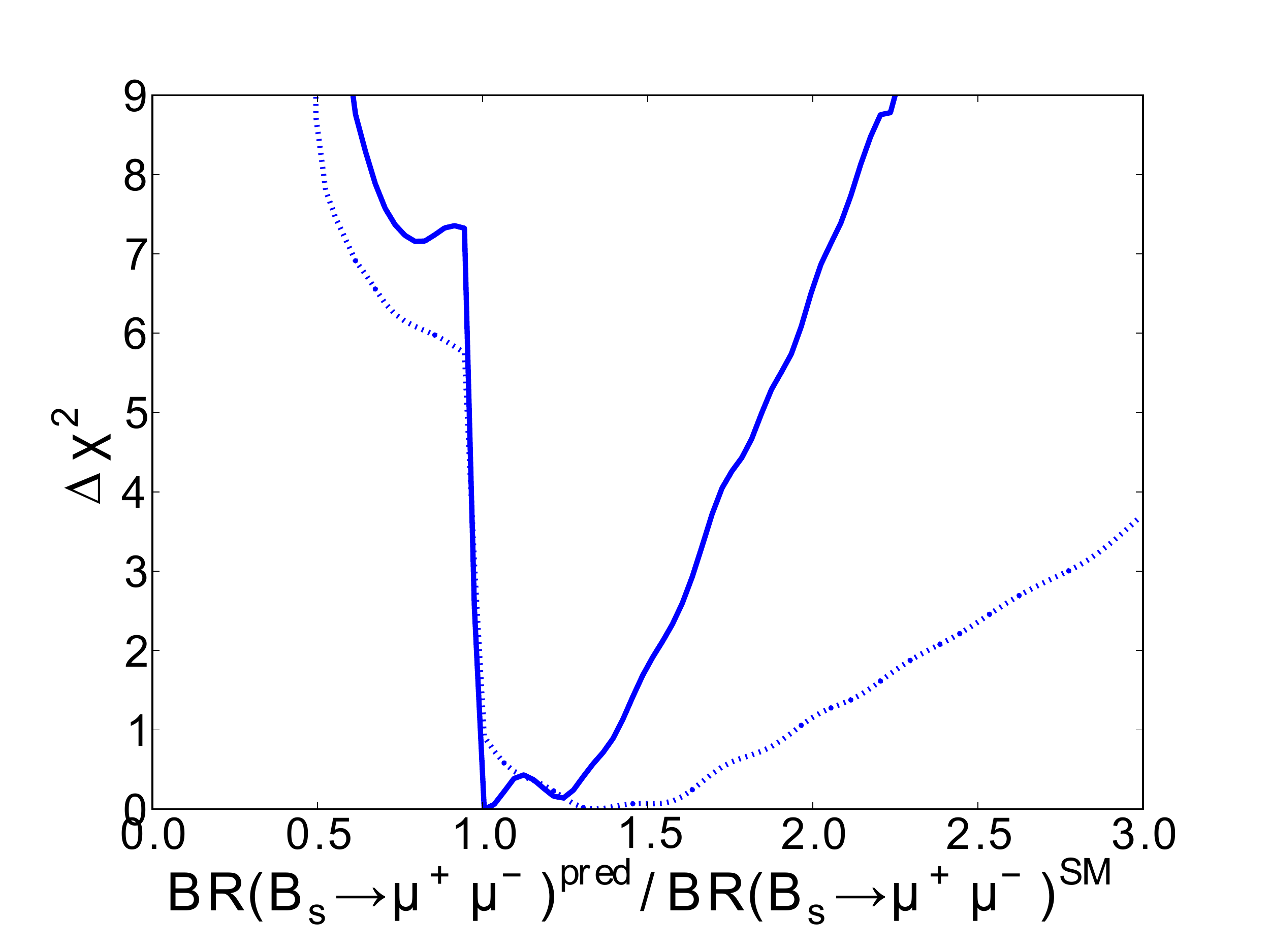}}
\resizebox{7.9cm}{!}{\includegraphics{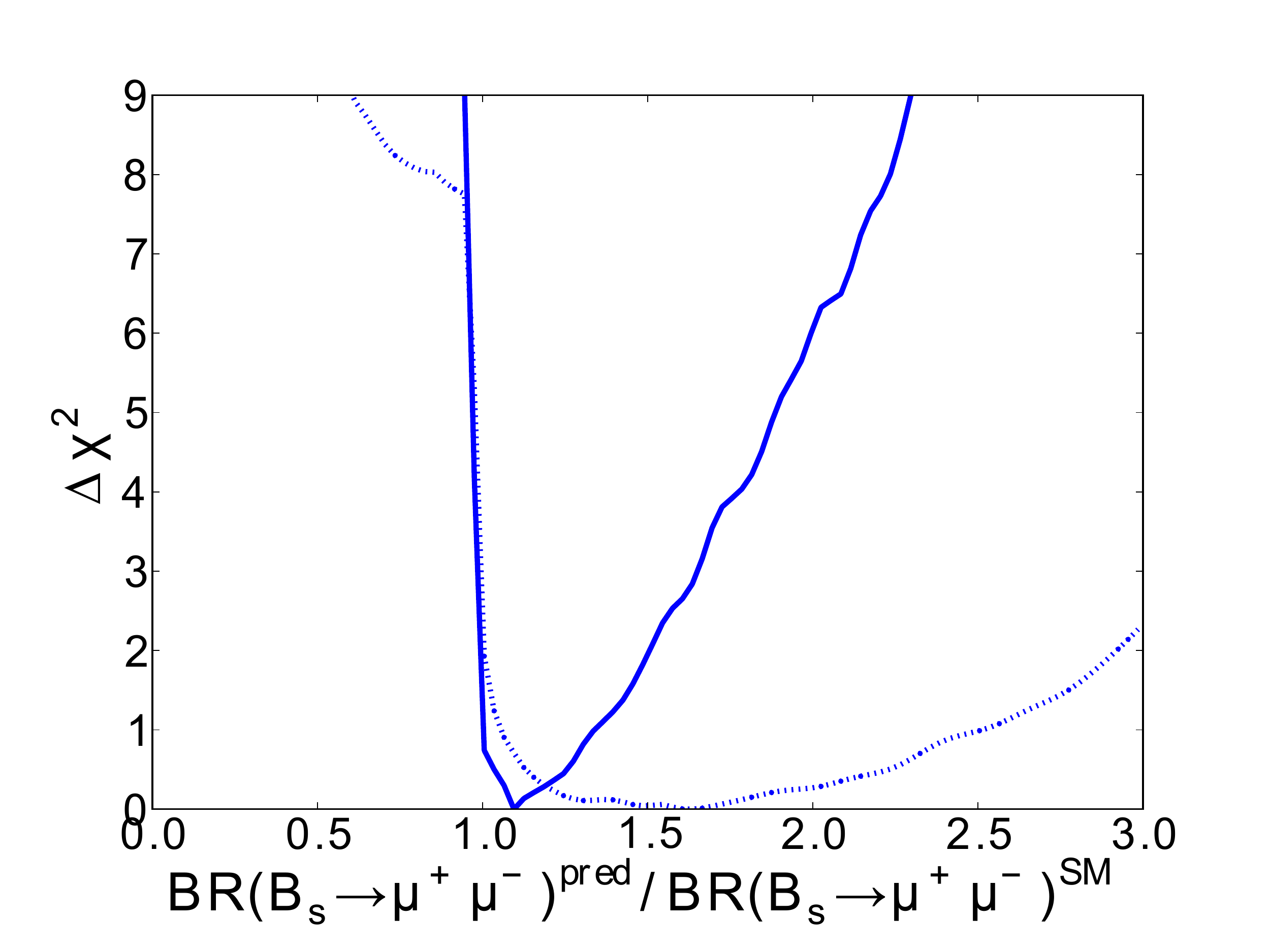}}
\caption{\it The $\chi^2$ likelihoods in the CMSSM (left panel) and
  NUHM1 (right panel) as functions of   \bmm\ based on global fits to the
  \lhcf, \bmm\ and new XENON100 data set (solid lines), and to the \lhco\ data set (dashed
  lines).} 
\label{fig:bmm}
\end{figure*}

Fig.~\ref{fig:bmm} displays the $\chi^2$ functions for \bmm\ in the
CMSSM (left panel) and the NUHM1 (right panel), for fits to the
\lhcf, \bmm\ and new XENON100 data set (solid lines) and the \lhco\ data (dashed lines), 
as in Fig.~\ref{fig:mgl}. Comparing the solid and dotted lines,
we see the strong impact of the recent \bmm\ update. In both the
CMSSM and the NUHM1, values of \bmm\ above the SM value are
preferred over values below the SM value. We see that, in either case, a
measurement of \bmm\ at the level of 30\%, which may soon be possible,
would greatly restrict the preferred ranges of both the CMSSM and NUHM1
parameter spaces.

\subsection*{\it \ssi\ }

\begin{figure*}[htb!]
\resizebox{8cm}{!}{\includegraphics{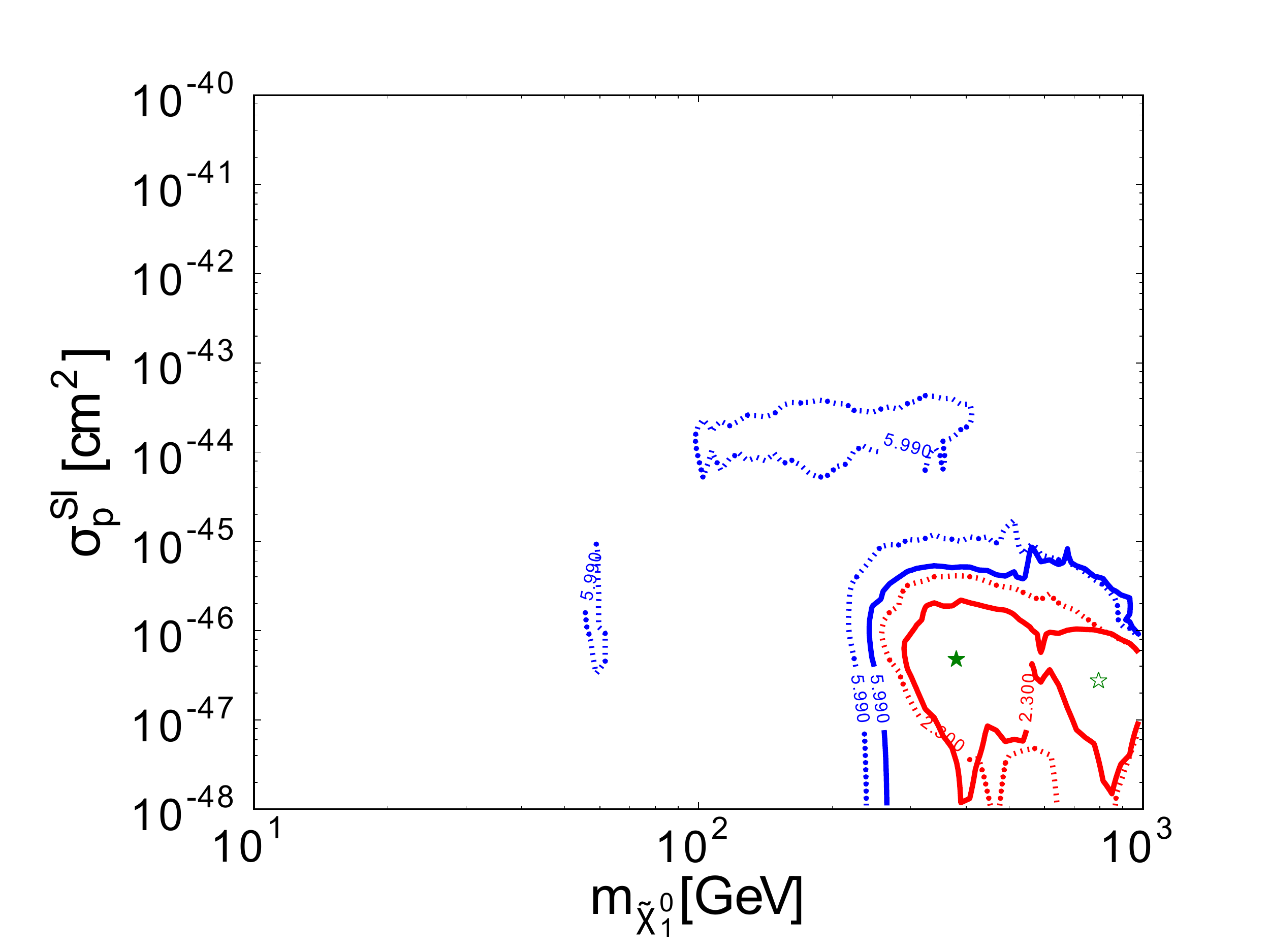}}
\resizebox{8cm}{!}{\includegraphics{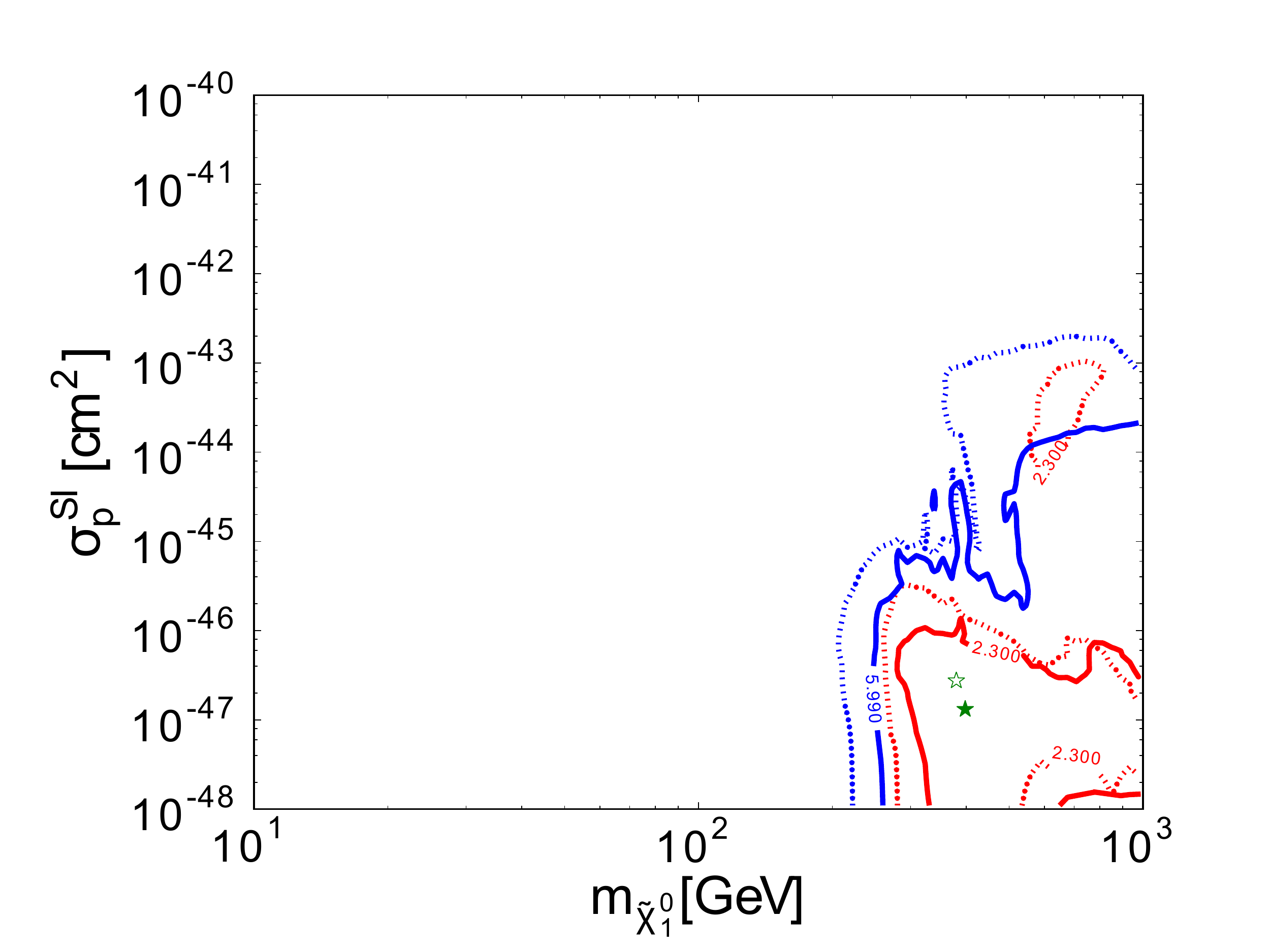}}
\vspace{-1cm}
\caption{\it The $(\mneu{1}, \ssi)$ planes in the CMSSM (left panel) and the NUHM1
(right panel). The $\Delta \chi^2 = 2.30 (5.99)$ contours, corresponding to the 68(95)\%~CL
are coloured red (blue). The solid (dashed) lines are for global fits to the \lhcf, \bmm\ and new XENON100 (\lhco) data, and
the corresponding best-fit points are indicated by solid (open) green stars.}
\label{fig:mchissi}
\end{figure*}

Fig.~\ref{fig:mchissi} displays the best-fit points and the 68\% and
95\%~CL contours (red and blue, respectively) in the $(\mneu{1}, \ssi)$
planes in the CMSSM (left panel) and and the NUHM1 (right panel). The
solid lines are for the global fit to the \lhcf\, \bmm\ and new XENON100 data set, and the
dashed lines for the previous \lhco\ fit, and the solid (open) green
stars mark the corresponding best-fit points~\footnote{We display results for
$\mneu{1} \le 1 \tev$ only, because results from XENON100 are not published for
larger masses.}. We see that the
region of the $(\mneu{1}, \ssi)$ favoured in the CMSSM is now
more restricted, in particular at small $\mneu{1}$.
The main effect of the new XENON100 constraint has been to remove focus points with
large \ssi\ that were previously allowed at the 95\% CL.

\section{Comparison with other Analyses}

We are not aware of any global analyses incorporating the new XENON100 constraint~\cite{newXENON100}.
The two global analyses including LHC 5/fb data that are most similar to ours
are~\cite{Fittino5} and~\cite{BayesFits5}, and we take this opportunity to
comment on the relations and differences between those and the present work.

As in this work, Ref.~\cite{Fittino5} incorporates the constraints imposed by the
LHC$_{\rm 5/fb}$ data using an implementation of \delphes\ that is reported to reproduce
the 95\%~CL limits published previously by ATLAS using 165/pb and 1/fb of data and
shown to reproduce approximately the 95\%~CL limits for $\tb = 10$ and $A_0$ found in the ATLAS analysis
used here and the CMS razor
analyss with 5/fb of data~\cite{CMS5}. However, the \atlasfive\ analysis has apparently not been modelled
in detail, and full information is not available on its validation for other values of the CMSSM parameters, nor on its
extension to the NUHM1. In contrast to previous papers by the same collaboration that used the
{\tt MasterCode}, Ref.~\cite{Fittino5} uses only a reduced set of precision electroweak and
flavour observables, a topic we comment on below. In contrast, Ref.~\cite{Fittino5} uses
{\tt AstroFit}, which provides input from both direct and indirect dark matter searches, whereas
we use only the direct XENON100 search. However, Ref.~\cite{Fittino5} finds that the indirect
dark matter searches have negligible impact at the present time. When the constraint
$\Mh = 126 \pm 2 \pm 3 \gev$ is imposed in~\cite{Fittino5}, the results for the CMSSM are relatively similar
to ours, whereas the results for the NUHM1 are only qualitatively similar.

Ref.~\cite{BayesFits5} implements the CMS razor analysis using 5/fb of data,
and provides more details of its efficiency and likelihood maps, and also compares
with the 95\%~CL limit obtained in the \atlasfive\ analysis. This paper favours values of
the CMSSM mass parameters $(m_0, m_{1/2})$ that are considerably larger than in 
our analysis and in~\cite{Fittino5}. However, Ref.~\cite{BayesFits5} does favour the stau 
coannihilation strip and the rapid-annihilation funnel regions of the CMSSM, disfavouring
the focus-point region that had been advocated in previous fits by the same group.
Ref.~\cite{BayesFits5} does not discuss the NUHM1, but does consider the possibilities
that $\mu < 0$ and of dropping the \gmt\ constraint. 

\section{Conclusions}

We have presented in this paper new global fits to the CMSSM and NUHM1
based on the 7 TeV \atlasfive\ constraint and the latest global combination of
limits on \bmm, the new XENON100 constraint, a Higgs mass measurement $\Mh \sim 125 \gev$,
and minor updates for other observables. 
Before the advent of the LHC, low-energy data, in
particular \gmt, gave hope that the scales of supersymmetry breaking in
these models might be within reach of the LHC within its first year of
operation at 7~TeV in the centre of mass. This 
has not been the case, and the $p$-values of the CMSSM and NUHM1
have declined to 8.5 and 9.1\%, respectively, prompting a re-evaluation of
their status. 

Two possibilities remain open at this point: either supersymmetry is relatively
light but manifests itself differently from the CMSSM or NUHM1, or the
supersymmetry-breaking mass scale is heavier, possibly even beyond the
reach of the LHC. In the first case, several possibilities suggest
themselves (see, e.g., Ref.~\cite{AbdusSalam:2011fc}):
other models of supersymmetry breaking, light third-generation squarks,
${\cal F}$-SU(5)~\cite{LMNW5}, 
the pMSSM, R-violating models, etc., and a discussion of these possibilities
lies beyond the scope of this work. As for the second case, the global fits
presented here in the CMSSM and NUHM1 give limited guidance on the
possible supersymmetric mass scale.

Perhaps surprisingly, the \atlasfive\ data are not much more problematic for the
CMSSM and the NUHM1 than were the \lhco\ data in combination with
  the Higgs mass constraint. The latter two had already
constrained the CMSSM and NUHM1 parameters to lie beyond the range
where they could fully `solve' the \gmt\ problem, and the contribution of the
\atlasfive\ constraint to the global $\chi^2$ function is not large, even at the
local best-fit point within the low-mass CMSSM `island'. A new feature
of this analysis is the impact of the latest \bmm\ constraint,
and we stress the importance of the constraint from the search for
spin-independent dark matter scattering by the XENON100 Collaboration~\cite{newXENON100}.

The most important upper limit on the supersymmetry-breaking mass
scales in the CMSSM and the NUHM1 is now provided by the cold
dark matter density constraint, and indications on the possible sparticle
masses within the range allowed by this constraint are now quite weak.
The \gmt\ constraint still offers some preference for low sparticle masses,
but this is largely counterbalanced by other constraints such as \bmm.
As we have seen, any value of $\mgl \in (1500, 5000) \gev$ is allowed
at the $\Delta \chi^2 < 2$ level. The lower part of this range may be
explored with LHC operating at 8~TeV, and the middle part with the LHC
at 13 or 14~TeV, but the upper part of the allowed range of $\mgl$
would lie beyond the reach of the LHC.

On the other hand, one must keep in mind that the increasing
tension within the CMSSM and NUHM1 between the low-energy data, mainly
\gmt\, and the \lhcf\ and \bmm\ data on the other side, has resulted
over time in a monotonously-increasing $\chi^2$ and a correspondingly
lower fit probability. Consequently, looking beyond the CMSSM and NUHM1,
to models with a different connection between  the coloured and
uncoloured sector, not only seems timely now, but mandatory.
We expect that some model with such a different connection will 
give a substantially better fit to {\em all} existing experimental data
and present more favorable expectations for subsequent LHC runs or a future LC.


\subsubsection*{Acknowledgements}

The work of O.B., M.C., J.E., J.M., S.N., K.A.O.\ and K.J.de V.\ is
supported in part by the London Centre for Terauniverse Studies (LCTS),
using funding from the European Research Council 
via the Advanced Investigator Grant 267352. 
The work of S.H.\ is supported 
in part by CICYT (grant FPA 2010--22163-C02-01) and by the
Spanish MICINN's Consolider-Ingenio 2010 Program under grant MultiDark
CSD2009-00064. The work of K.A.O.\ is supported in part by DOE grant
DE-FG02-94ER-40823 at the University of Minnesota. We thank Robert Fleischer
for discussions.


\end{document}